\documentclass[12pt]{article}
\usepackage{relsize}
\usepackage{array}
\usepackage{hyperref}
\usepackage{fullpage}
\usepackage{graphicx}
\usepackage{psfrag}
\usepackage{epsf}
\usepackage{amsfonts}
\usepackage{pstricks}
\usepackage{color}
\usepackage{setspace}
\usepackage{srcltx}
\usepackage[novbox]{pdfsync}
\input epsf.sty
\usepackage{amsmath}
\usepackage{amssymb}
\usepackage{graphicx}

\def\timesbox{\hbox{$\scriptscriptstyle\times$}}
\def\ant{ {{\lower 1ex  \timesbox} \atop {\raise 1.5ex  \timesbox}}}
\def\f{\frac}

\def\non{\nonumber\\}
\def\o{\omega}

\def\t{\tau}

\newcommand{\mathsym}[1]{{}}
\newcommand{\unicode}[1]{{}}

\newcommand {\expect}[1]{\left\langle #1 \right\rangle}

\newcommand\ZZZ{{\hbox{ Z\kern-1.6mm Z}}}
\newcommand{\Iop}{\relax{\rm I\kern-.18em I}}
\newcommand{\Lop}{\relax{\rm I\kern-.18em L}}
\newcommand{\dop}{\relax{\rm I\kern-.8em d}}
\newcommand{\one}{{\hbox{ 1\kern-1.2mm l}}}

\newcommand{\be}{\begin{equation}}
\newcommand{\ee}{\end{equation}}
\newcommand{\beqa}{\begin{eqnarray}}
\newcommand{\eeqa}{\end{eqnarray}}
\newcommand{\bsp}{\begin{split}}
\newcommand{\esp}{\end{split}}
\newcommand{\bgth}{\begin{gather}}
\newcommand{\egth}{\end{gather}}

\newcommand{\hf}{{1\over 2}}

\begin{document}

\title{
\begin{flushright}
\small{IMSc/2014/3/1}
\end{flushright}
{\textbf{BCS Instability and Finite Temperature Corrections to Tachyon Mass in Intersecting $D1$-Branes}}
\author{Sudipto Paul Chowdhury$^{a}$\footnote{sudiptopc@imsc.res.in}, 
Swarnendu Sarkar$^{b}\footnote{ssarkar@physics.du.ac.in}$,\\  and B. Sathiapalan$^{a}\footnote{bala@imsc.res.in}$\\
$^a$\small{{\em The Institute of Mathematical Sciences,
Taramani,}}\\  
\small{{\em Chennai 600113, India }} \\
$^b$\small{{\em Department of Physics and Astrophysics,
University of Delhi,}} \\ 
\small{{\em Delhi 110007, India}}\\}}
\maketitle

\abstract{ A holographic description of BCS superconductivity is given in  
{\cite{KalyanaRama:2011ny}}. This model was constructed by insertion of 
a pair of $D8$-branes on a $D4$-background. The spectrum of intersecting $D8$-branes has tachyonic 
modes indicating an instability which is identified with the 
BCS instability in superconductors. Our aim is to study the stability of the intersecting branes 
under finite temperature effects.
Many of the technical aspects of this problem are captured by a simpler problem of two 
intersecting $D1$-branes on flat background. 
In the simplified set-up we compute the one-loop finite temperature corrections to the 
tree-level tachyon mass-squared-squared using the 
frame-work of SU(2) Yang-Mills theory in $(1+1)$-dimensions. We show that the one-loop 
two-point functions are
ultraviolet finite due to cancellation of ultraviolet divergence between the amplitudes 
containing bosons and fermions in the loop. 
The amplitudes are found to be infrared divergent
due to the presence of massless fields in the loops.
We compute the finite temperature mass-squared correction to all the massless fields and use 
these temperature dependent masses-squared to compute the tachyonic 
mass-squared correction. We show numerically the existence of a transition temperature at 
which the effective 
mass-squared of the tree-level tachyons becomes zero, thereby stabilizing the brane configuration.} 
\newpage

\tableofcontents

\baselineskip=18pt

\section{Introduction}\label{intro}
There have been many applications of the AdS/CFT correspondence to understand condensed matter systems. 
When there are gapless modes present these systems are described by conformal field theories at 
low energies. 
This can happen at second order phase transitions, but also in metals 
where the excitations above the Fermi surface are gapless. For recent developments see \cite{LMV}-\cite{DHS}.

At low temperatures metals are unstable towards electron Cooper pair formation and an energy gap develops. 
This is the BCS instability. The Cooper pairs are charged
and so the condensate breaks the $U(1)$ of electromagnetism and the photon effectively becomes massive as a 
result of the Higgs phenomenon. The energy gap ensures
that at low frequencies there is no dissipation of energy when a current flows. The mass of the 
photon results 
in the exponential fall off 
of the magnetic field inside a superconductor. These are typical characteristics of superconductors. 
Studies of various types of superconductors using holographic techniques have been been the 
subject of research 
for the past few years. A partial list of references is \cite{G}-\cite{HHH}.

Inspired by this BCS phenomenon Nambu and Jona-Lasinio gave a description of chiral symmetry breaking 
in strong interactions \cite{Nambu}.
Their starting point was a non renormalizable model with four Fermi interactions.
The pairing between quarks and anti-quarks is analogous to Cooper pairing. The main point of difference 
(as summarized in \cite{KalyanaRama:2011ny}) 
is that due to the absence of a Fermi surface this instability in QCD happens only for large 
enough coupling. 
Another point of difference is that the resultant condensate breaks an 
axial symmetry  (rather than a vector symmetry as in BCS)- the $U(1)$ chiral symmetry that is present in QCD 
(in the absence of bare mass to quarks).

A holographic dual of 3+1 QCD was constructed in \cite{EW1} starting from M theory on $AdS_7\times S^4$. 
Many interesting calculations have been done 
with this model including calculation of the glueball mass spectrum - albeit at strong bare coupling 
\cite{KK}-\cite{AHJK}.
An extension of this model to include flavor degrees of freedom was constructed by Sakai and Sugimoto 
\cite{SS1}. Various aspects of this model was further explored in \cite{Hata}-\cite{Dhar}. 
The flavour branes are D8 branes hanging down from the boundary (where they intersect the D4 branes) 
and are wrapped on $S^4$. 
It was shown that when there are D8 branes and D8 anti branes, a stable configuration is described by 
the brane and anti-brane  
bending towards each other and joining to form a continuous U-shaped brane. Since the branes describe 
left handed quarks and anti-branes 
describe right handed quarks (or left handed anti quarks) this U configuration breaks chiral symmetry.

In \cite{KalyanaRama:2011ny} the Sakai-Sugimoto model was modified to describe BCS superconductivity. 
The Sakai Sugimoto model has unbroken vector like symmetries 
corresponding to the flavour group. Thus for two flavours there is a $U(2)$. In \cite{KalyanaRama:2011ny} 
it was shown that
in the presence of a finite chemical potential for the $U(1)$ embedded in $SU(2)$, a D8 brane and an anti 
D8 brane cross each other. 
Such a configuration is known to be tachyonic and it has been argued that the stable configuration to which 
this flows has a non zero
charged condensate that Higgses the $U(1)$ symmetry 
\cite{Hashimoto:2003xz,Hashimoto:2003pu,Epple:2003xt,BDL,HT,N,Jokela1,Jokela2}. In \cite{KalyanaRama:2011ny} 
analytical solutions were given for 
such systems by solving the Yang-Mills equations describing intersecting branes in flat space-time. 
Semi analytic and numerical solutions were also given in the 
curved background of this modified Sakai-Sugimoto model.

Being a strongly coupled system the expression for the gap in terms of the coupling and other parameters 
is different from weak coupling BCS. The gap and thus 
the transition temperature are expected to be larger here. For weak coupling the relation is
$\Delta \approx \epsilon_ce^{-{1\over g {dn\over d\epsilon}}}$, whereas for strong coupling one expects 
$\Delta \approx \epsilon_c g {dn\over d\epsilon}$. Here $\epsilon_c$ is some parameter of the metal 
that fixes the region around the Fermi surface that participates
in the pair formation.

In this paper we attempt to calculate the transition temperature for the model described in 
\cite{KalyanaRama:2011ny}. We do this in flat space-time for  simplicity.   The low energy theory on 
the brane can be described by the DBI action for the massless fields on the brane.
This is valid as long as only energies $<<{1\over \alpha'}$ are being probed. The DBI action describes 
the effect of "integrating out classically" (i.e. via equations of motion) the massive modes of the string. 
However even if the massive modes are integrated out in the quantum theory, the resultant action for the 
massless modes would look like supersymmetric Yang-Mills corrected by higher dimensional operators down by 
powers of $\alpha'$ - very similar to the DBI action. 

We can study this as a quantum theory with a cutoff $\Lambda < {1\over \sqrt \alpha'}$ and proceed to study 
the  corrections due to the massless mode quantum {\em and} thermal fluctuations. Since the Yang-Mills action
is renormalizable, we know that the effect of the irrelevant higher dimension operators is to make finite 
renormalizations of the lower dimensional operators. This is just a re-phrasing of the decoupling theorem: 
If the low energy theory is renormalizable the massive  modes decouple and further the ambiguities associated 
with the physics at and above the cutoff scale can be absorbed into a renormalization of the parameters. 
While this is a consistent procedure, this is not good enough for us because if we want to estimate the 
finite thermal corrections to the action, the finite part should be unambiguous and cannot be the 
finite part of an infinite term. Fortunately, because of supersymmetry, the mass-squared corrections are 
finite and therefore calculable in principle. More precisely if we calculate the corrections as a 
power series in $\Lambda$ (with $\Lambda << {1\over \sqrt \alpha'}$), one expects that terms that diverge 
when $\Lambda \to \infty$ (i.e. positive powers and logarithms) are absent. Thus all corrections are finite 
and at most of order ${E\over \Lambda} < \sqrt {\alpha'} E$ where $E$ is a typical energy scale. 
Supersymmetry in fact can ensure this even if the theory is not renormalizable as in $Dp$-branes with $p>3$. 

The physical quantity we are interested is the temperature correction to the tachyon mass-squared. 
The tachyon mass-squared is $O({\theta\over \alpha'}) =- q$, where $\theta$ is the angle of intersection 
of the $D$-branes and $q$ is defined more precisely later.  Thus we would like to keep this finite 
as ${1\over \alpha'}\to \infty$. This can be achieved by taking the limit 
$\theta \to 0, {1\over \alpha'}\to \infty$ such that $q$ is fixed. Thus in this limit we can use 
the supersymmetric Yang-Mills theory on the brane. We also simplify further the problem
by studying $D1$ branes. Classically the solutions we are considering  depend only on one coordinate. 
So the solutions are the same
for all $Dp$ branes. Quantum mechanically the fluctuations will be different and the momentum 
integrals in Feynman diagrams will be different. However many of the techniques used for $D1$ branes 
should go through since the mass-squared corrections are finite due to supersymmetry. 

Even with these simplifications the calculations are already quite involved. The main reason is that
the background configuration about which quantum corrections need to be calculated is space dependent. 
The intersecting $D$-brane configuration is described by one of the adjoint scalars having a value that 
is linear in $x$, in the form $\phi = qx$ . Thus in the $x$-direction one cannot 
use plane waves as a basis. One has to work with eigenfunctions that are essentially harmonic oscillator 
wave functions. One should then calculate the effective potential 
at finite temperature and then obtain the transition temperature. This is a rather difficult calculation. 
In this paper as a first step we adopt the simpler 
procedure of calculating the corrections to the tachyon mass-squared and finding the temperature at which 
this turns positive. This is not the same thing because positive $(mass)^2$ only ensures local stability.  
In any case with these simplifications the calculation becomes tractable. Even so, some of the calculations 
have to be done numerically. 

There are two technical issues that become complicated because of using Hermite polynomials instead of 
plane waves. One is that of showing UV finiteness. 
As mentioned above, if the theory has divergent mass-squared corrections that need to be renormalized 
then one cannot {\em calculate} the transition temperature from first principles  
because there is always an arbitrary parameter corresponding to finite mass renormalization. 
Especially when calculations have to be done numerically,  one needs to be sure that the series being 
summed is convergent. This demonstration is made difficult, 
once again because we cannot use a plane wave basis.  In this paper we show  UV finiteness at one loop.  
This check is also useful because it ensures that the degrees of freedom counting has been done correctly.

The second complication is that there are many massless modes. This results in infrared divergences. 
The correct solution to this problem is to use a 
renormalization group and integrate out high momentum modes first. This should typically induce 
mass-squared corrections to the massless modes (unless they are 
protected) and the final solution to the RG equations should not have any IR divergences. The full RG 
is difficult to implement. However what can be done 
is to first do a one loop integral where the internal lines have only  modes that do not generate 
IR divergences. In this step one can calculate the mass-squared 
correction to all the massless modes. At the next step one includes the remaining unintegrated modes in 
the tachyon mass-squared correction, with corrected massive propagators and now, because there are no 
massless modes, there are no 
IR divergences. \footnote{At this point one is going beyond the one loop approximation and one has in 
effect summed an infinite number of diagrams.} 
In practice one needs to calculate mass-squared corrections only for those modes that are needed for the 
tachyon mass-squared correction. Thus this procedure 
takes care of the infrared divergences. 

Once these problems are taken care of one can proceed to a calculation of the finite temperature 
correction to the tachyon mass-squared.
Both, the temperature independent finite quantum correction to the masses-squared and the temperature 
dependent 
finite corrections are calculated. However in the final calculation of the tachyon mass-squared, which is 
done numerically, only the total correction is calculated and plotted. Thus we are able to calculate 
numerically the transition temperature at which the tachyon becomes massive.
 
As mentioned above this calculation needs to be generalized to higher branes. Also instead of calculating 
the correction to the mass-squared
one should do a more complete calculation and calculate the effective potential. Finally one should 
generalize to curved space time.
This last may not however be very important because most of the dynamics takes place locally at the 
intersection point and one should be able to 
make a simple extrapolation to curved space locally using the equivalence principle. 

This paper is organized as follows. Section (\ref{spectrum}) is dedicated to the study of mass spectrum 
of intersecting $D1$-branes at zero temperature
in the Yang-Mills approximation. We choose a background given by the vev of one of the scalar field 
components.
We compute the normalizable eigenfunctions for all the bosonic fields. Those fields which couple to 
the chosen background become massive at the tree-level
and have a discrete mass spectrum. Those fields which do not couple to the background remain massless 
with a continuous spectrum.
Apart from the massive modes there are also massless modes in the expansion of some of the bosonic fields 
which are accompanied with eigenfunctions with zero mass 
eigenvalue. The lowest lying modes in the bosonic mass spectrum are the tachyons.   
In section (\ref{finiteT}) we present the finite temperature analysis with a single scalar field using 
background field method.
In section (\ref{d1}) we present the finite temperature analysis for the intersecting $D1$-branes. 
In section (\ref{d1b}) we present the computations for 
the one-loop bosonic corrections to the tree-level tachyon mass-squared at finite temperature. The 
fermionic eigenfunctions and their contributions to the one-loop 
corrections are presented in section (\ref{D1ferm}). In section (\ref{uvir}) we discuss the problems 
of ultraviolet and infrared divergences of the amplitudes
which arise due to the presence of massless fields in the loop. While the amplitudes containing bosons 
in the loop are both IR and UV divergent the amplitudes containing fermions in the loop are only UV 
divergent. 
In section (\ref{uvtach}) we present the computations 
for the cancellation of the UV divergences of the amplitude for the tachyonic mode. To tackle the IR problem 
we first compute the 
one-loop corrections to all the massless fields. Thus we have all massive fields at one-loop level 
which then allows us to compute the finite two-point functions 
for the tachyonic modes. The one-loop mass-squared corrections to all the massless fields at finite 
temperature 
are computed in the sections (\ref{zeroamp}) 
(\ref{phi1amp}),(\ref{phiIamp}) and (\ref{A3xamp}). We also discuss their UV finiteness. In section 
(\ref{numerics2}) we present the numerical computation of the 
transition temperature and an analytical estimate of the behaviour of the masses-squared with varying 
temperature. 
For large values of temperature, the masses-squared 
are found to grow linearly with temperature. We present relevant details of the computation 
in the appendices.

\section{Tree-level spectrum}\label{spectrum}

In this section we study the classical mass spectrum for an intersecting $D1$ brane configuration at 
zero temperature in the Yang-Mills approximation.
The action for two coincident $D1$ branes with gauge group $SU(2)$ has been worked out in 
appendix \ref{dimr}. The action (\ref{2action}) is the dimensional reduction 
of $10$ dimensional ${\cal N}=1$ Super Yang-Mills. 
The equations of motion in $1+1$ dimensions have a solution $\Phi^3_1=qx$ and $A=0$. This corresponds 
to an intersecting brane configuration with slope $q$. 
Considering this background solution a fluctuation analysis was done in \cite{Hashimoto:2003xz} 
to analyze the spectrum of the theory. To review this analysis we 
first write down 
the relevant bosonic part of the action (\ref{2action}).

\beqa
\label{act1}
S^1_{1+1}&=&\f{1}{g^2}\mbox{tr} \int d^2x \left[-\f{1}{2}F_{\mu\nu}F^{\mu\nu}
+ D_{\mu}\Phi_I D^{\mu}\Phi_I+\f{1}{2}\left[\Phi_I,\Phi_J\right]^2\right]
\eeqa

The Bosonic Lagrangian up to the quadratic order in fluctuations separates into various decoupled sets. 
In the $A^a_0=0 \mbox{~} (a=1,2,3)$ gauge we write the 
decoupled parts separately below.

\be
\label{Lagrangian1}
\begin{split}
{\cal L}(A_x^2,\Phi_1^1)=&-\hf A_x^2 \partial_0^2 A_x^2-\hf \Phi_1^1  \partial_0^2  \Phi_1^1
+\hf  \Phi_1^1   \partial_x^2 \Phi_1^1-q^2 x^2\hf (A_x^2)^2\\
&+ q A_x^2 \Phi_1^1-qx \partial_x \Phi_1^1 A_x^2.
\end{split}
\ee

\be
\label{Lagrangian2}
\begin{split}
{\cal L}(A_x^1,\Phi_1^2)=&-\hf A_x^1 \partial_0^2 A_x^1-\hf \Phi_1^2  \partial_0^2  \Phi_1^2
+\hf  \Phi_1^2   \partial_x^2 \Phi_1^2-q^2 x^2 \hf (A_x^1)^2\\
&-q A_x^1 \Phi_1^2+qx \partial_x \Phi_1^2 A_x^1
\end{split}
\ee

\be
\label{Lagrangian3}
\begin{split}
{\cal L}(\Phi_I,A_x^3)=&-\hf \Phi_I^a \partial_0^2  \Phi_I^a+\hf  \Phi_I^a   
\partial_x^2 \Phi_I^a-\hf q^2 x^2 (\Phi_I^1)^2-\hf q^2 x^2 (\Phi_I^2)^2\\
&-\hf A_x^3 \partial_0^2  A_x^3 \mbox{~~~~~(for all $I\ne1$)}
\end{split}
\ee

We thus have various decoupled sets of equations at this quadratic order. 
The solutions of these equations give the wave functions corresponding to the normal modes.
The first two terms ${\cal L}(A_x^2,\Phi_1^1)$ and ${\cal L}(A_x^1,\Phi_1^2)$ implies that we have 
a two coupled sets of equations for $(A_x^2,\Phi_1^1)$ and $(A_x^1,\Phi_1^2)$.

To compute eigenfunctions and the spectrum we first consider the equations for $(A_x^2,\Phi_1^1)$ fields. 
The eigenvalue equation for the spatial part is,

\beqa
\label{eigenvalue1}
\left(\begin{array}{cc}
m^2-q^2x^2&-qx\partial_x +q\\
2q+qx\partial_x&m^2+\partial_x^2\end{array}
\right)\left(\begin{array}{c}A_x^2\\ \Phi_1^1\end{array}\right)=0
\eeqa

where $m^2$ is the eigenvalue. Eliminating, $A_x^2$ from the above equations gives the following equation 
for $\Phi_1^1$,

\beqa\label{ph1}
\partial_x^2\Phi_1^1+\left[P(x)-Q(x)\right]\Phi_1^1-x P(x)\partial_x\Phi_1^1=0
\eeqa

where,

\beqa
P(x)=\f{2q^2}{Q(x)}~~~;~~~Q(x)=q^2x^2-m^2
\eeqa

The asymptotic $( x\rightarrow \infty)$ form of the equation (\ref{ph1}) is,

\beqa
\left[\partial_x^2+m^2-q^2x^2\right]\Phi_1^1=0
\eeqa

which is the Schroedinger's equation for a harmonic oscillator. The ground state wave function is 
$e^{-qx^2/2}$.
Thus writing,

\beqa
A_x^2= e^{-qx^2/2}A ~~~;~~~\Phi_1^1=e^{-qx^2/2}\phi
\eeqa

we get the following equations,

\beqa
\label{eigenvalue2}
&&(m^2-q^2x^2)A+(q+q^2x^2-qx\partial_x)\phi=0\\
&&(-q^2x^2+qx\partial_x+2q)A+(m^2-q+q^2x^2-2qx\partial_x +\partial_x^2)\phi=0
\eeqa

Now assuming series solution of the form,

\beqa\label{ser}
A=\sum_k a_k (x\sqrt{q})^k ~~~;~~~ \phi=\sum_k b_k (x\sqrt{q})^k
\eeqa

we get the following recursion relations,

\beqa\label{r1}
\f{\left[(2k-1)-\f{m^2}{q}\right]}{\left[k-\f{m^2}{q}\right]}b_k-
\f{(k+1)(k+2)}{\left[(k+2)-\f{m^2}{q}\right]}b_{k+2}=0
\eeqa

and

\beqa\label{r2}
a_k=b_k-\f{(k+1)(k+2)}{\left[(k+2)-\f{m^2}{q}\right]}b_{k+2}
\eeqa

The quantization condition on $m^2$ is obtained by demanding that the series (\ref{ser}) 
terminates for some value of $k$ that is $n$. This implies
that the numerator of the first term of (\ref{r1}) vanishes for this value of $k$. 
We thus have the spectrum given by $m^2=(2n-1)q$. The lowest mode of mass spectrum given by $n=0$ 
is tachyonic. 
To solve for the eigenfunctions 
we simply need to compute the various coefficients $(a_k, b_k)$ using the recursion relations.  

For $n=0,2,4 \cdots$,

\begin{gather}
\label{eveneignfunc}
a_k=\f{(-1)^{k/2} 2^{k/2}}{(2n-1) k!}n (n-2) \cdots (n-k+2)(k-1)\\
b_k=\f{(-1)^{k/2} 2^{k/2}}{(2n-1) k!}n (n-2) \cdots (n-k+2)(2n-k-1)
\end{gather}

For $n=3,5,7 \cdots$,

\begin{gather}
\label{oddeignfunc}
a_k=\f{(-1)^{(k-1)/2} 2^{(k-1)/2}}{(2n-1) k!} (n-1)(n-3) \cdots (n-k+2)(k-1)\\
b_k=\f{(-1)^{(k-1)/2} 2^{(k-1)/2}}{2(n-1) k!} (n-1)(n-3) \cdots (n-k+2)(2n-k-1)
\end{gather}

Putting these back in (\ref{ser}), the eigenfunctions are,

\beqa\label{serf}
A_n=\sum_{k\le n}a_k (x\sqrt{q})^k ~~~;~~~ \phi_n=\sum_{k\le n} b_k (x\sqrt{q})^k
\eeqa

The normalized eigenfunctions $\{ A_n(x), \phi_n(x)\}$ for both odd and even $n$ can be combined into 
the following expressions

\begin{gather}
\label{bosegnfunc}
A_n(x) =   {\cal N}(n)e^{- q x^2/2} \left(H_n (\sqrt{q} x) + 2 n H_{n-2} (\sqrt{q} x) \right) \\
\phi_n(x) = {\cal N}(n)e^{- q x^2/2} \left(H_n (\sqrt{q} x) - 2 n H_{n-2} (\sqrt{q} x) \right)
\end{gather} 

where $H_n(\sqrt{q}x)$ are Hermite polynomials and the normalization,\\ 
${\cal N}(n)=1/\sqrt{\sqrt{\pi} 2^n (4 n^2-2n) (n-2)!}$. Let us define,

\beqa\label{cfunctions}
\zeta_n(x)=\left(\begin{array}{c}
A_n(x)\\
\phi_n(x)
\end{array}\right).
\eeqa

$\zeta_n(x)$ then satisfies the orthogonality condition

\beqa\label{ortho1}
\sqrt{q}\int dx \zeta^{\dagger}_n(x)\zeta_{n^{'}}(x)=\delta_{n,n^{'}}.
\eeqa

There is also a set of infinitely many degenerate eigenfunctions with $m_n^2=0$. The 
normalized eigenfunctions for the zero eigenvalues can also be written in 
terms of Hermite polynomials as

\begin{gather}
\label{zerobosegnfunc}
\tilde{A}_n(x) =  \tilde{{\cal N}}(n) e^{- q x^2/2} \left(H_n (\sqrt{q} x) - 2 (n-1) H_{n-2} (\sqrt{q} x) \right) \\
\tilde{\phi}_n(x) = \tilde{{\cal N}}(n) e^{- q x^2/2} \left(H_n (\sqrt{q} x) + 2 (n-1) H_{n-2} (\sqrt{q} x) \right)
\end{gather}

where the normalization, $\tilde{{\cal N}}(n)=1/\sqrt{\sqrt{\pi} 2^n (4 n-2)(n-1)!}$. 
We define a different set,

\beqa
\tilde{\zeta}_n(x)=\left(\begin{array}{c}
\tilde{A}_n(x)\\
\tilde{\phi}_n(x)
\end{array}\right).
\eeqa

$\tilde{\zeta}_n(x)$ again satisfies the orthogonality condition as (\ref{ortho1}). So that, 

\beqa\label{tortho1}
\sqrt{q}\int dx \tilde{\zeta}^{\dagger}_n(x)\tilde{\zeta}_{n^{'}}(x)=\delta_{n,n^{'}}.
\eeqa

Along with this we also have,

\beqa\label{ortho2}
\sqrt{q}\int dx \zeta^{\dagger}_n(x)\tilde{\zeta}_{n^{'}}(x)=0 \mbox{~~ for all~~$n$~and~$n^{'}$~~}.
\eeqa

Unlike the non-zero eigenvalue sector, in the zero eigenvalue 
sector we have normalizable eigenfunction for $n=1$, which is simply $H_1(\sqrt{q} x)$. There is however 
no normalizable eigenfunctions for $n=0$ in 
this sector. The spectrum for $m_n^2=0$ is completely degenerate. Henceforth in this paper we shall refer 
the eigenfunctions for $m_n^2=0$ as the ``zero-eigenfunctions''.

From the equations of motion for $(A_x^1,\Phi_1^2)$ obtained from (\ref{Lagrangian2}), their eigenfunctions 
are simply $(-A_n(x), \phi_n(x))$, 
and $(-\tilde{A}_n(x), \tilde{\phi}_n(x))$ for $m_n^2=(2n-1)q$ and $m_n^2=0$ respectively.
There is thus a two fold degeneracy for this spectrum of the theory.

The term ${\cal L}(\Phi_I,A_x^3)$ gives decoupled equations for $\Phi_I^a$ for each value of $I\ne 1$ and 
the gauge index $a$ and another equation for $A_x^3$
alone.The equation of motion for $\Phi_I^1$ is 

\beqa
\left(-\partial_{0}^2+\partial_x^2-q^2 x^2\right)\Phi_I^1=0
\eeqa

The same equation for  $\Phi_I^2$. The spatial part of the equation is the wave function equation for a 
Harmonic oscillator. The time independent eigenfunctions will thus 
be given by  $\mathcal{N}^{'}(n)e^{-qx^2/2} H_n(\sqrt{q} x)$ where $H_n(\sqrt{q} x)$ are Hermite polynomials. 
The normalization $\mathcal{N}^{'}(n)=1/\sqrt{\sqrt{\pi}2^n n!}$.The corresponding eigenvalues 
are $\gamma_n=(2n+1)q$.
The equations of motion for $\Phi_I^3$ and $A_1^3$ are,

\beqa
\left(-\partial_{0}^2+\partial_x^2\right)\Phi_I^3=0 \mbox{~~;~~} \partial_{0}^2 A_x^3=0
\eeqa

This means that the time independent part of $\Phi_I^3$ is just a plane wave $e^{ilx}$. Tables \ref{t1} and \ref{t2} in appendix \ref{table} summarize 
the various dimensionfull parameters, normalizations and the eigenfunctions.

At this point we should note that the only tachyons that arise in the spectrum are those as discussed 
above. There are no other tachyons.
The presence of the tachyon signals an instability. As noted in \cite{KalyanaRama:2011ny} and the 
introduction, this instability corresponds to the onset of superconducting phase transition of the 
baryons in the boundary theory. In the brane picture, the tachyon condenses and at the end of the 
process we are left with a smoothened out brane configuration \cite{Hashimoto:2003xz}.  
In the following sections we would like to study the quantum theory at finite temperature. 
The main aim is to find the critical temperature at which the tachyonic instability vanishes.

\section{Finite temperature analysis with one scalar: Warm up exercise}\label{finiteT}

In this section we first study a simplified model consisting of only one adjoint scalar. 
The purpose of this section is to outline the basic idea involved
in the computation of the mass-squared corrections of the tachyon due to finite temperature effects. 
With only one scalar (namely $\Phi_1^1$) we have the equations resulting from (\ref{Lagrangian1}), 
up to the quadratic order. By doing a one-loop integral over the fluctuations we will find an effective 
action 
for the tachyonic mode. 
The coefficient of the quadratic part of the effective action gives the  mass-squared of the tachyon as 
function of 
the temperature. We thus start by doing a fluctuation analysis with the doublet of fields 
($\Phi_1^1$, $A_x^2$) fields.

\begin{equation}
\label{bkgndfluc1}
A_x^2=A_B+\delta A ~~~~\Phi_1^1=\Phi_B+\delta \Phi
\end{equation}

To do a finite temperature analysis, we define the Euclidean coordinate $\tau=it$. $\tau$ is 
periodic with period $\beta$, so that the integration limits over $\tau$ are from $0$ to $\beta$.   

We now denote the background field and the fluctuations as, 

\begin{equation}
\label{bkgndfluc2}
\zeta(x,\t)=\left(\begin{array}{c}
A_B(x, \t)\\
\Phi_B(x, \t)
\end{array}\right)~~~,~~~
\delta\zeta(x,\t)=\left(\begin{array}{c}
\delta A(x, \t)\\
\delta\Phi(x, \t)
\end{array}\right)
\end{equation}

The quadratic background part of the action can be written as,
\beqa
S_B=\f{1}{2g^2}\int d\tau dx \zeta^{\dagger}(x,\t){\cal O}_0(x,\t)\zeta(x,\t)
\eeqa

where

\be
\label{eqnfluc2}
{\cal O}_0(x,\t)=\left(\begin{array}{cc}
\partial_{\t}^2-q^2x^2&-qx\partial_x +q\\
2q+qx\partial_x&\partial_{\t}^2+\partial_x^2\end{array}
\right)
\ee

The mode expansions for the background fields is,

\be
\label{bosonmodes}
\begin{split}
\zeta(x,\t)&=N^{1/2}\sum_{w,k}\left(C_{w,k}\zeta_k(x) + \tilde{C}_{w,k}\tilde{\zeta}_k(x)\right)e^{i\o_w\t}\\
&=N^{1/2}\sum_{w,k} \left(C_{w,k} \left(\begin{array}{c}
A_k(x)\\
\phi_k(x)
\end{array}\right) + \tilde{C}_{w,k} \left(\begin{array}{c}
\tilde{A}_k(x)\\
\tilde{\phi}_k(x)
\end{array}\right)\right)e^{i\o_w\t}
\end{split}
\ee

\noindent
Where $\omega_w=2\pi w/\beta$. The normalization constant $N$ is equal to $\sqrt q /\beta$. $\zeta_k(x)$ 
and $\tilde\zeta_k(x)$ are defined in (\ref{cfunctions}) ans (\ref{zerobosegnfunc}) respectively. 
The corresponding eigenvalues are $-\lambda_k=-(2k-1)q$ of the first set of eigenfunctions and 
$\tilde{\lambda}_k=0 \mbox{~}(\mbox{for~all~} k)$ of the second set. The reality of $\zeta(x,\t)$ 
means that $C_{-w,k}=C_{w,k}^{*}$ and $\tilde{C}_{-w,k}=\tilde{C}_{w,k}^{*}$.

With these observations, and using the orthogonality properties (\ref{ortho1}), (\ref{tortho1}) 
and (\ref{ortho2}) the quadratic background part of the action can then be written as,

\be
\label{classical}
S_B=-\f{1}{2g^2}\sum_{w,k}\left(|C_{w,k}|^2 (\omega_w^2+\lambda_k) + |\tilde{C}_{w,k}|^2 \omega_w^2\right)
\ee

\noindent
The mass spectrum now consists of a tower of tachyons of mass-squared $(\omega_w^2-q)/g^2$. Of 
these the zero mode, $w=0$ has the lowest value of $mass^2$.
\\

\noindent
We now study the fluctuations about the above background. The part of the Lagrangian containing the 
fluctuations is given by,

\beqa
S_{\delta}=\f{1}{2g^2}\int d\tau dx \delta\zeta^{\dagger}(x,\t){\cal O}_{\delta}(x,\t)\delta\zeta(x,\t)
\eeqa

where the operator ${\cal O}_{\delta}(x,\t)$ is,

\be
\label{flcn}
\begin{split}
{\cal O}_{\delta}(x,\t)=&{\cal O}_0(x,\t)+{\cal O}_B(x,\t)\\
=&\left(\begin{array}{cc}
\partial_{\t}^2-q^2x^2-\Phi^2_B(x,\t)&-qx\partial_x +q-2A_B(x,\t)\Phi_B(x,\t)\\
2q+qx\partial_x-2A_B(x,\t)\Phi_B(x,\t)&\partial_{\t}^2+\partial_x^2-A^2_B(x,\t)\end{array}
\right)
\end{split}
\ee

There will also be terms linear in the fluctuations. However since these terms do not contribute to 
the $1PI$ effective action, we have dropped them here. 

The mode expansions for the fluctuations is,

\beqa
\label{modeflcn}
\delta\zeta(x,\t) &=& N^{1/2}\sum_{m,n}\left(D_{m,n}\delta\zeta_n(x)+\tilde{D}_{m,n}
\delta\tilde{\zeta}_n(x)\right)e^{i\o_m\t}\\
&=& N^{1/2}\sum_{m,n} \left(D_{m,n} \left(\begin{array}{c}\delta A_n(x)\\
 \delta \phi_n(x)\end{array}\right)
+\tilde{D}_{m,n} \left(\begin{array}{c}\delta \tilde{A}_n(x)\\
\delta \tilde{\phi}_n(x) \end{array}\right)\right)e^{i\o_m\t}\nonumber
\eeqa

where $\delta\zeta_n(x)$ and $\delta\tilde{\zeta}_n(x)$ are now eigenfunctions of the $\t$-independent 
part of the operator ${\cal O}_{\delta}(x,\t)$. Let us assume that the corresponding eigenvalues 
are $-\Lambda_n$ and $-\tilde{\Lambda}_n$ respectively. 
Again since $\delta\zeta(x,\t)$ is real, $D_{-m,n}=D_{m,n}^{*}$ and $\tilde{D}_{-m,n}=\tilde{D}_{m,n}^{*}$. 
The partition function for the fluctuations is thus,

\be
\label{partfunc}
\begin{split}
Z(\beta,q)&=\int {\cal D}[D_{m,n}][\tilde{D}_{m,n}]e^{S_{\delta}}\\
&=\int {\cal D}[D_{m,n}][\tilde{D}_{m,n}]e^{-\f{1}{2g^2}\sum_{m,n}\left[|D_{m,n}|^2 (\omega_m^2+\Lambda_n)
+|\tilde{D}_{m,n}|^2 (\omega_m^2+\tilde{\Lambda}_n)\right]}\\
&=\prod_{m,n\ne 0}\left[\f{1}{(2\pi g^2)^2}(\omega_m^2+\Lambda_n)(\omega_m^2+\tilde{\Lambda}_n)\right]^{-1/2}
\end{split}
\ee

The eigenvalues $\Lambda_n$ and  $\tilde{\Lambda}_n$  are yet to be determined, for which we use 
perturbation theory 
by assuming that the background field modes are small. We already know the time independent eigenfunctions 
and the corresponding 
eigenvalues for the operator ${\cal O}_0$. We can now treat the background fields in (\ref{flcn}) as 
perturbations and find the
corrections. The background fields can be expanded in terms of the $C_{w,k}$ modes. 
Since we are only interested in the quadratic contribution in $C_{w,k}$'s, we do not need beyond the 
leading correction. This is because the perturbation matrix ${\cal O}_B$ contains two powers of 
background fields giving rise to terms quadratic in the $C_{w,k}$ modes.

\be
\label{eignvl}
\Lambda_n=\Lambda_n^{(0)}+\Lambda_n^{(1)}+.... ~~~~~{\rm with}~~~~~
\Lambda_n^{(0)}=\lambda_n=(2n-1)q
\ee

\be
\label{eignvl1}
\tilde{\Lambda}_n=\tilde{\Lambda}_n^{(0)}+\tilde{\Lambda}_n^{(1)}+.... ~~~~~{\rm with}~~~~~
\tilde{\Lambda}_n^{(0)}=\tilde{\lambda}_n=0
\ee
and,

\be
\label{flcpertrb}
\delta\zeta_n(x)=\delta\zeta_n^{(0)}(x)+\delta\zeta_n^{(1)}(x)+.... ~~~~~{\rm with}~~~~~
\delta\zeta_n^{(0)}(x)=\zeta_n(x)
\ee

\be
\label{flcpertrb1}
\delta\tilde{\zeta}_n(x)=\delta\tilde{\zeta}_n^{(0)}(x)+\delta\tilde{\zeta}_n^{(1)}(x)+.... ~~~~~
{\rm with}~~~~~
\delta\tilde{\zeta}_n^{(0)}(x)=\tilde{\zeta}_n(x)
\ee
so that,

\be
\label{eignvl2}
\Lambda_n^{(1)}=-\int dx d\t\delta\zeta_n^{(0)\dagger}(x){\cal O}_B(x,\t)\delta\zeta_n^{(0)}(x)
\ee

\be
\label{eignvl21}
\tilde{\Lambda}_n^{(1)}=-\int dx d\t\delta\tilde{\zeta}_n^{(0)\dagger}(x){\cal O}_B(x,\t)
\delta\tilde{\zeta}_n^{(0)}(x)
\ee
with this,

\be
\label{partfunc2}
\begin{split}
\log Z(\beta,q)&=-\hf\sum_{m,n\ne 0}\left[\log\left(\f{\o_m^2+(2n-1)q+\Lambda_n^{(1)}}{2\pi g^2}\right)
+\log\left(\f{\o_m^2+\tilde{\Lambda}_n^{(1)}}{2 \pi g^2}\right)\right]\\
&=-\hf\sum_{m,n\ne 0}\left[\log\left(1+\frac{\Lambda_n^{(1)}\beta^2}{(2\pi m)^2+(2n-1)q\beta^2}\right)
+\log\left(1+\frac{\tilde{\Lambda}_n^{(1)}\beta^2}{(2\pi m)^2}\right)\right]
\end{split}
\ee

where in the last line we have omitted the field independent terms. For small value of $\Lambda_n^{(1)}$
the leading term which gives the quadratic correction to the effective action of the background $C_{w,k}$ 
fields is,

\beqa
\label{twopt}
\log Z(\beta,q)=-\hf\sum_{m,n\ne 0}\left[\frac{\Lambda_n^{(1)}\beta^2}{(2\pi m)^2+(2n-1)q\beta^2}
+\frac{\tilde{\Lambda}_n^{(1)}\beta^2}{(2\pi m)^2}\right]
\eeqa

To find the form of the effective action due to the perturbations we first compute $\Lambda^{(1)}_n$ 
given in equation (\ref{eignvl2}). The expression for  $\Lambda^{(1)}_n$ after putting in the mode 
expansions for the background $\Phi_B$ and $A_B$ fields reads, 

\beqa\label{lambda1}
\Lambda^{(1)}_n=\sum_{w,k,k^{'}}\left[C_{w,k}C^{*}_{w,k^{'}}F_1(k,k^{'},n,n)+
\tilde{C}_{w,k}\tilde{C}^{*}_{w,k^{'}}F^{'}_1(k,k^{'},n,n)
+2 C_{w,k}\tilde{C}^{*}_{w,k^{'}}F^{''}_1(k,k^{'},n,n)\right]\non
\eeqa

\beqa
F_1(k,k^{'},n,n)&=&\sqrt{q}\int dx \left[\phi_k(x)\phi_{k^{'}}(x)A_n(x)A_n(x)
+2A_k\phi_{k^{'}}(x)\phi_n(x)A_n(x)\right. \non
&+& \left. 2A_{k^{'}}\phi_{k}(x)\phi_n(x)A_n(x)+ A_k(x)A_{k^{'}}(x)\phi_n(x)\phi_n(x)\right]
\eeqa

\beqa
F^{'}_1(k,k^{'},n,n)&=&\sqrt{q}\int dx \left[\tilde{\phi}_k(x)\tilde{\phi}_{k^{'}}(x)A_n(x)A_n(x)
+2\tilde{A}_k\tilde{\phi}_{k^{'}}(x)\phi_n(x)A_n(x)\right. \non
&+& \left. 2\tilde{A}_{k^{'}}\tilde{\phi}_{k}(x)\phi_n(x)A_n(x)
+ \tilde{A}_k(x)\tilde{A}_{k^{'}}(x)\phi_n(x)\phi_n(x)\right]
\eeqa

\beqa
F^{''}_1(k,k^{'},n,n)&=&\sqrt{q}\int dx \left[{\phi}_k(x)\tilde{\phi}_{k^{'}}(x)A_n(x)A_n(x)
+2{A}_k\tilde{\phi}_{k^{'}}(x)\phi_n(x)A_n(x)\right. \non
&+& \left. 2\tilde{A}_{k^{'}}{\phi}_{k}(x)\phi_n(x)A_n(x)+ {A}_k(x)\tilde{A}_{k^{'}}(x)\phi_n(x)\phi_n(x)\right]
\eeqa

Similarly expanding $\tilde{\Lambda}_n^{(1)}$ given in (\ref{eignvl21}) gives,

\beqa\label{tlambda1}
\tilde{\Lambda}^{(1)}_n =\sum_{w,k,k^{'}}\left[C_{w,k}C^{*}_{w,k^{'}}\tilde{F}_1(k,k^{'},n,n)
+\tilde{C}_{w,k}\tilde{C}^{*}_{w,k^{'}}\tilde{F}^{'}_1(k,k^{'},n,n)
+2 C_{w,k}\tilde{C}^{*}_{w,k^{'}}\tilde{F}^{''}_1(k,k^{'},n,n)\right]\non
\eeqa

\beqa
\tilde{F}_1(k,k^{'},n,n)&=&\sqrt{q}\int dx \left[\phi_k(x)\phi_{k^{'}}(x)\tilde{A}_n(x)\tilde{A}_n(x)
+2A_k\phi_{k^{'}}(x)\tilde{\phi}_n(x)\tilde{A}_n(x)\right. \non
&+& \left. 2A_{k^{'}}\phi_{k}(x)\tilde{\phi}_n(x)\tilde{A}_n(x)
+ A_k(x)A_{k^{'}}(x)\tilde{\phi}_n(x)\tilde{\phi}_n(x)\right]
\eeqa

\beqa
\tilde{F}^{'}_1(k,k^{'},n,n)&=&\sqrt{q}\int dx \left[\tilde{\phi}_k(x)\tilde{\phi}_{k^{'}}(x)
\tilde{A}_n(x)\tilde{A}_n(x)+2\tilde{A}_k\tilde{\phi}_{k^{'}}(x)\tilde{\phi}_n(x)\tilde{A}_n(x)\right. \non
&+& \left. 2\tilde{A}_{k^{'}}\tilde{\phi}_{k}(x)\tilde{\phi}_n(x)\tilde{A}_n(x)
+ \tilde{A}_k(x)\tilde{A}_{k^{'}}(x)\tilde{\phi}_n(x)\tilde{\phi}_n(x)\right]
\eeqa

\beqa
\tilde{F}^{''}_1(k,k^{'},n,n)&=&\sqrt{q}\int dx \left[{\phi}_k(x)\tilde{\phi}_{k^{'}}(x)
\tilde{A}_n(x)\tilde{A}_n(x)+2{A}_k\tilde{\phi}_{k^{'}}(x)\tilde{\phi}_n(x)\tilde{A}_n(x)\right. \non
&+& \left. 2\tilde{A}_{k^{'}}{\phi}_{k}(x)\tilde{\phi}_n(x)\tilde{A}_n(x)+ {A}_k(x)\tilde{A}_{k^{'}}(x)
\tilde{\phi}_n(x)\tilde{\phi}_n(x)\right]
\eeqa

Putting these expansions (\ref{lambda1}) and (\ref{tlambda1}) in (\ref{twopt}) we now collect 
terms containing two $C_{w,k}$ fields, two $\tilde{C}_{w,k}$ or one $C_{w,k}$ and one $\tilde{C}_{w,k}$ 
modes separately. A general expression for (\ref{twopt}) finally is,

\beqa
\log Z(\beta,q)&=&-\sum_{w,k,k^{'}}\left[C_{w,k}C_{w,k^{'}}^*\Sigma^{2}(k,k^{'},\beta,q)
+\tilde{C}_{w,k}\tilde{C}_{w,k^{'}}^*\tilde{\Sigma}^{2}(k,k^{'},\beta,q)\right.\non
&+&\left.2 C_{w,k}\tilde{C}_{w,k^{'}}^*\Sigma^{'2}(k,k^{'},\beta,q)\right]
\eeqa

The coefficient of the $C_{w,k}C_{w,k^{'}}^*$ term, $\Sigma^{2}(k,k^{'},\beta,q)$ is given by,
\be
\label{mass}
\Sigma^{2}(k,k^{'},\beta,q)=\hf\sqrt{q}\beta\sum_{m,n\ne 0}\left[\frac{F_1(k,k^{'},n,n)}{(2\pi m)^2
+(2n-1)q\beta^2}+
\frac{\tilde{F}_1(k,k^{'},n,n)}{(2\pi m)^2}\right]
\ee

This is the one-loop correction to the two point amplitude for the $C_{w,k}$ modes.
Similarly,

\be
\label{mass1}
\tilde{\Sigma}^{2}(k,k^{'},\beta,q)=\hf\sqrt{q}\beta\sum_{m,n\ne 0}\left[\frac{F^{'}_1(k,k^{'},n,n)}{(2\pi m)^2
+(2n-1)q\beta^2}+
\frac{\tilde{F}^{'}_1(k,k^{'},n,n)}{(2\pi m)^2}\right]
\ee

and

\be
\label{mass21}
\Sigma^{'2}(k,k^{'},\beta,q)=\hf\sqrt{q}\beta\sum_{m,n\ne 0}\left[\frac{F^{''}_1(k,k^{'},n,n)}{(2\pi m)^2
+(2n-1)q\beta^2}+
\frac{\tilde{F}^{''}_1(k,k^{'},n,n)}{(2\pi m)^2}\right]
\ee

The fields $C_{w,k}$ and $\tilde{C}_{w,k}$ for different $k$ and same $w$ are all coupled to each other 
at the quadratic order. 
This is due to the broken translational along the $x$ direction. To compute the mass-squared correction 
of any of the modes we should
compute the mass matrix. However this matrix is infinite dimensional. We will be interested in the 
mass-squared corrections of the mode
$C_{0,0}$. This we do numerically. In this paper for numerical simplicity we just compute the correction 
term $\Sigma^{2}(0,0,\beta,q)$, 
reserving a more detailed numerical analysis for the future.

Now, after doing the sum over $m$ in the first term of (\ref{mass}), 
\be
\label{mass2}
\begin{split}
\Sigma^2(k,k^{'},\beta,q)&=\hf\sum_{n\ne 0}\left[\f{F_1(k,k^{'},n,n)}{\sqrt{(2n-1)}}
\left(\hf+\f{1}{e^{\sqrt{(2n-1)q}\beta}-1}\right)
+\frac{\tilde{F}_1(k,k^{'},n,n)}{(2\pi m)^2}\right]\\
&= \Sigma^2_{vac}+\Sigma^2_{\beta}
\end{split}
\ee

$\Sigma^2_{vac}$ is the temperature independent piece. This term is potentially ultraviolet divergent. 
In the following sections we shall consider the the theory on the intersecting $D1$ branes. 
This theory is obtained from a finite ${\cal N}=8$ SYM in two dimensions by giving an expectation value 
to one of the scalars $\Phi_1^3=qx$. Supersymmetry in the intersecting $D1$ brane theory is completely 
broken by the background. 
However the action is supersymmetric and finiteness of the ${\cal N}=8$ theory implies that the the theory 
on the intersecting $D1$ branes must also be ultraviolet finite.
We will see that for the two point functions computed later the ultraviolet divergences cancel between 
the contributions from the boson and the fermion loops.
   
There is also an infrared divergence in (\ref{mass2}) that comes from the second term for $m=0$. 
The IR divergence is due to the massless $\tilde{D}_{m,n}$ modes in the loop. To treat these divergences 
we shall follow the procedure outlined in the introduction.

A similar computation leading to (\ref{mass}), (\ref{mass1}) and (\ref{mass21}) can also be performed 
as follows. 
Expanding both the background and the fluctuation fields using same basis functions that are the 
eigenfunctions of ${\cal O}_0$ {\it i.e}. 
to the lowest order the fluctuation wave functions, 
$\delta A_n(x)=A_n(x)$, $\delta \phi_n(x)=\phi_n(x)$, $\delta \tilde{A}_n(x)=\tilde{A}_n(x)$ 
and $\delta \tilde{\phi}_n(x)=\tilde{\phi}_n(x)$,

\be
\label{acttot}
\begin{split}
S&=S_B+S_{\delta}\\
&=-\f{1}{2g^2}\sum_{w,k}\left(|C_{w,k}|^2 (\omega_w^2+\lambda_k) + |\tilde{C}_{w,k}|^2 \omega_w^2 \right)
-\f{1}{2g^2}\sum_{m,n}\left(|D_{m,n}|^2 (\omega_m^2+\lambda_n)+ |\tilde{D}_{m,n}|^2 \omega_m^2 \right)\\
&+ I + {\rm background~fields~of~quartic~order}
\end{split}
\ee
where the interaction term $I$, is
\be
\label{interaction}
\begin{split}
 I=&
-\f{N}{2g^2}\sum_{m,m^{'},n,n^{'}}\sum_{w,w^{'},k,k^{'}}\left(C_{w,k}C_{w^{'},k^{'}}
D_{m,n}D_{m^{'},n^{'}}F_1(k,k^{'},n,n^{'})\right.\\
&+\left.C_{w,k}C_{w^{'},k^{'}}\tilde{D}_{m,n}\tilde{D}_{m^{'},n^{'}}\tilde{F}_1(k,k^{'},n,n^{'})
+\tilde{C}_{w,k}\tilde{C}_{w^{'},k^{'}}{D}_{m,n}{D}_{m^{'},n^{'}}{F^{'}}_1(k,k^{'},n,n^{'})\right.\non 
&+\left.\tilde{C}_{w,k}\tilde{C}_{w^{'},k^{'}}\tilde{D}_{m,n}\tilde{D}_{m^{'},n^{'}}\tilde{F^{'}}_1(k,k^{'},n,n^{'})
+2 C_{w,k}\tilde{C}_{w^{'},k^{'}}{D}_{m,n}{D}_{m^{'},n^{'}}{F^{''}}_1(k,k^{'},n,n^{'})\right.\non 
&+ \left. 2 C_{w,k}\tilde{C}_{w^{'},k^{'}}\tilde{D}_{m,n}\tilde{D}_{m^{'},n^{'}}\tilde{F^{''}}_1(k,k^{'},n,n^{'})\right)
\delta_{m+{m^{'}}+{w}+{w^{'}}, 0}
\end{split}
\ee

where, $N=\sqrt{q}/\beta$. Here again we have dropped the terms linear in fluctuations as 
they do not contribute 
to the $1PI$ effective action. The terms cubic in fluctuations have also not been included as 
they do not contribute 
at the one-loop order. The tree-level $D_{m,n}$ and $\tilde{D}_{m,n}$ propagators are then

\be
\label{corrlnD}
\expect{D_{m,n}D_{m^{'},n^{'}}}=g^2\f{\delta_{m,-m^{'}}\delta_{n,n^{'}}}{\left[\o_m^2+\lambda_n\right]}
\mbox{~~;~~}\expect{\tilde{D}_{m,n}\tilde{D}_{m^{'},n^{'}}}=g^2\f{\delta_{m,-m^{'}}\delta_{n,n^{'}}}{\o_m^2}
\ee

With this the one-loop two point amplitudes are same as equations ({\ref{mass}), (\ref{mass1}) and 
(\ref{mass21}).
Henceforth in the following sections we will denote both the background and the fluctuation modes 
as $C_{w,k}$.

\section{Intersecting D1 branes at finite temperature}\label{d1}
We have computed the effective mass-squared of the tachyons as function of temperature in section 
(\ref{finiteT}) for a simplified theory 
with only one scalar field and no fermions. The corrections are infrared divergent due to the presence 
of the tree-level massless modes in the loops. 
One way of dealing with the problem is by computing the mass-squared corrections to the tree-level 
massless modes $\tilde{C}_{w,k}$,
at finite temperature at the one-loop level. With the mass-squared corrections, the tree-level massless
modes become massive at one-loop level at finite temperature. The temperature-dependent effective 
mass-squared of the tree-level massless modes shift their propagator 
by $\o_m^{-2} \rightarrow (\o^2_m + m^2(\beta))^{-1}$. Now we can use this shifted propagator to 
calculate the finite effective mass-squared 
of the tree-level tachyons at finite temperature. 

The first amplitude in (\ref{mass2}) has a purely temperature independent part and a purely temperature
dependent part. The temperature dependent part is exponentially damped, hence finite even for large 
values of the momentum $n$. The temperature independent part
however has a non-convergent sum over the momentum $n$ and hence give rise to Ultraviolet divergence. 
The problem of UV divergence can be dealt with by introducing 
suitable regularization scheme but this in turn renders the finite answer for the mass-squared 
corrections regularization-dependent. Hence there is no unique answer 
for the effective mass-squared of the tachyon. This indicates that the calculation should be done in a 
framework where cancellation of UV divergences are possible.
Hence we work in the framework of a originally supersymmetric theory and should consider the contributions 
from the bosonic as well as fermionic degrees of freedom.
In this set-up the UV divergences are expected to cancel among the amplitudes containing bosons and 
fermions in the loop.             

In this section we thus study the finite temperature effects for the full $D1$ brane theory (\ref{2action}). 
As seen in section \ref{spectrum} an intersecting brane configuration with only one non-zero angle is 
given by the background solution 
$\Phi_1^3=qx$ and $A=0$. In the following we will study the tachyon mass-squared as  a function of the 
temperature for the theory (\ref{2action}) 
including all the other bosonic fields and the fermions. We will compute contribution from Bosons 
and the Fermions towards the tachyon mass-squared correction separately below.

\subsection{Bosons}\label{d1b}

In this section we compute the one-loop correction to the tree-level tachyon mass-squared due to the 
Bosons in the loop. To do this we must first write the mode expansions for 
the individual fields. The mode expansion for $(A^2_x,\Phi_1^1)$ is given in (\ref{bosonmodes}) 
in section \ref{finiteT} using the eigenfunctions that have been 
worked out in section \ref{spectrum}. The $(A^1_x,\Phi_1^2)$ fields satisfy the same mode expansion. 
The only distinction between this and the earlier 
mode expansion is the sign in front of the eigenfunctions $A_k(x)$. Thus

 \be
\label{bmodeexpn}
\begin{split}
\zeta^{'}(x,\t)&=N^{1/2}\sum_{w,k}\left(C^{'}_{w,k}\zeta_k(x)e^{i\o_w\t} 
+ \tilde{C}^{'}_{w,k}\tilde{\zeta}_k(x)e^{i\o_w\t}\right)\\
&=N^{1/2}\sum_{w,k} \left(C^{'}_{1w,k} \left(\begin{array}{c}
-A_k(x)\\
\phi_k(x)
\end{array}\right)e^{i\o_w\t} + \tilde{C}^{'}_{w,k} \left(\begin{array}{c}
-\tilde{A}_k(x)\\
\tilde{\phi}_k(x)
\end{array}\right)e^{i\o_w\t}\right)
\end{split}
\ee
where $\zeta^{'}(x,\t)=\left(\begin{array}{c}
A^1_x(x, \t)\\
\Phi_1^2(x, \t)
\end{array}\right)$.

Similarly since $\Phi^1_I$ and $\Phi^2_I$ $(I\ne 1)$ are harmonic oscillators in the $x$ 
direction and $A^3_x$ and $\Phi_I^3$ (for all $I$) are just plane waves, 
we have the following mode expansions

\begin{gather}
\label{phiImode}
\Phi_I^1(x, \t)=N^{1/2} \mathcal{N}^{'}(n)\sum_{m,n}\Phi_I^1(n, m) e^{-qx^2/2} 
H_n(\sqrt{q}x)e^{i\o_m\t},~~ \{I\ne1\}\\
\label{phi3mode}
\Phi_I^3(x, \t)=\f{N^{1/2}}{\sqrt{q}} \sum_{m} \int\f{dl}{2\pi} \Phi_I^3(l, m)e^{i(\o_m\t+lx)}, \{I=1,\cdots,8\}\\
\label{A3mode}
A^3_x(x,\t)= \f{N^{1/2}}{\sqrt{q}} \int \f{dl}{2\pi}\sum_{m} A^3_x(m,l)e^{i\o_m\t+ilx}\\
\end{gather}

where $H_{n}(\sqrt{q}x)$ are the Hermite polynomials and $e^{-qx^2/2}H_n(\sqrt{q}x)$ 
are the harmonic oscillator wave functions with 
eigenvalue $\gamma_n=(2n+1)q$. The normalization $\mathcal{N}^{'}(n)=1/\sqrt{\sqrt{\pi}2^n n!}$.

The propagators and the interaction vertices are listed in the appendix (\ref{bosons}). 
With these we now write down the contribution to the one loop mass-squared 
corrections to the background fields. The bosonic 
contributions to the one loop mass-squared corrections at finite temperature can be collected 
in two groups, namely the ones coming from the four-point vertices 
and the ones coming from the three-point vertices.

The bosonic four-point vertices listed in (\ref{bosons}) together with their corresponding propagators give,
\beqa
\label{masscorc1}
\Sigma^1(w,w^{'},k,k^{'},\beta,q)&=& \hf  N \sum_m \left[\sum_n 
\left(\f {F_1(k,k^{'},n,n)}{\o_m^2 + \lambda_n} + \f {\tilde{F}_1(k,k^{'},n,n)}{\o_m^2} 
+ \f {7 F_2(k,k^{'},n,n)}{\o_m^2 + \gamma_n}\right)\right. \nonumber\\
&+& \left. \int \f{dl}{(2\pi \sqrt{q})}\left(\f{7 F^{'}_2(k,k^{'},l,-l)}{\o_m^2 + l^2}
+\f{F^{'}_3(k, k^{'},l,-l)}{\o_m^2+l^2}\right)\right.\nonumber\\
&+&\left. \int \f{dl}{2 \pi \sqrt{q}} \f{F_3(k,k^{'},l,-l)}{\o_m^2}\right] \delta_{w+w^{'}}
\eeqa
The Feynman diagrams that constitute the correction (\ref{masscorc1}) are given in figure \ref{bosfourfeyn}. 
In (\ref{masscorc1}) the first term  represented  by the Feynman diagram in 
figure \ref{bosfourfeyn}(a) consists of the three-point vertex 
(\ref{V1}) and receives contributions from the $C_{m,n}$ modes 
with the propagators (\ref{propzeta}) while the second term whose Feynman diagram is given is 
figure \ref{bosfourfeyn}(b) comes from the four-point vertex 
(\ref{V1t}) and bears contributions from 
the massless modes $\tilde{C}_{m,n}$ in the loop with propagator (\ref{propzetat}).  
The third term with the Feynman diagram figure \ref{bosfourfeyn}(c) comprises of the 
four-point vertex (\ref{V2})  having the seven massive fields 
$\Phi^{1,2}_I(m,n)$ for $I\ne1$ with propagators given in (\ref{propphi1I}).
\begin{figure}[h]
\begin{center}
\begin{psfrags}
\psfrag{c1}[][]{$C_{w,k}$}
\psfrag{c2}[][]{$C_{w,k^{'}}$}
\psfrag{c3}[][]{$C_{m,n}$}
\psfrag{c4}[][]{$\tilde{C}_{m,n}$}
\psfrag{c5}[][]{$\Phi^{(1,2)}_I(m,n)$}
\psfrag{c6}[][]{$\Phi^3_I(m,l)$}
\psfrag{c7}[][]{$A^3_x(m,l)$}
\psfrag{c8}[][]{$\Phi^3_1(m,l)$}
\psfrag{v1}[][]{$V_1$}
\psfrag{v2}[][]{$\tilde{V}_1$}
\psfrag{v3}[][]{$V_2$}
\psfrag{v4}[][]{$V^{'}_2$}
\psfrag{v5}[][]{$V_3$}
\psfrag{v6}[][]{$V^{'}_3$}
\psfrag{a1}[][]{(a)}
\psfrag{a2}[][]{(b)}
\psfrag{a3}[][]{(c)}
\psfrag{a4}[][]{(d)}
\psfrag{a5}[][]{(e)}
\psfrag{a6}[][]{(f)}
\includegraphics[ width= 10cm,angle=0]{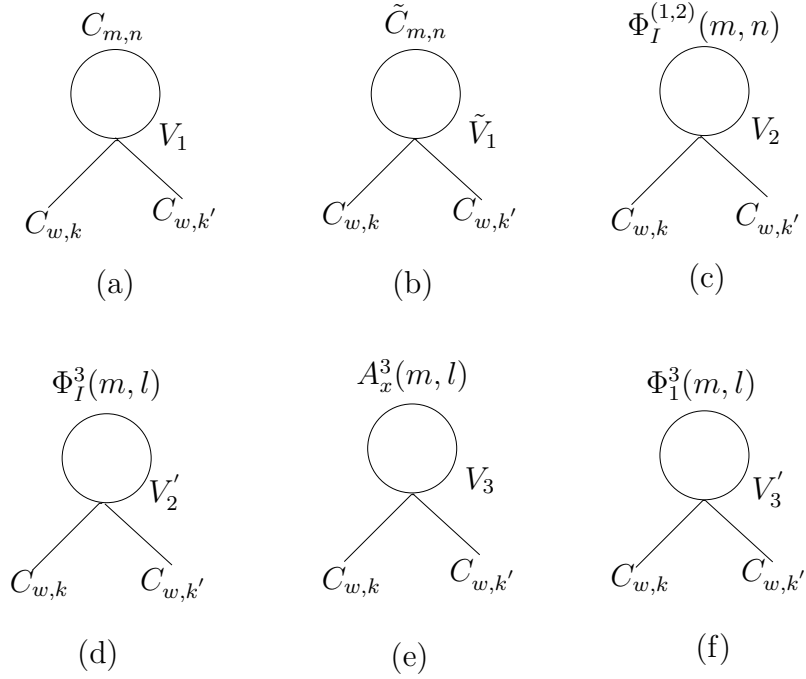}
\end{psfrags}
\caption{{Feynman diagrams for the amplitudes with four-point vertices.}}
\label{bosfourfeyn}
\end{center}
\end{figure} 
The fourth term in (\ref{masscorc1}) is represented by the Feynman diagram figure 
\ref{bosfourfeyn}(d) and is the amplitude for the vertex (\ref{V2p}) which comprise of the 
seven fields $\Phi^3_I$, for $I\ne1$, with propagator (\ref{propphi3I}). The fifth term have 
contributions from the fields $\Phi^3_1$ with propagators  
(\ref{propphi3I}) and is depicted in the Feynman diagram figure \ref{bosfourfeyn}(f) 
while the sixth term bears the massless gauge field $A^3_x(m,l)$ in the loop
with propagator (\ref{propA31}) and the relevant Feynman diagram given in figure \ref{bosfourfeyn}(e). 
Similarly, the three-point bosonic vertices listed in (\ref{bosons}) combined with their 
respective propagators constitute the one-loop bosonic mass-squared corrections 
at finite temperature, {\it{viz.}}   
\beqa
\label{masscorc2}
\Sigma^2(w,w^{'},k,k^{'},\beta,q)&=&-\hf qN \sum_{m,n}\left[
\int \f{dl}{2 \pi \sqrt{q}}\f{F_4(k,l,n)F^{*}_4(k^{'},l,n)}{(\o_m^2+\lambda_n)\o_{m^{'}}^2} 
+ \int \f{dl}{2 \pi \sqrt{q}}\f{\tilde{F}_4(k,l,n)\tilde{F}^{*}_4(k^{'},l,n)}
{\o_m^2 \o_{m^{'}}^2}\right.\nonumber\\
&+&\left.\int \f{dl}{2\pi\sqrt{q}}
\left(\f{7 F_5(k,l,n)F^{*}_5(k^{'},-l,n)}{(\o_m^2+\gamma_n)(\o_{m^{'}}^2+l^2)}
+\f{F^{'}_5(k,l,n) F^{'*}_5(k^{'},-l,n)}{(\o_m^2+\lambda_n)(\o_{m^{'}}^2+l^2)}
\right)\right.\nonumber\\ 
&+&\left.\int \f{dl}{2\pi\sqrt{q}}\f{\tilde{F}^{'}_5(k,l,n) \tilde{F}^{'*}_5(k^{'},-l,n)}
{(\o_m^2)(\o_{m^{'}}^2+l^2)}\right]
\delta_{w+w^{'}}
\eeqa

where, $w=m+m^{'}$.
In (\ref{masscorc2}), the first amplitude gets contributions from the fields $C_{m,n}$ with 
propagator (\ref{propzeta}) and $A^3_x(m^{'},l)$
with propagator (\ref{propA31}) in the loop with the three-point vertex $F_4(k,l,n)$ given 
in (\ref{F4}). The corresponding Feynman diagram is given in  
figure \ref{bosthreefeyn}(a). In the second amplitude, whose Feynman diagram is given by 
figure \ref{bosthreefeyn}(b) comprises of   
the three-point vertex $\tilde{F}_4(k,l,n)$ given in (\ref{F4t}) which gets contributions 
from the massless modes $\tilde{C}_{m,n}$  with propagator
(\ref{propzetat}) and the massless gauge 
fields $A^3_x(m,l)$.       
\begin{figure}[h]
\begin{center}
\begin{psfrags}
\psfrag{c1}[][]{$C_{w,k}$}
\psfrag{c2}[][]{$C_{w,k^{'}}$}
\psfrag{c3}[][]{$C_{m,n}$}
\psfrag{c3p}[][]{$A^3_x(m,l)$}
\psfrag{c4}[][]{$\tilde{C}_{m,n}$}
\psfrag{c4p}[][]{$A^3_x(m,l)$}
\psfrag{c5}[][]{$\Phi^{(1,2)}_I(m,n)$}
\psfrag{c5p}[][]{$\Phi^3_I(m,l)$}
\psfrag{c6}[][]{$C_{m,n}$}
\psfrag{c6p}[][]{$\Phi^3_1(m,l)$}
\psfrag{c7}[][]{$\tilde{C}_{m,n}$}
\psfrag{c7p}[][]{$\Phi^3_1(m,l)$}
\psfrag{a1}[][]{(a)}
\psfrag{a2}[][]{(b)}
\psfrag{a3}[][]{(c)}
\psfrag{a4}[][]{(d)}
\psfrag{a5}[][]{(e)}
%\psfrag{c8}[][]{$\phi^3_1(m,l)$}
\psfrag{v1}[][]{$V_4$}
\psfrag{v2}[][]{$V^*_4$}
\psfrag{v3}[][]{$\tilde{V}_4$}
\psfrag{v4}[][]{$\tilde{V}^{*}_4$}
\psfrag{v5}[][]{$V_5$}
\psfrag{v6}[][]{$V^{*}_5$}
\psfrag{v7}[][]{$\tilde{V}_5$}
\psfrag{v8}[][]{$\tilde{V}^{*}_5$}
\psfrag{v9}[][]{$\tilde{V}^{'}_5$}
\psfrag{v10}[][]{$\tilde{V}^{'*}_5$}
\includegraphics[ width= 12cm,angle=0]{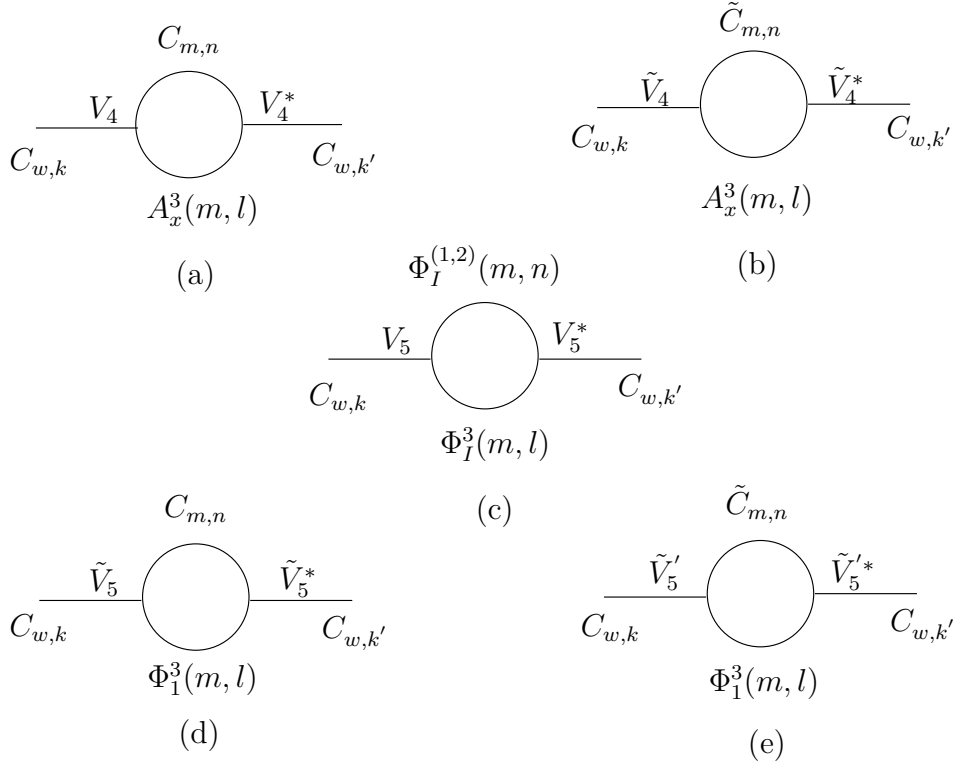}
\end{psfrags}
\caption{{Feynman diagrams for the amplitudes with three-point vertices.}}
\label{bosthreefeyn}
\end{center}
\end{figure} 

The third amplitude in (\ref{masscorc2}) with the vertices $F_5(k,l,n)$, has contributions 
from the pairs of fields
$\Phi^{(1,2)}_I(m,n)$ with propagator (\ref{propphi1I}) and $\Phi^3_I(m,l)$ with propagator 
(\ref{propphi3I}) and the Feynman diagram drawn in figure \ref{bosthreefeyn}(c). 
The fourth amplitude consisting of the three-point vertex $F^{'}_5(k,l,n)$ receives contributions 
in the loop from $C_{m,n}$. 
and $\Phi^3_1(m,l)$ with propagator (\ref{propphi3I}), while the fifth one with 
vertex $\tilde{F}^{'}_5(k,l,n)$ comprises 
of the loop fields $\tilde{C}_{m,n}$ and $\Phi^3_1(m,l)$. The Feynman diagrams for the 
fourth and fifth terms in (\ref{masscorc2}) are given in figure \ref{bosthreefeyn}(d)
and figure \ref{bosthreefeyn}(e) respectively. 

We are interested in the two point function for the $w=w^{'}=k=k^{'}=0$ 
mode because this mode is the tachyon with the lowest $mass^2$ value. The finite temperature 
correction involves sum over the Matsubara frequency. Upon performing the Matsubara sum 
(except on the massless modes), the correction (\ref{masscorc1}) for  $w=w^{'}=0$ can be written as

\beqa
\label{masscorc3}
\Sigma^1(0,0,k,k^{'},\beta,q)
&=&\hf \left[\sum_n\f{F_1(k,k^{'},n,n)}{\sqrt{(2n-1)}} 
\left(\hf+\f{1}{e^{\beta\sqrt{(2n-1)q}}-1}\right)\right.\nonumber\\
&+& N\left.\sum_m \left(\sum_n\f{\tilde{F}_1(k,k^{'},n,n)}{\o^2_m} 
+ \int \f{dl}{2 \pi \sqrt{q}}\f{F_3(k,k^{'},l-l)}{{\o^2_m}}\right)\right.\nonumber\\
&+& \left.\sum_n\left(\f{7 F_2(k,k^{'},n,n)}{\sqrt{(2n+1)}}
\left(\hf+\f{1}{e^{\beta\sqrt{(2n+1)q}}-1}\right)\right)\right.\nonumber\\
&+& \left.\left(\int \f{dl}{2\pi\sqrt{q}}\f{(7+1/2)\delta_{k,k^{'}}}{(l/\sqrt{q})}
\left(\hf+\f{1}{e^{\beta l}-1}\right)\right)\right]
\eeqa
The correction given in (\ref{masscorc2}), after the Matsubara sum 
(leaving out the massless modes) assumes the form 
\beqa
\label{masscorc4}
&&\Sigma^2(0,0,k,k^{'},\beta,q)\nonumber\\
&=& -\hf\sum_{n}\left[
\int \f{dl}{2 \pi \sqrt{q}} \f{F_4(k,l,n)F^{*}_4(k^{'},l,n)}{(2n-1)}\left[\left(\sum_m
\f{\sqrt{q}}{\beta\o_m^2} -\f{1}{\sqrt{2n-1}}\left(\f{1}{2} 
+ \f{1}{e^{\sqrt{(2n-1)q}\beta}-1}\right)\right)\right]\right.\nonumber\\ 
&+&\left. qN\int \f{dl}{2 \pi \sqrt{q}} \sum_{m}\f{\tilde{F}_4(k,l,n)
\tilde{F}^{*}_4(k^{'},l,n)}{\o_m^4}\right.\nonumber\\
&+&\left.\int \f{dl}{2 \pi\sqrt{q}}\left[\f{7 F_5(k,l,n) F^{*}_5(k^{'},-l,n)}
{(l/\sqrt{q})^2-(2n+1)} \left(\f{1}{\sqrt{2n+1}}\left(\hf + \f{1}{e^{\sqrt{(2 n+1)q}\beta}-1}
\right)\right.\right.\right.\nonumber\\
&-& \left.\left.\left.\f{1}{(l/\sqrt{q})}\left(\hf 
+ \f{1}{e^{l\beta}-1}\right)\right)\right]\right.\nonumber\\
&+&\left.\int \f{dl}{2 \pi\sqrt{q}}\left[\f{F^{'}_5(k,l,n) 
F^{'*}_5(k^{'},-l,n)}{(l/\sqrt{q})^2-(2n-1)} \left(\f{1}{\sqrt{2n-1}}\left(\hf + \f{1}{e^{\sqrt{(2 n-1)q}\beta}-1}
\right)\right.\right.\right.\nonumber\\
&-& \left.\left.\left.\f{1}{(l/\sqrt{q})}\left(\hf 
+ \f{1}{e^{l\beta}-1}\right)\right)\right]\right.\nonumber\\
&+&\left.\int \f{dl}{2\pi\sqrt{q}}\f{\tilde{F}^{'}_5(k,l,n) 
\tilde{F}^{'*}_5(k^{'},-l,n)}{(l/\sqrt{q})^2}\left(\sum_m\f{\sqrt{q}}{\beta \o_{m}^2}
-\f{1}{(l/\sqrt{q})}\left(\hf+\f{1}{e^{l\beta}-1}
\right)\right)\right]\nonumber\\
\eeqa

In both (\ref{masscorc3}) and (\ref{masscorc4}), the Matsubara sums give rise to two different kinds 
of terms, the zero temperature quantum corrections
and the temperature dependent part. While the zero-temperature parts are  independent of $q$, 
the temperature dependent parts are functions of both
$\beta$ and $q$. The temperature dependent terms are damped by exponential factors and are 
hence finite. The temperature independent parts however
have problems of divergences. We shall discuss these problems in the section (\ref{uvir}).    

\subsection{Fermions} \label{D1ferm}

We will now compute the contribution to the tachyon two point amplitude due to fermionic fluctuations. 
The fermions in this contribution only appear in the internal loops. Consider first the 
free and the part of the
action (\ref{2action}) in appendix \ref{dimr} that couples to $\Phi^3_1$. The corresponding terms are,

\be
\label{fermlag}
\begin{split}
{\cal L}^{2'}_{1+1}=&\f{1}{2}\left(\psi^{aT}_L \partial_0 \psi^{a}_L+\psi^{aT}_R \partial_0 \psi^a_R+\psi^{aT}_L 
\partial_x \psi^{a}_L-\psi^{aT}_R \partial_x \psi^a_R\right)\\
& +\Phi^3_1\left(\psi^{1T}_R\alpha^T_1\psi^{2}_L-\psi^{2T}_R\alpha^T_1\psi^{1}_L\right)
\end{split}
\ee

We will now call the components of $\psi^a_L$ and those of $\psi^a_R$ as $L_i^{a}$ and $R_i^{a}$ 
respectively, where $a$ is the gauge index and $i=1,\cdots,8$ 
is the fermion index. In this notation,

\be
\label{fermlag2}
\begin{split}
{\cal L}^{2'}_{1+1}=&\f{1}{2}\left(L^a_i \partial_0 L_i^a+ R^a_i \partial_0 R_i^a+ L_i^a \partial_x L_i^a-R_i^a \partial_x R_i^a\right)\\
& +\Phi^3_1\left(\psi^{1T}_R\alpha^T_1\psi^{2}_L-\psi^{2T}_R\alpha^T_1\psi^{1}_L\right)
\end{split}
\ee

With the background value of $\Phi^3_1=qx$ and putting in the value of $\alpha_1$ from (\ref{gamma}), 
we proceed to diagonalize the action. This amounts to solving for the
eigenfunctions of  $L_i^{a}$ and $R_i^{a}$. We have the following sets of equations from (\ref{fermlag2}).

\begin{gather}
\label{eqnferm1a}
(\partial_0+\partial_x )L_1^1+qx R_8^2=0\\
\label{eqnferm1b}
(-\partial_0+\partial_x) R_8^2+qx L_1^1 =0
\end{gather}

\begin{gather}
\label{eqnferm2a}
(\partial_0+\partial_x) L_1^2-qx R_8^1=0\\
\label{eqnferm2b}
(-\partial_0+\partial_x) R_8^1-qx L_1^2 =0
\end{gather}

There are sixteen such sets of decoupled equations. The eight sets 
\beqa
\label{fermset1}
 (L_1^1,R_8^2), (L_4^1,R_5^2), (L_6^1,R_3^2), (L_7^1,R_2^2),(L_2^2,R_7^1), (L_3^2,R_6^1), 
(L_5^2,R_4^1), (L_8^2,R_1^1)
\eeqa

satisfy identical coupled equations like (\ref{eqnferm1a}, \ref{eqnferm1b}). The remaining eight sets 
\beqa
\label{fermset2}
(L_1^2,R_8^1), (L_4^2,R_5^1), (L_6^2,R_3^1), (L_7^2,R_2^1),
(L_8^1,R_1^2), (L_5^1,R_4^2), (L_3^1,R_6^2), (L_2^1,R_7^2)
\eeqa
 satisfy coupled equations like (\ref{eqnferm2a}, \ref{eqnferm2b}). For future convenience 
we denote the left and right fermions of 
(\ref{fermset1})as $L(x,t)$ and $R(x,t)$ and those of (\ref{fermset2}) as $\hat{L}(x,t)$ 
and $\hat{R}(x,t)$. Now we solve the differential equations. 
The set of coupled differential equations (\ref{eqnferm1a}, \ref{eqnferm1b}) satisfied by 
the set of fermionic fields in (\ref{fermset1}) can be 
promoted to the status of second order differential equations as 
\begin{gather}
\label{fermeqnL}
(- \partial^2_0 + \partial^2_x) L(x,t) + q R(x,t) - q^2 x^2 L(x,t) = 0\\
\label{fermeqnR}
(- \partial^2_0 + \partial^2_x) R(x,t) + q L(x,t) - q^2 x^2 R(x,t) = 0
\end{gather} 
It is important to note that the set of fermions in (\ref{fermset2}) also satisfy the same 
set of equations as (\ref{fermeqnL}, \ref{fermeqnR}) 
with $L(x,t)$ and $R(x,t)$being replaced by $\hat{L}(x,t)$ and $- \hat{R}(x,t)$ respectively. 
Let us discuss the solutions to the equations 
(\ref{fermeqnL}, \ref{fermeqnR}). Adding (\ref{fermeqnL}) and (\ref{fermeqnR}) 
and subtracting (\ref{fermeqnR}) from (\ref{fermeqnL}) we get 
the following set of equations
\begin{gather}
\label{fermeqnF}
(- \partial^2_0 + \partial^2_x) F(x,t) + q F(x,t) - q^2 x^2 F(x,t) = 0\\
\label{fermeqnG}
(- \partial^2_0 + \partial^2_x) G(x,t) - q G(x,t) - q^2 x^2 G(x,t) = 0
\end{gather}
where $F(x,t) = L(x,t) + R(x,t)$ and $G(x,t) = L(x,t) - R(x,t)$. In this context let us 
point out that one can construct similar combinational functions 
with the fields in (\ref{fermset2}) viz. $\hat{F}(x,t) = \hat{L}(x,t) + \hat{R}(x,t)$ 
and $\hat{G}(x,t) = \hat{L}(x,t) - \hat{R}(x,t)$, where $\hat{F} = G$ 
and $\hat{G} = F$.  As in the case of bosonic fields, the fermionic differential 
equations can also be analyzed in the asymptotic limit and the fermionic 
eigenfunctions can be written as
\begin{gather}
\label{fermeigenL}
L^a_i(t,x) = e^{- \frac{q x^2}{2}} \tilde{L}^a_i (x,t) \\
\label{fermeigenR}
R^a_i(t,x) = e^{- \frac{q x^2}{2}} \tilde{R}^a_i (x,t)
\end{gather}

Note that although there are two different sets of coupled differential equations; 
one being (\ref{eqnferm1a}, \ref{eqnferm1b}) satisfied by 
(\ref{fermset1}) and the other (\ref{eqnferm2a}, \ref{eqnferm2b}) satisfied by 
(\ref{fermset2}), both sets of equations when recombined give 
rise to the same differential equations as (\ref{fermeqnL}) for the left moving 
fermions and (\ref{fermeqnR}) for the right moving fermions.
The eigenfunctions from (\ref{fermeqnL}) and (\ref{fermeqnR}) that also satisfies 
the first order equations (\ref{eqnferm1a}) are given by
\begin{equation}
\label{fermdouble}
\psi_n(x) = \left(\begin{array}{c}
 L_n(x)\\R_n(x)
\end{array}\right)
\end{equation}

The corresponding eigenvalue is $=-i\sqrt{\lambda^{'}}=-i\sqrt{2nq}$.
Similarly for the set of fermions given in (\ref{fermset2}) and obeying the equations 
of motion (\ref{eqnferm2a}), the eigenfunctions can be obtained by repeating the above procedure and we get 
\begin{equation}
\label{fermdouble2}
\hat{\psi}_n(x) = \left(\begin{array}{c}
\hat{L}_n(x)\\\hat{R}_n(x)
\end{array}\right)= \left(\begin{array}{c}
 L_n(x)\\-R_n(x)
\end{array}\right)
\end{equation} 
where,
\begin{equation}
\label{fermsolnLR}
\begin{split}
&L_n(x) = \hat{L}_n(x) = {\cal N}_F e^{- \frac{q x^2}{2}}\left(- \frac{i}{\sqrt{2n}} H_{n}(\sqrt{q} x) 
+  H_{n-1} (\sqrt{q} x)\right)\\
&R_n(x) = - \hat{R}_n(x) ={\cal N}_F e^{- \frac{q x^2}{2}}\left(- \frac{i}{\sqrt{2n}} H_{n}(\sqrt{q} x) 
-  H_{n-1} (\sqrt{q} x)\right). 
\end{split}
\end{equation}

$H_n(\sqrt{q}x)$ are the Hermite Polynomials. The normalization ${\cal N}_F=\sqrt{\sqrt{\pi}2^{n+1} (n-1)!}$.
We now list some important relations satisfied by the eigenfunctions

\begin{gather}
\label{fermorthocon}
\sqrt{q}\int dx~\psi^{\dagger}_n(x) \psi_{n^{'}}(x) = \sqrt{q}\int dx \left(L^*_n(x) 
L_{n^{'}}(x) + R^*_n(x) R_{n^{'}}(x)\right) = \delta_{n,n^{'}}.\\
\label{fermreln1}
\sqrt{q}\int dx~L^*_n(x) L_{n^{'}}(x) = \sqrt{q}\int dx~R^*_n(x) R_{n^{'}}(x) = \hf \delta_{n,n^{'}}\\
\label{fermreln2}
\sqrt{q}\int dx~\psi^{T}_n(x) \psi_{n^{'}}(x)=\sqrt{q}\int dx \left(L_n(x) L_{n^{'}}(x) 
+ R_n(x) R_{n^{'}}(x)\right) = 0\\
\label{fermreln3}
\sqrt{q}\int dx~\psi^{\dagger}_n(x) \psi^{*}_{n^{'}}(x)=\sqrt{q}\int dx \left(L^{*}_n(x) L^{*}_{n^{'}}(x) 
+ R^{*}_n(x) R^{*}_{n^{'}}(x)\right) = 0    
\end{gather}

With the eigenfunctions as defined above, we
can now write down the mode expansions for the sixteen pairs defined in (\ref{fermset1}) 
and (\ref{fermset2}). For example we write,

\beqa
\label{fermmode}
\left(\begin{array}{c}L_1^1(x,\t)\\R_{8}^2(x,\t)\end{array}\right)
= N^{3/4}\sum^{\infty}_{n,m = \infty}
\left(\theta_1(m,n)e^{i\omega_m\tau} \left(\begin{array}{c}L_n(x)\\R_n(x)\end{array}\right) 
+ \theta^*_1(m,n)e^{-i\omega_m\tau} 
\left(\begin{array}{c}L^*_n(x)\\R^*_n(x)\end{array}\right)\right)
\eeqa

where we have used $\tau= it$ and $N=\sqrt{q}/{\beta}$. For each doublet 
in  (\ref{fermset1}) and (\ref{fermset2}) we have a corresponding set of 
modes $(\theta_j(m,n),\theta_j^{*}(m,n))$. So the index $j$ on the $\theta$'s run from $1\cdots 16$.

Since $L_i^3$ and $R_i^3$ do not couple to the $\Phi^3_1 = qx$ background, 
they just satisfy the plane wave equations,

\be
\label{fermeqn}
(\partial_0+\partial_x )L_i^3=0\mbox{~~~;~~~}
(-\partial_0+\partial_x) R_i^3=0
\ee

Hence their mode expansions are

\be 
\label{fermLR3mode}
\begin{split}
&L_i^3(x,\tau)=N^{3/4}\sum_{m}\frac{1}{\sqrt{q}}\int  \f{dk}{(2\pi)}L^3_{i}(m,k)e^{i(kx+\omega_{m}\tau)} \\
&R_i^3(x,\tau)=N^{3/4}\sum_{m}\frac{1}{\sqrt{q}}\int  \f{dk}{(2\pi)}R^3_{i}(m,k)e^{i(kx+\omega_{m}\tau)} 
\rm{~~~}\mbox{for all $i=1,\cdots,8$}
\end{split}
\ee

where $L^{3*}_i(m,k) = L^{3}_i(m,k)$.

Using the orthogonality relations and the mode expansions the quadratic part of the 
fermionic action can thus be written as,

\be
\label{quadraticact}
\begin{split}
S_f=&\frac{N^{1/2}}{g^2}\left[\sum_{m,n,j=1}^{j=16}\theta_j(m,n)(i\omega_m-\sqrt{\lambda^{'}_n})\theta_j^*(m,n)
\right.\\
&+\left. \frac{1}{2\sqrt{q}}\int\frac{dk}{2\pi}\sum_{m,i=1}^{i=8}L^3_i(m,k)(i \omega_m+k)L^{3*}_i(m,k)\right.\\
&+\left. \frac{1}{2\sqrt{q}}\int\frac{dk}{2\pi}\sum_{m,i=1}^{i=8}R^3_i(m,k)(i\omega_m-k)R^{3*}_i(m,k)\right]
\end{split}
\ee

With these we can write down the fermionic propagators as listed in the appendix. 
We now turn to the interaction terms. These are the terms in the fermionic action 
(\ref{fermlag2}) that couple to
the background fields $\Phi^1_1$ and $A^2_x$ that we simply call $\phi_B$ and $A_B$ respectively as before. 
\be
\label{ferminteract}
\mathcal{L}_{f} = \phi_B\psi^{2T}_R\alpha_1^{T}\psi^{3}_L-\phi_B\psi^{3T}_R\alpha_1^{T}\psi^{2}_L
+A_B\psi_L^{1T}\psi_L^3-A_B\psi_R^{1T}\psi_R^3
\ee

The corresponding vertices have been worked out in appendix. We thus have the following 
contribution to the tachyon 
two-point amplitude.

\beqa
\label{fermmasscorc}
\Sigma^3(w,w^{'},k,k^{'},\beta,q)&=& -(8 N)\sum_{n,m,m^{'}}\int \frac{dl}{2\pi \sqrt{q}}\frac{1}{(i\o_m-
\sqrt{\lambda_n^{'}})}\nonumber\\
&\times&\left[\f{F^R_6(k,n,l)F^{R*}_6(k^{'},n,l)}{(i\o_{m^{'}}+l)}+
\f{F^L_6(k,n,l)F^{L*}_6(k^{'},n,l)}{(i\o_{m^{'}}-l)}\right.\nonumber\\
&+&\left.\f{F^L_7(k,n,l)F^{L*}_7(k^{'},n,l)}{(i\o_{m^{'}}+l)}+
\f{F^R_7(k,n,l)F^{R*}_7(k^{'},n,l)}{(i\o_{m^{'}}-l)}\right.\nonumber\\
&+&\left. \f{F^R_6(k,n,l)F^{L*}_7(k^{'},n,l) 
+ F^{R*}_6(k,n,l)F^L_7(k^{'},n,l)}{(i\o_{m^{'}}+l)}\right.\nonumber\\
&+&\left.\f{F^L_6(k,n,l)F^{R*}_7(k^{'},n,l) 
+ F^{L*}_6(k,n,l)F^R_7(k^{'},n,l)}{(i\o_{m^{'}}-l)}\right]\delta_{w+w^{'}}\nonumber\\
\eeqa

where $w=m+m^{'}$.
In (\ref{fermmasscorc}) the Matsubara frequency $\o_m =m\pi/\beta$ and $m$ is an odd integer due 
to anti-periodic boundary conditions 
along the time-cycle for the fermions and $\o_{-m}=-\o_{m}$. The various diagrams contributing 
to the amplitude is shown in figures 
\ref{fermthree} and \ref{fermthreecross}. Figure \ref{fermthreecross} shows the contractions 
between $R^2_i$-$L^1_{9-i}$ and $L^2_i-R^1_{9-i}$ 
as they are expanded with the same $\theta$'s. The first term within 3rd bracket in 
(\ref{fermmasscorc}) is represented by the Feynman diagram in figure 
\ref{fermthree}(a), while the second term by figure \ref{fermthree}(b). The third and 
fourth terms are represented by figure \ref{fermthree}(c) and figure 
\ref{fermthree}(d) respectively.

\begin{figure}[h]\label{ferampfig}
\begin{center}
\begin{psfrags}
\psfrag{c1}[][]{$C_{w,k}$}
\psfrag{c2}[][]{$C_{w,k^{'}}$}
\psfrag{c3}[][]{$R^2_i(m,n)$}
\psfrag{c4}[][]{$L^3_{9-i}(m,l)$}
\psfrag{c5}[][]{$L^2_i(m,n)$}
\psfrag{c6}[][]{$R^3_{9-i}(m,l)$}
\psfrag{c7}[][]{$R^1_i(m,n)$}
\psfrag{c8}[][]{$R^3_{9-i}(m,l)$}
\psfrag{c9}[][]{$L^1_i(m,n)$}
\psfrag{c10}[][]{$L^3_{9-i}(m,l)$}
\psfrag{v1}[][]{$V^R_{6}$}
\psfrag{v2}[][]{$V^{R*}_{6}$}
\psfrag{v3}[][]{$V^{L}_{6}$}
\psfrag{v4}[][]{$V^{L*}_{6}$}
\psfrag{v5}[][]{$V^{R}_{7}$}
\psfrag{v6}[][]{$V^{R*}_7$}
\psfrag{v7}[][]{$V^{L}_7$}
\psfrag{v8}[][]{$V^{L*}_7$}
\psfrag{a1}[][]{(a)}
\psfrag{a2}[][]{(b)}
\psfrag{a3}[][]{(c)}
\psfrag{a4}[][]{(d)}
%\psfrag{v9}[][]{$\tilde{V}^{'}_5$}
%\psfrag{v10}[][]{$\tilde{V}^{'*}_5$}
\includegraphics[ width= 10cm,angle=0]{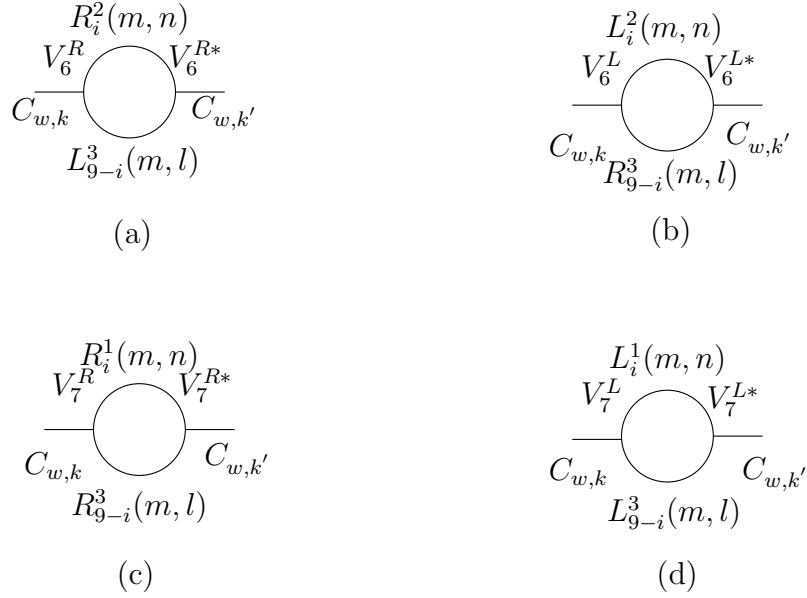}
\end{psfrags}
\caption{\tiny{Feynman diagrams for the amplitudes with three-point fermionic vertices $V^{R/L}$  
and their complex conjugates $V^{*{R/L}}$}}
\label{fermthree}
\end{center}
\end{figure} 
The fifth and sixth terms in (\ref{fermmasscorc}) results from the ``cross''-contraction 
between the right-moving and left-moving fermions and are depicted in the Feynman 
diagrams in figures \ref{fermthreecross}(a) \ref{fermthreecross}(b) respectively.
The functions $F^{R/L}_6$ and $F^{R/L}_7$ are all given in eqns (\ref{FL6}), (\ref{FR6}) and (\ref{FRL7}). 
The massive fermions namely $L^{(1,2)}_i$ and $R^{(1,2)}_i$ have their propagators given by (\ref{propferm1}). 
The propagators for the massless fermions which are the 3rd gauge component of the fermionic fields 
namely $L^3_i$ and $R^3_i$ are given in (\ref{propferm2}). 
\begin{figure}[h]
\begin{center}
\begin{psfrags}
\psfrag{c1}[][]{$C_{w,k}$}
\psfrag{c2}[][]{$C_{w,k^{'}}$}
\psfrag{c3}[][]{$R^2_i$}
\psfrag{c4}[][]{$L^1_{9-i}$}
\psfrag{c5}[][]{$L^3_{9-i}$}
\psfrag{c6}[][]{$L^3_{9-i}$}
\psfrag{c7}[][]{$L^2_i$}
\psfrag{c8}[][]{$R^1_{9-i}$}
\psfrag{c9}[][]{$R^3_{9-i}$}
\psfrag{c10}[][]{$R^3_{9-i}$}
\psfrag{v1}[][]{$V^R_{6}$}
\psfrag{v2}[][]{$V^{L*}_{7}$}
\psfrag{v3}[][]{$V^{L}_{6}$}
\psfrag{v4}[][]{$V^{R*}_{7}$}
\psfrag{a1}[][]{(a)}
\psfrag{a2}[][]{(b)}
%\psfrag{v9}[][]{$\tilde{V}^{'}_5$}
%\psfrag{v10}[][]{$\tilde{V}^{'*}_5$}
\includegraphics[ width= 14cm,angle=0]{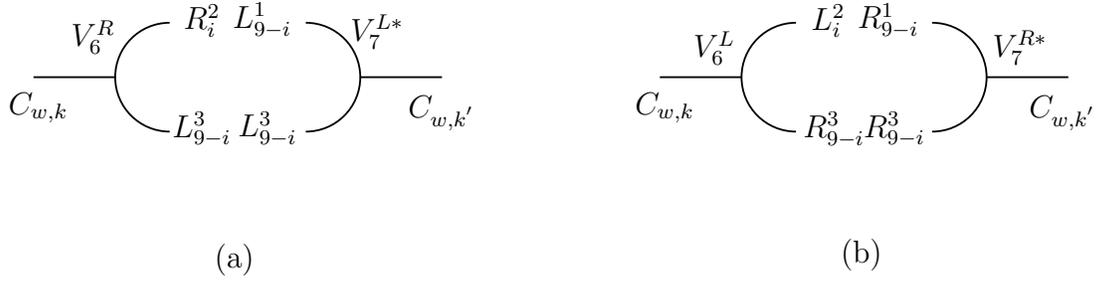}
\end{psfrags}
\caption{Feynman diagrams showing the cross terms in the amplitude with fermions in the loop.}
\label{fermthreecross}
\end{center}
\end{figure} 

After performing the Matsubara sum in (\ref{fermmasscorc}), the fermionic contribution to the 
one-loop mass-squared 
corrections for the tree-level tachyon can be written as 
\beqa
\label{fermmasscorc1}
&&\Sigma^3(0,0,k,k^{'},\beta,q)=\nonumber\\
&&(8 N)\sum_{n}\left[\int \frac{dl}{2\pi \sqrt{q}} 
\left(\f{-\beta \tanh \left(\frac{\beta l}{2}\right)-
\beta \tanh \left(\f{1}{2} \beta \sqrt{2 n q}\right)}
{2 \left(l+\sqrt{2 n q}\right)}\right)\right.\nonumber\\
&&\left.\left[F^R_6(k,n,l)F^{R*}_6(k^{'},n,l) + F^L_7(k,n,l)F^{L*}_7(k^{'},n,l)\right.\right. \nonumber\\ 
&&\left.\left.+ F^R_6(k,n,l)F^{L*}_7(k^{'},n,l) + F^{R*}_6(k,n,l)F^L_7(k^{'},n,l)\right]\right.\nonumber\\
&+& \left. \int \f{dl}{2 \pi \sqrt{q}}
\left(\f{-\beta \tanh \left(\frac{\beta l}{2}\right)+
\beta \tanh \left(\f{1}{2} \beta \sqrt{2 n q}\right)}{2 \left(l-\sqrt{2 n q}\right)}\right)\right.\nonumber\\
&&+ \left.\left[F^L_6(k,n,l)F^{L*}_6(k^{'},n,l) + F^R_7(k,n,l)F^{R*}_7(k^{'},n,l)\right.\right.\nonumber\\ 
&&\left.\left.+ F^L_6(k,n,l)F^{R*}_7(k^{'},n,l) + F^{L*}_6(k,n,l)F^R_7(k^{'},n,l)\right]\right]
\eeqa

As found in the case of bosonic amplitudes the fermionic counterpart given by 
(\ref{fermmasscorc1}) also can be regrouped into two different parts
the zero-temperature quantum corrections and the finite temperature pieces. The amplitudes
 containing fermions in the loop are infrared finite because of anti-periodic
boundary conditions imposed on the fermions whereby they pick up temperature dependent mass at 
tree-level. However the fermionic amplitudes (\ref{fermmasscorc1}) are ultraviolet divergent. 
The divergence come from the temperature independent pieces in (\ref{fermmasscorc1}). 
We shall discuss this problem in details in the following section.    

\section{The Ultraviolet and Infrared Problems}\label{uvir}
Each integral as well as the terms bearing the contribution from the massless modes $C_{w,k}$ in 
(\ref{masscorc3}) and (\ref{masscorc4}) are infrared divergent for $(\o_m=0, l=0)$. 
Moreover the sums over the momentum $n$  do not converge and integral over the momentum $l$ are 
log divergent. 
These give rise to ultraviolet divergence in each term in the two-point functions in 
(\ref{masscorc3}), (\ref{masscorc4}) and (\ref{fermmasscorc1}). We deal with the ultraviolet problem first. 

\subsection{Ultraviolet finiteness of Tachyonic amplitudes}\label{uvtach}
As mentioned above, every term in the two-point functions (\ref{masscorc1}), 
(\ref{masscorc2}) and (\ref{fermmasscorc}) is ultraviolet divergent. 
In the present scenario supersymmetry is 
completely broken by choice of background, namely, $\langle \Phi^3_1 \rangle = q x$, 
however the equality in the number of bosonic and fermionic degrees of freedom 
still holds good in the intersecting brane configuration. In the ultraviolet limit 
the degeneracy in the masses of the bosons and fermions is restored and the 
ultraviolet divergences from the bosonic terms cancel with that from the fermionic terms. 
We now proceed to show this cancellation.

After performing the Matsubara sum (see Appendix F), the temperature independent part of 
the bosonic propagators can be written as

\begin{gather}
\label{tempindprop1}
\f{1}{\o_m^2 + \lambda_n} \rightarrow \f{1}{2 \sqrt{\lambda_n}}\\
\label{tempindprop2}
\f{1}{\o_m^2 + \gamma_n} \rightarrow \f{1}{2 \sqrt{\gamma_n}}\\
\label{tempindprop3}
\f{1}{\o_m^2 + l^2} \rightarrow \f{1}{2 l}\\
\label{tempindprop4}
\f{1}{(\o_m^2 + \lambda_n)(\o_m^2 + \gamma_{n^{'}})} \rightarrow 
\f{1}{\gamma_{n^{'}}\sqrt{\lambda_n} (\gamma_{n^{'}}+\sqrt{\lambda_n})}\\
\end{gather}

Let us now look at the various four-point and three-point vertices. 
We compute the UV limit of the amplitudes for external momentum $k = 0 = k^{'}$. 
This computation gives rise to Gamma functions summed over their arguments. 
For UV behaviour we take asymptotic expansion of the Gamma functions.   
We first compute the four-pont vertices in the bosonic corrections (\ref{masscorc1}) 
in the limit $n \rightarrow \infty$. 
For one-loop calculation $n=n^{'}$. We use the following properties of $\Gamma(*)$ functions:
\begin{gather}
\label{gammaprop}
\lim_{n \rightarrow \infty}\Gamma\left(n+1\right) \sim \left({\f{n}{e}}\right)^n \sqrt{2 \pi n}\\
\Gamma\left(n + \hf\right) = {\f{2n!}{4^n n!}}\sqrt{\pi}
\end{gather}
Also the asymptotic expansion of the Hermite Polynomials for $n\rightarrow \infty$ gives
\begin{equation}
\label{hermasym}
e^{-\f{x^2}{2}} H_n(x) \sim \f{2^n}{\sqrt{\pi}} \Gamma\left(\f{n+1}{2}\right)\cos(\sqrt{2n}x - n \f{\pi}{2}) 
\end{equation}
 
With this asymptotic expansions at our disposal, we can now compute the four-point bosonic vertices. 
The four-point vertex $F_1(0,0,n,n)$ 
can be written as  (using the results of Appendix B,C)
\beqa
\label{f1ex}
F_1(0,0,n,n) = \f{\mathcal{N}^2(n)}{2 \sqrt{\pi}}\int^\infty_\infty dx~e^{-2 \sqrt{q}x^2}
\left(6 H_n(\sqrt{q}x)H_n(\sqrt{q}x)- 8 n^2 H_{n-2}(\sqrt{q}x)H_{n-2}(\sqrt{q}x)\right)\non
\eeqa
Using the asymptotic expansion of the Hermite polynomials given in (\ref{hermasym}), we get         
\begin{eqnarray}
\label{vf1}
F_1(0,0,n,n) &=& \f{\mathcal{N}^2(n) \sqrt{q}}
{2 \sqrt{\pi}} \int^\infty_{-\infty} dx~e^{-2 \sqrt{q} x^2} \left(6(H_n(\sqrt{q} x))^2 - 
8 n^2 (H_{n-2}(\sqrt{q} x))^2\right) \nonumber\\
&=& \f{2^{2n+1}\mathcal{N}^2(n) \Gamma^2\left(\f{n+1}{2}\right)}{2 \pi}\sqrt{q}
\int^\infty_{-\infty} dx~e^{- \sqrt{q} x^2} \cos^2(\sqrt{2 n q}x - n \f{\pi}{2})\nonumber\\
&=& \left(1+ \f{(-1)^n}{e^{2n}}\right)\f{2^{2n} \left(\Gamma\left(\f{n+1}{2}\right)\right)^2}{\pi^2 2^n 
(4n-2)n(n-2)!}.
\end{eqnarray}
In the ultraviolet limit the vertex function $F_1(0,0,n,n)$ reduces to 
\beqa
F_1(0,0,n,n)|_{n\rightarrow \infty}= 
\left(\left(1+ \f{(-1)^n}{e^{2n}}\right)\f{2^{2n} \left(\Gamma\left(\f{n+1}{2}\right)\right)^2}
{\pi^2 2^n (4n-2)n(n-2)!}\right)|_{n\rightarrow \infty} 
= \f{1}{2 \pi \sqrt{2 n}}.\nonumber\\
\eeqa
The vertex $\tilde{F}_1(0,0,n,n)$ has a similar form and in the ultraviolet limit also reduces to 

\beqa
\label{vf1t}
\tilde{F}_1(0,0,n,n)|_{n \rightarrow \infty}= 
\left(\left(1+ \f{(-1)^n}{e^{2n}}\right)\f{2^{2n} \left(\Gamma\left(\f{n+1}{2}\right)\right)^2}
{\pi^2 2^n (4n-2)(n-1)!}\right)|_{n\rightarrow \infty} 
= \f{1}{2 \pi \sqrt{2 n}}.
\eeqa
Putting the external momenta $k= k^{'}=0$, the four-point vertex function $F_2(0,0,n,n)$  
in the limit $n\rightarrow \infty$ can be written as   
\beqa
\label{vf2}
F_2(0,0,n,n) &=&\f{1}{2^{n+1} \sqrt{\pi} n!}\sqrt{q} \int^\infty_{-\infty} dx~e^{-2 \sqrt{q} x^2} 
(H_n(\sqrt{q} x))^2 \\
&=& \left(1+ \f{(-1)^n}{e^{2n}}\right)\f{2^{2n} \left(\Gamma\left(\f{n+1}{2}\right)\right)^2}
{\pi^2 2^{n+1} n!}
\eeqa

In the UV limit the vertex function (\ref{vf2}) reduces to 
\begin{equation}
\label{vf2uv}
F_2(0,0,n,n)|_{n\rightarrow \infty} = 
\left(1+ \f{(-1)^n}{e^{2n}}\right)\f{2^{2n} \left(\Gamma\left(\f{n+1}{2}\right)\right)^2}
{\pi^2 2^{n+1} n!}|_{n\rightarrow \infty}
=\f{1}{\pi \sqrt{2n}}. 
\end{equation}
The remaining four-point bosonic vertex functions namely 
$F^{'}_2(0,0,l,-l)$, $F_3(0,0)$ and $F^{'}_3(0,0,l,-l)$ 
are independent of $n$. Therefore for 
a fixed value of the external momenta, namely $k=k^{'}=0$ they can be exactly computed and found to be, 
\begin{eqnarray}
\label{vf2p}
F^{'}_2(0,0,l,-l) =1\\
\label{vf3p}
F^{'}_3(0,0,l,-l) = \hf\\
\label{vf3}
F_3(0,0,l,-l) = \hf
\end{eqnarray}

The three-point bosonic vertices contain both the continuous momentum $l$ coming from 
the massless fields as well as the discrete momentum 
$n$ coming from the fields coupled to the background $ \expect{\phi^3_1} = q x$.  
The UV-limit must be taken unambiguously for each term in the 
amplitude $\Sigma^2(0,0,0,0,\beta,q)$. Let us first 
try to compute the amplitude for the three-point vertices $F_4(0,l,n)$ and 
$\tilde{F}_4(0,l,n)$ with external momenta $k=0$. 
The three-point vertex $F_4(0,l,n)$ can be written as 
\beqa
\label{vf4}
F_4(0,l,n) &=& \f{\mathcal{N}(n)}{\sqrt{2\sqrt{\pi}}} \sqrt{q} \int dx~e^{- qx^2} e^{i l x} 4nH_n(\sqrt{q}x)
\eeqa    

where $\mathcal{N}(n)$ are the normalization factors for the eigenvectors $\zeta_n(x)$ and the 
factor $e^{ilx}$ can be attributed to the presence of the massless field
$A^3_x(m,l)$ in the loop. Hence the integral in (\ref{vf4}) is simply the Fourier 
transform of the Hermite polynomials weighted by Gaussian factor. 

The Fourier transform of a single Hermite Polynomial is given by
\begin{equation}
\label{furherm}
\int^\infty_{-\infty} dx  (e^{ilx} e^{- x^2} H_n( x)) = (-1)^n i^n \sqrt{\pi} e^{-\f{l^2}{4}} l^n  
\end{equation}

In the following analysis we will set $q=1$ and will restore factors of $q$ only in the final expressions.
Using (\ref{furherm}) in (\ref{vf4}) the three-point vertex $F_4(0,l,n)$ can be written as  

\begin{equation}
\label{vf41}
F_4(0,l,n)= \sqrt{\pi} (-1)^{n-1} i^{n-1} \left(\f{ 4n e^{-\f{l^2}{4}} l^{n-1}}{\sqrt{2^{n+1} 
\pi (4n^2-2n) (n-2)!}}\right)
\end{equation}

We decompose the corresponding propagator into partial fraction
(for reference see the first term in (\ref{masscorc2}) and the Feynman diagram in fig. 
\ref{bosthreefeyn}(a)).

\beqa
\label{propdec}
\f{1}{\o_m^2(\o_m^2 + \lambda_n)} = \f{1}{\lambda_n} \left(\f{1}{\o_m^2} - \f{1}{(\o_m^2 + \lambda_n)}\right)
\eeqa

The amplitude comprising of the vertex $F_4(0,l,n)$ given in (\ref{masscorc2}) is given by 

\beqa
\label{ampf4}
\sum_{m,n} \int \f{dl}{2\pi } \left(\f{16 n^2e^{-\f{l^2}{2}} l^{2n-2}}{2^{n+1}  (4n^2-2n) (n-2)!}
\f{1}{\lambda_n}\f{1}{\o^2_m}
-\f{16 n^2e^{-\f{l^2}{2}}l^{2n-2}}{2^{n+1} (4n^2-2n) (n-2)!}
\f{1}{\lambda_n(\o^2_m + \lambda_n)}\right)\nonumber\\
\eeqa

In the first term in (\ref{ampf4}) we perform the sum over $n$ and take the UV limit $l \rightarrow \infty$. 
In the second term we first compute the integration over $l$ and then expand the resulting 
expression asymptotically about $n= \infty$.
The leading order contribution to the UV divergence obtained from this amplitude is thus

\begin{equation}
\sum_{m}\int \f{dl}{2\pi\sqrt{q}}\f{1}{2 \o^2_m} - \sum_{m,n}\f{1}{2 \pi \sqrt{2 n}} 
\f{1}{\o^2_m + \lambda_n}. 
\end{equation}
     
The three-point vertex $\tilde{F}_4(0,l,n)$ vanishes for all $n$. Hence the corresponding 
amplitude vanishes. 
The two-point functions corresponding to the three-point vertices $F_5(0,l,n)$, $F^{'}_5(0,l,n)$ 
and $\tilde{F}^{'}_5(0,l,n)$ 
(obtained from (\ref{V5}), (\ref{V5p}) and (\ref{V5tp})) contain propagators with momentum $l$ 
as well as those containing the momentum $n$. 
The amplitude bearing the three-point vertex 
$F_5(0,l,n)$ contains contributions from the fields $\Phi^{1,2}_I(m,n)$ and the massless fields 
$\Phi^3_I(m,l)$ in the loop. 
Therefore the propagators contain both the mass-squared eigenvalues $\gamma_n = (2 n + 1) q$ 
of the basis functions of the fields $\Phi^{1,2}_I$ 
and the momentum $l$ of the massless field $\Phi^3_I$. 
Similarly $F^{'}_5(0,l,n)$ has contributions from the fields $C_{m,n}$ and the massless fields 
$\Phi^3_1(m,l)$. Therefore the propagators in the corresponding two-point function has both 
the mass-squared eigenvalues $\lambda_n = (2n-1)q$ and the
momentum $l$. In both these amplitudes  we first 
perform the integration over $l$ and then asymptotically expand the resulting 
expressions about $n= \infty$. As for the two-point function for the three-point vertex 
$\tilde{F}^{'}_5(0,l,n)$
the fields participating in the loop are the massless modes $\tilde{C}_{m,n}$ as well as 
$\Phi^3_1(m,l)$. In this case we 
first decompose the mixed propagator into two parts.
We then sum over $n$ and then take the limit $l \rightarrow \infty$.  
Throughout the computations the external momentum is kept fixed at $k=0$.        
All the three-point vertices $F_5(0,l,n)$, $F^{'}_5(0,l,n)$ and $\tilde{F}^{'}_5(0,l,n)$  has the 
factor $e^{ilx}$ due to the presence of massless fields in the loop.
As in the cases of $F_4(0,l,n)$, the integrals over the world-volume coordinate $x$ in these 
vertex functions amount to evaluating the 
the Fourier transform of the various Hermite polynomials constituting the vertex functions.

Using the result from (\ref{furherm}) in computing the vertex functions (\ref{V5}), (\ref{V5p}) 
and (\ref{V5tp}) for $k=0$, 
we arrive at the following expressions  
\begin{equation}
\label{vf5}
F_5(0,l,n)= -(-1)^{n+1}i^{(n+1)}e^{-\f{l^2}{4}}\f{\left(l^{(n+1)} + 2n l^{(n-1)}\right)}{\sqrt{2^{n+1} \pi n!}}
\end{equation}
\begin{equation}
\label{vf5p}
F^{'}_5(0,l,n)= (-1)^{n+1}i^{(n+1)}e^{-\f{l^2}{4}}\f{3 l^{(n+1)} + 4 n (n-2) l^{(n-3)}}
{\sqrt{2^{n+2} \pi (4 n^2 - 2 n) (n-2)!}}
\end{equation}
\begin{equation}
\label{vf5tp}
\tilde{F}^{'}_5(0,l,n)= (-1)^{n+1}i^{(n+1)}e^{-\f{l^2}{4}}\f{3 l^{(n+1)} - 4 n (n-2) l^{(n-3)}}
{\sqrt{2^{n+2} \pi (4 n - 2) (n-1)!}}
\end{equation}

The amplitude for the three-point vertex $F_5(0,l,n)$ can be written as 

\beqa
\label{ampf5}
&&\sum_{m,n} \int \f{dl}{2\pi} \f{F_5(0,l,n) F^{*}_5(0,l,n)}{(\o^2_m + \gamma_n)(\o^2_m+ l^2)}\\
&=& \sum_{m,n} \int \f{dl}{2\pi} \f{e^{-\f{l^2}{2}}}{(\o^2_m + \gamma_n)(\o^2_m+ l^2)}
\f{\left(l^{(n+1)} + 2n l^{(n-1)}\right)^2}{2^{n+1} \pi n!}
\nonumber\\
\eeqa
where the second line in (\ref{ampf5}) is obtained by plugging in (\ref{vf5}) in the first line.
After performing the integral over $l$ , we asymptotically expand the result about $n = \infty$ 
This results into
\beqa
\label{lintf5}
&&\sum_{m,n} \int \f{dl}{2\pi} \f{F_5(0,l,n) F^{*}_5(0,l,n)}{(\o^2_m + \gamma_n)(\o^2_m+ l^2)}\nonumber\\
&=& \hf \sum_{m,n} \f{1}{(\o^2_m + \gamma_n)}\left[{\left(\frac{2 \sqrt{2} \sqrt{\frac{1}{n}}}{\pi}
-\frac{\left(\left(-7+4\o_m^2\right)\right) \left(\frac{1}{n}\right)^{3/2}}{2 \left(\sqrt{2}
\pi \right)}+\mathcal{O}\left(\frac{1}{n}\right)^2\right)}\right.\nonumber\\
&+&\left.{2^{-n} \left(\frac{e}{n}\right)^n 
\left(\o_m^2\right)^n \sec(n \pi) \left(-\frac{e^{\frac{\o_m^2}{2}} \sqrt{\frac{2}{\pi }} n^{3/2}}{\o_m^3}
+\frac{e^{\frac{\o_m^2}{2}}
\left(24\o_m^2+2\right) \sqrt{n}}{12 \sqrt{2 \pi}\o_m^3}
+\mathcal{O}\left(\frac{1}{n}\right)^0\right)}\right]\nonumber\\
\eeqa 

In the above expression the terms with odd power of $\o_m$ vanishes under summation over $m$ 
over $\{-\infty, \infty\}$.
The leading order term in the amplitude contributing to UV divergence is 
\begin{equation}
\label{leadf5}
\sum_{m,n} \f{4}{2\pi\sqrt{2 n}}\f{1}{(\o^2_m + \gamma_n)}
\end{equation}

Similarly using the result of Fourier Transform in (\ref{vf5p}), 
the amplitude for the three-point vertex $F^{'}_5$ can be written as

\beqa
\label{ampf5p}
&&\sum_{m,n} \int \f{dl}{2\pi } \f{F^{'}_5(0,l,n) F^{'*}_5(0,l,n)}{(\o^2_m + \lambda_n)(l^2+\o^2_m)}\\
&=& \sum_{m,n} \int \f{dl}{2\pi} \f{e^{-\f{l^2}{2}}}{(\o^2_m + \lambda_n)(l^2+\o^2_m)}
\f{\left(3 l^{(n+1)} + 4 n (n-2) l^{(n-3)}\right)^2}{2^{(n+2)} \pi (4 n^2 - 2 n) (n-2)!}\nonumber\\
\eeqa
After integrating the amplitude in (\ref{ampf5p}) over $l$ and expanding about $n=\infty$ 
we get the expansion   
\begin{eqnarray}
\label{lintf5p}
&&\sum_{m,n} \int \f{dl}{2\pi } \f{F^{'}_5(0,l,n) F^{'*}_5(0,l,n)}{(\o^2_m + \lambda_n)(l^2+\o^2_m)}\nonumber\\
&=&\hf \sum_{m,n}\f{1}{(\o^2_m + \lambda_n)}\left[\left(\frac{2 \sqrt{2} \sqrt{\frac{1}{n}}}{\pi }
+\mathcal{O}\left(\frac{1}{n}\right)^1\right)\right.\nonumber\\
&+&\left. 2^{-n} \left(\frac{e}{n}\right)^n \left(\frac{1}{\o_m^2}\right)^{-n}
\sec(n \pi) \left(-\frac{e^{\frac{\o_m^2}{2}} \sqrt{\frac{2}{\pi }} n^{7/2}}{\o_m^7}
+{\mathcal{O}\left(\frac{1}{n}\right)^3}\right)\right]
\end{eqnarray}
The terms with odd powers of $\o_m$ vanishes under the sum over $m$ over $\{-\infty, \infty\}$.
The leading order term in the expansion of the amplitude (\ref{lintf5p}) 
that contributes to the UV divergence is  
\begin{equation}
\label{leadf5p}
\sum_{m,n} \f{4}{2\pi\sqrt{2 n}}\f{1}{(\o^2_m + \lambda_n)}
\end{equation}

As for the amplitude containing the three-point vertex $\tilde{F}^{'}_5(0,l,n)$ we have  
\beqa
\label{ampf5tp}
&&\sum_{m,n} \int \f{dl}{2\pi } \f{\tilde{F}^{'}_5(0,l,n) \tilde{F}^{'*}_5(0,l,n)}{\o^2_m (\o^2_m+l^2)}\\
&=& \sum_{m,n} \int \f{dl}{2\pi} \f{e^{-\f{l^2}{2}}}{\o^2_m (\o^2_m+l^2)}
\f{\left(3 l^{(n+1)} - 4 (n-1)(n-2) l^{(n-3)}\right)^2}{2^{(n+2)} \pi (4 n - 2) (n-1)!}\nonumber
\eeqa

We rewrite the propagators as

\beqa
\label{propdec1}
\f{1}{\o_m^2(\o_m^2 + l^2)} = \f{1}{l^2} \left(\f{1}{\o_m^2} - \f{1}{(\o_m^2 + l^2)}\right)
\eeqa

we thus have the following expression,

\beqa
\sum_{m} \int \f{dl}{2\pi}~\f{e^{-\f{l^2}{2}}}{l^2}\left(\f{1}{\o^2_m} - \f{1}{\o^2_m+l^2}\right)
\f{\left(3 l^{(n+1)} - 4 (n-1)(n-2) l^{(n-3)}\right)^2}{2^{(n+2)} \pi (4 n - 2) (n-1)!}
\eeqa    

In the first term of the above expression we perform the integral over $l$ and then take 
the $n\rightarrow \infty$ limit. 
The first term gives
\beqa
\sum_{m,n} \f{1}{2\pi \sqrt{2n}}\f{1}{\o_m^2}
\eeqa
In the second term we perform the sum over $n$, which gives

\beqa
&-&\sum_{m} \int \f{dl}{2\pi}~\f{e^{-\f{l^2}{2}}}{l^2}\left(\f{1}{\o^2_m+l^2}\right)\nonumber\\
&&\f{\left(-128+96 l^4-18 l^8-9 l^{10}+8 e^{\frac{l^2}{2}} \left(16-8 l^2-10 l^4+6 l^6+l^8\right)\right)}{16
l^6}\nonumber\\
\eeqa    

The divergent piece in the last line is 
\beqa
-\hf\sum_{m} \int \f{dl}{2\pi\sqrt{q}}\left(\f{1}{\o^2_m + l^2}\right) 
\eeqa   

Let us now look at the fermionic amplitudes in (\ref{fermmasscorc}). We first note that the
massless fermionic fields namely $R^3_i$ and $L^3_i$ has propagators 
$(i\o_m \pm l)^{-1}$ where the ``$+$''-sign stands for $R^3_i$ and the ``$-$''-sign for $L^3_i$. 
However while computing the two-point functions one needs to perform 
integrations with respect the momentum $l$ over the entire range of $\{-\infty, \infty\}$. 
Hence to analyze the UV behaviour it suffices to consider only one one sign for the 
propagators. For our purpose we consider the following propagator and decomposed it into

\beqa
\label{fermprop}
\f{1}{(i\o_m - \sqrt{\lambda^{'}_n})(i\o_m + l)}= \hf \left(-\f{1}{\o_m^2 + \lambda^{'}_n} - \f{1}
{\o_m^2 + l^2} + \f{l^2+ \lambda^{'}_n}{(\o_m^2 + \lambda^{'}_n)(\o_m^2 + l^2)}\right)\nonumber\\   
\eeqa

where we have dropped the terms with odd powers of $\o_m$ and $l$, because they are odd 
functions of $\o_m$ and $l$ and will vanish
with respect to sum over $m$ over $\{-\infty, \infty\}$ as well as integral over $l$ 
over the same interval.  
The fermionic vertices are all three-point vertices with contributions from the 
massless fermions $L^3_{i}$ and $R^3_i$ respectively in the loop.
As found in the bosonic amplitude the integrals in (\ref{FR6}), (\ref{FL6}) and (\ref{FRL7}) 
at $k=0$  also amounts to computing 
the Fourier transform the Hermite polynomials from the massive fermions which in turn produce 
the vertex functions in terms of $l$ and $n$. The
fermionic vertices an thus be written as   
\begin{gather}
\label{fl67}
F^L_6(0,n,l)=F^L_7(0,n,l)= (-1)^{n+1}i^{(n+1)}e^{-\f{l^2}{4}}\f{(\f{l^{n}}{\sqrt{2n}} - l^{(n-1)})}
{\sqrt{2\sqrt{\pi}} \sqrt{2^{n+1}\sqrt{\pi}(n-1)!}}\\
\label{fr67}
F^R_6(0,n,l)= F^R_7(0,n,l)=- (-1)^{n+1} i^{(n+1)}e^{-\f{l^2}{4}}\f{(\f{l^{n}}{\sqrt{2n}} + l^{(n-1)})}
{\sqrt{2\sqrt{\pi}} \sqrt{2^{n+1}\sqrt{\pi}(n-1)!}}
\end{gather}

Combining the equations (\ref{fermprop}) with (\ref{FL6}), (\ref{FR6}) and (\ref{FRL7}), 
the total fermion two-point functions for the tree-level tachyon at
finite temperature is given by   

\begin{eqnarray}
\label{fermampuv}
&&\Sigma^3(0,0, 0,0,\beta,q)\nonumber\\
&&= (8) \sum_{m,n} \int \f{dl}{2 \pi }\hf \left(-\f{1}{\o_m^2 + \lambda^{'}_n} - \f{1}
{\o_m^2 + l^2} + \f{l^2+ \lambda^{'}_n}{(\o_m^2 + \lambda^{'}_n)(\o_m^2 + l^2)}\right)\nonumber\\ 
&&\left(e^{-\f{l^2}{2}}\f{2 (\f{l^{n}}{\sqrt{2n}} - l^{(n-1)})^2 + 2 (\f{l^{n}}{\sqrt{2n}} 
+ l^{(n-1)})^2 + 4 (\f{l^{n}}{\sqrt{2n}} - l^{(n-1)})
(\f{l^{n}}{\sqrt{2n}} + l^{(n-1)})}
{2^{n+2}\pi(n-1)!}\right)\nonumber\\
\end{eqnarray}

Combining all the  fermionic vertices, eqn.(\ref{fermampuv}) can be finally written down as  
\begin{eqnarray}
\label{fermampuv1}
&&\Sigma^3(0,0, 0,0,\beta,q) = (8) \sum_{m,n} \int \f{dl}{2 \pi } 
\hf \left[-\f{1}{\o_m^2 + \lambda^{'}_n} \left(e^{-\f{l^2}{2}}\f{l^{2 n}}
{2^{n+1}\pi n!}\right)\right.\nonumber\\ 
&&\left.- \f{1}{\o_m^2 + l^2} \left(e^{-\f{l^2}{2}}\f{l^{2 n}}
{2^{n+1}\pi n!}\right)
+ \f{l^2 + \lambda^{'}_n}
{(\o_m^2 + \lambda^{'}_n)(\o_m^2 + l^2)} 
\left(e^{-\f{l^2}{2}}\f{l^{2 n}}
{2^{n+1}\pi n)!}\right)\right]\nonumber\\
\end{eqnarray}

The first term in (\ref{fermampuv1}) upon integration over the momentum $l$ and in the   
$n \rightarrow \infty$ limit yields the leading order fermionic contribution  
\begin{equation}
- \hf(16) \sum_{m,n} \f{1}{ 2 \pi \sqrt{2n}} \f{1}{\o^2_m + \lambda^{'}_n}
\end{equation}

In the second term in (\ref{fermampuv1}) we sum over $n$ and then take the $l \rightarrow \infty$ limit
and extract the leading order term as

\beqa
-\hf (8) \int \f{dl}{2 \pi \sqrt{q}} \sum_{m} \f{1}{\o^2_m + l^2}
\eeqa

In the 3rd and last term we integrate over $l$ and the leading order term in the 
large $n$-expansion is given by 

\begin{equation}
\hf(8) \sum_{m,n} \f{4}{ 2 \pi \sqrt{2n}} \f{1}{\o^2_m + \lambda^{'}_n}
\end{equation}

The total leading order contribution to the UV divergence from the bosonic side is 

\beqa
\label{bosonlead}
&&\underbrace{\hf \sum_{m,n} \f{1}{ 2 \pi \sqrt{2n}} \f{1}{\o^2_m 
+ \lambda_n} + \hf \sum_{m,n} \f{7 \times 2}
{ 2 \pi \sqrt{2n}} \f{1}{\o^2_m + \gamma_n}}_{\text{amplitudes involving $F_1(0,0,n,n)$ and $F_2(0,0,n,n)$}}
+ \underbrace{\sum_{m,n}\f{1}{2\pi\sqrt{2n}} \f{1}{2 \o^2_m}}_{\text{ $\tilde{F}_1(0,0,n,n)$}}\nonumber\\
&-&\underbrace{\int \f{dl}{2 \pi \sqrt{q}}\sum_{m}\f{1}{2 \o^2_m} 
+ \hf \sum_{m,n} \f{1}{ 2 \pi \sqrt{2n}} \f{1}{\o^2_m 
+ \lambda_n}}_{\text{amplitude involving ${F}_4(0,l,n)$}} 
+ \underbrace{\int \f{dl}{2\pi \sqrt{q}}\sum_m \f{1}{2 \o^2_m}}_{\text{ $F_3(0,0,l,-l)$}}\nonumber\\
&+&\underbrace{\left(\hf(7) \int \f{dl}{2\pi \sqrt{q}}\sum_m \f{1}{\o^2_m + l^2} 
+ \hf \times\hf \int \f{dl}
{2\pi \sqrt{q}}\sum_m \f{1}{\o^2_m + l^2}\right)}_{\text{amplitudes involving $F^{'}_2(0,0,n,n)$ and $F^{'}_3(0,0,n,n)$}}
\nonumber\\
&+&\underbrace{\hf \times \hf \int \f{dl}{2\pi \sqrt{q}}
\sum_m \f{1}{\o^2_m + l^2}-
\sum_{m,n}\f{1}{2\pi\sqrt{2n}}\f{1}{2 \o^2_m}}_{\text{amplitude involving $\tilde{F}^{'}_5(0,l,n)$}}\nonumber\\
&-&\underbrace{\left(\hf(7) \sum_{m,n} \f{4}{ 2 \pi \sqrt{2n}} \f{1}{\o^2_m + \gamma_n} 
+ \hf \sum_{m,n} \f{4}{2 
\pi \sqrt{2n}} \f{1}{\o^2_m + \lambda_n}\right)}_{\text{amplitudes involving $F_5(0,l,n)$ and $F^{'}_5(0,l,n)$}}\nonumber\\
\eeqa

The total leading order contribution to the UV divergence from the amplitudes containing fermions 
in the loop is 

\beqa
\label{fermlead}
&&-\underbrace{\hf(16) \sum_{m,n} \f{1}{ 2 \pi \sqrt{2n}} \f{1}{\o^2_m 
+ \lambda^{'}_n}}_{\text{1st term in $\Sigma^3(0,0,0,0,\beta, q)$}} 
- \underbrace{\hf(8) \int \f{dl}{2 \pi \sqrt{q}}\sum_m \f{1}{\o^2_m 
+ l^2}}_{\text{2nd term in $\Sigma^3(0,0,0,0,\beta, q)$}}\\
&&+\underbrace{\hf(8) \sum_{m,n} \f{4}{ 2 \pi \sqrt{2n}} \f{1}{\o^2_m 
+ \lambda^{'}_n}}_{\text{3rd term in $\Sigma^3(0,0,0,0,\beta, q)$}}
\eeqa
As $n \rightarrow \infty$, we see that $\gamma_n=\lambda_n= \lambda^{'}_n = 2 n q$.
Comparing (\ref{bosonlead}) with (\ref{fermlead}), we see that the leading order terms 
from bosonic sides cancel with that from the fermionic side.
In this method of proving UV finiteness of the finite temperature corrections to the 
tree-level tachyon mass-squared
the UV divergence in $l$ or $n$ is thus softened by the fact that the Matsubara sum is 
left untouched. The large $n$ expansion is valid under the
assumption that $m<n$. This assumption restricts our proof to a corner in the phase space. 
However the counting of the degrees of freedom on the bosonic and 
fermionic sides still match which in turns forces the divergences to cancel out. We expect 
the finiteness of the two-point
functions to hold for large values of $m$ also because higher order terms in the expansion 
is heavily suppressed by Gaussian factors.      

\subsection{Infrared problem}\label{ir}

We now address the problem of infrared divergences. The appearance of IR divergence is 
due to the presence of massless fields namely $\tilde{C}_{w,k}$, $A_x^3$ and $\Phi_I^3$ 
(for $I=1\cdots 8$) in the loop. To compute the IR-finite two-point $C_{w,k}$ amplitude we shall 
follow a two step procedure as mentioned in the introduction. In the first step we 
compute the temperature corrected masses-squared of the massless fields by integrating 
over the modes in the internal lines with an IR cutoff.  
The next step is to introduce these masses in the propagators for the massless fields. 
This is equivalent to summing over an infinite set of diagrams which is illustrated in 
Figure \ref{selfconsistent}. In the figure, the $\chi$ field stands for the modes with 
tree-level mass zero. The bold line is the corrected propagator for the $\chi$ field 
due to the sum of an infinite set of diagrams on the right.

\begin{figure}[t]
\begin{center}
\begin{psfrags}
\psfrag{s}[][]{$\mathlarger{\mathlarger{\sum_{n=0}^{\infty}}}$}
\psfrag{c}[][]{$\chi$}
\psfrag{1}[][]{$1$}
\psfrag{2}[][]{$2$}
\psfrag{n}[][]{$n$}
\psfrag{=}[][]{$\mathlarger{\mathlarger{=}}$}
\psfrag{v2}[][]{$V^1_{2}$}
\psfrag{v3}[][]{$\tilde{V}^1_{2}$}
\psfrag{a3}[][]{$C_{w,k}$}
\psfrag{a4}[][]{$C_{w,k^{'}}$}
\includegraphics[width= 9cm,angle=0]{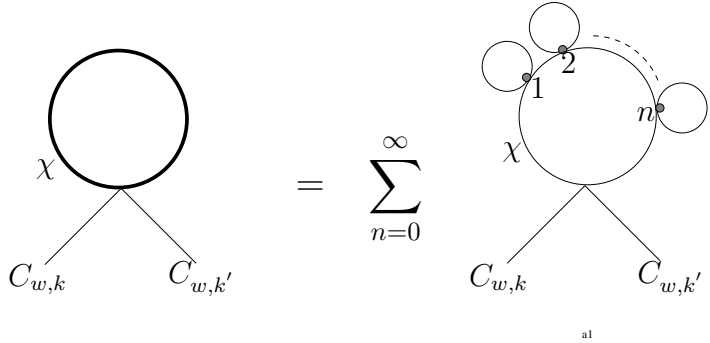}
\end{psfrags}
\caption{Diagram showing the correction to the propagator for the $\chi$ field due to mass insertions.}
\label{selfconsistent}
\end{center}
\end{figure} 

At each temperature the sums and the integrals are now IR-finite because all fields 
in the loops are now massive. 
Thus we can compute the mass-squared corrections and obtain the temperature corrected 
effective masses-squared of the tree-level tachyon as a function of $q$ and $\beta$.  

In the following few sections we compute the two-point functions 
for the massless modes of the doublet of fields $(\Phi^1_1, A^2_x)$ namely the 
$\tilde{C}_{w,k}'s$ and the other massless fields 
namely $\Phi^3_1, \Phi^3_I, A^3_x$ at finite temperature.

\subsection{Two-point functions for the $\tilde{C}_{w,k}$ modes}\label{zeroamp}

In this section we compute the two point function for the $\tilde{C}_{w,k}$ modes. 
During these computations we note that there are no normalizable eigenfunctions 
$\tilde{A}_n(x)$ and $\tilde{\phi}_n(x)$ 
(see eqn (\ref{zerobosegnfunc})) for $n=0$. 

The amplitudes with bosons in the loop and consisting of the four-point vertices 
for the one-loop masses-squared of the massless modes are given by     

\beqa
\label{masscorc5}
\Sigma^1_H(w,w^{'},k,k^{'},\beta,q)&=& \hf  N \sum_m 
\left[\sum_n \left(\f {H_1(k,k^{'},n,n)}{\o_m^2 + \lambda_n} + \f {\tilde{H}_1(k,k^{'},n,n)}{\o_m^2} 
+ \f {7 H_2(k,k^{'},n,n)}{\o_m^2 + \gamma_n}\right)\right. \nonumber\\
&+& \left. \int \f{dl}{(2\pi \sqrt{q})}\left(\f{7 H^{'}_2(k,k^{'},l,-l)}
{\o_m^2 + l^2}+\f{H^{'}_3(k, k^{'},l,-l)}{\o_m^2+l^2}\right)\right.\nonumber\\
&+&\left. \int \f{dl}{2 \pi \sqrt{q}}\f{H_3(k,k^{'}l,-l)}{\o_m^2}\right] \delta_{w+w^{'}},
\eeqa
where $w=m+m^{'}$. Here `$H$' denotes the vertices corresponding to the finite temperature 
two-point functions for the massless modes $\tilde{C}_{w,k}$'s.
The Feynman diagrams comprising the four-point vertices that contribute to the mass-squared 
corrections to the  massless (at tree-level) modes $\tilde{C}_{w,k}$ are depicted in  
figure \ref{bosfourfeynH}.

\begin{figure}[h]
\begin{center}
\begin{psfrags}
\psfrag{c1}[][]{$\tilde{C}_{w,k}$}
\psfrag{c2}[][]{$\tilde{C}_{w,k^{'}}$}
\psfrag{c3}[][]{$C_{m,n}$}
\psfrag{c4}[][]{$\tilde{C}_{m,n}$}
\psfrag{c5}[][]{$\Phi^{(1,2)}_I(m,n)$}
\psfrag{c6}[][]{$\Phi^3_I(m,l)$}
\psfrag{c7}[][]{$A^3_x(m,l)$}
\psfrag{c8}[][]{$\Phi^3_1(m,l)$}
\psfrag{v1}[][]{$V_{H_1}$}
\psfrag{v2}[][]{$\tilde{V}_{H_1}$}
\psfrag{v3}[][]{$V_{H_2}$}
\psfrag{v4}[][]{$V^{'}_{H_2}$}
\psfrag{v5}[][]{$V_{H_3}$}
\psfrag{v6}[][]{$V^{'}_{H_3}$}
\psfrag{a1}[][]{(a)}
\psfrag{a2}[][]{(b)}
\psfrag{a3}[][]{(c)}
\psfrag{a4}[][]{(d)}
\psfrag{a5}[][]{(e)}
\psfrag{a6}[][]{(f)}
\includegraphics[ width= 10cm,angle=0]{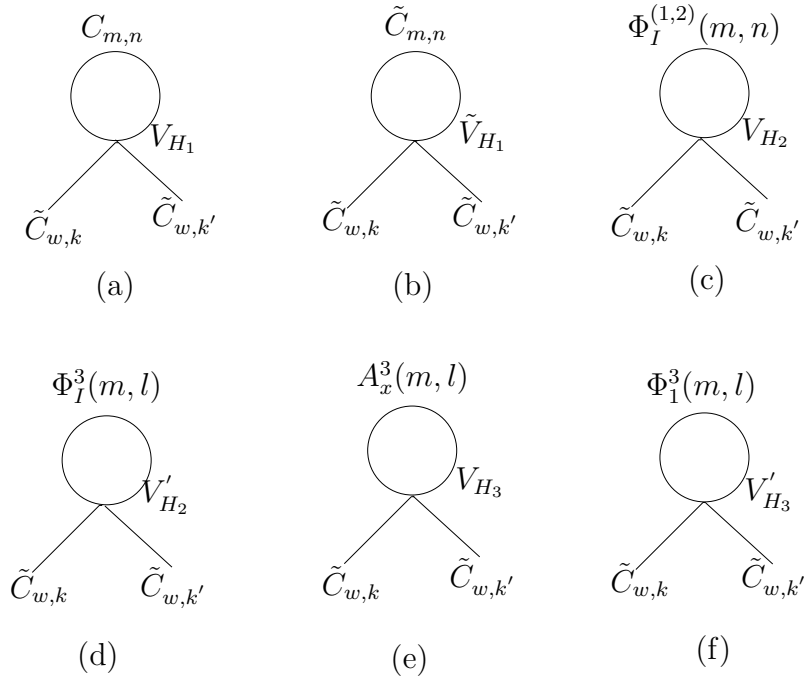}
\end{psfrags}
\caption{Feynman diagrams for the two-point $\tilde{C}_{w,k}$ amplitudes with four-point vertices.}
\label{bosfourfeynH}
\end{center}
\end{figure} 

The amplitudes arising from three-point interaction vertices for the massless modes 
$\tilde{C}_{w,k}$ are collected into

\beqa
\label{masscorc6}
\Sigma^2_H(w,w^{'},k,k^{'},\beta,q)&=&-\hf qN \sum_{m,n}\left[
\int \f{dl}{2 \pi \sqrt{q}}\f{H_4(k,l,n)H^{*}_4(k^{'},l,n)}{(\o_m^2+\lambda_n)\o_{m^{'}}^2} 
+ \int \f{dl}{2 \pi \sqrt{q}}\f{\tilde{H}_4(k,l,n)\tilde{H}^{*}_4(k^{'},l,n)}
{\o_m^2 \o_{m^{'}}^2}\right.\nonumber\\
&+&\left.\int \f{dl}{2\pi\sqrt{q}}
\left(\f{7 H_5(k,l,n)H^{*}_5(k^{'},-l,n)}{(\o_m^2+\gamma_n)(\o_{m^{'}}^2+l^2)}
+\f{H^{'}_5(k,l,n) H^{'*}_5(k^{'},-l,n)}{(\o_m^2+\lambda_n)(\o_{m^{'}}^2+l^2)}
\right)\right.\nonumber\\ 
&+&\left.\int \f{dl}{2\pi\sqrt{q}}\f{\tilde{H}^{'}_5(k,l,n) \tilde{H}^{'*}_5(k^{'},-l,n)}
{(\o_m^2)(\o_{m^{'}}^2+l^2)}\right]
\delta_{w+w^{'}}
\eeqa
where $w=m+m^{'}$. The Feynman diagrams for the three-point interactions are given 
in figure \ref{bosthreefeynH}.
\begin{figure}[h]
\begin{center}
\begin{psfrags}
\psfrag{c1}[][]{$\tilde{C}_{w,k}$}
\psfrag{c2}[][]{$\tilde{C}_{w,k^{'}}$}
\psfrag{c3}[][]{$C_{m,n}$}
\psfrag{c3p}[][]{$A^3_x(m,l)$}
\psfrag{c4}[][]{$\tilde{C}_{m,n}$}
\psfrag{c4p}[][]{$A^3_x(m,l)$}
\psfrag{c5}[][]{$\Phi^{(1,2)}_I(m,n)$}
\psfrag{c5p}[][]{$\Phi^3_I(m,l)$}
\psfrag{c6}[][]{$C_{m,n}$}
\psfrag{c6p}[][]{$\Phi^3_1(m,l)$}
\psfrag{c7}[][]{$\tilde{C}_{m,n}$}
\psfrag{c7p}[][]{$\Phi^3_1(m,l)$}
%\psfrag{c8}[][]{$\phi^3_1(m,l)$}
\psfrag{v1}[][]{$V_{H_4}$}
\psfrag{v2}[][]{$V^*_{H_4}$}
\psfrag{v3}[][]{$V_{H_4}$}
\psfrag{v4}[][]{$V^{*}_{H_4}$}
\psfrag{v5}[][]{$V_{H_5}$}
\psfrag{v6}[][]{$V^{*}_{H_5}$}
\psfrag{v7}[][]{$\tilde{V}_{H_5}$}
\psfrag{v8}[][]{$\tilde{V}^{*}_{H_5}$}
\psfrag{v9}[][]{$\tilde{V}^{'}_{H_5}$}
\psfrag{v10}[][]{$\tilde{V}^{'*}_{H_5}$}
\psfrag{a1}[][]{(a)}
\psfrag{a2}[][]{(b)}
\psfrag{a3}[][]{(c)}
\psfrag{a4}[][]{(d)}
\psfrag{a5}[][]{(e)}
\includegraphics[ width= 14cm,angle=0]{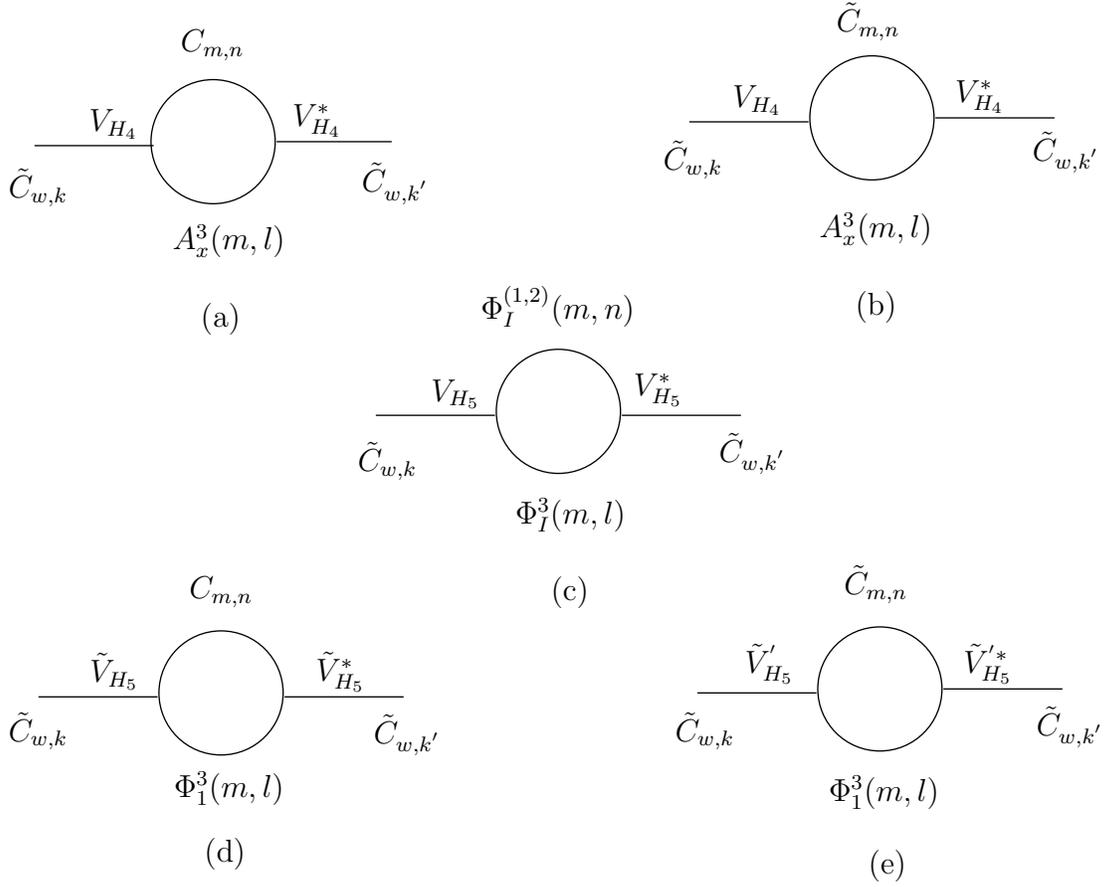}
\end{psfrags}
\caption{{Feynman diagrams for the two-point $\tilde{C}_{w,k}$ amplitudes with three-point vertices}}
\label{bosthreefeynH}
\end{center}
\end{figure} 
The various four-point  and three-point vertices are given in appendix (\ref{zerobosvertex}). 
Note that the one-loop mass-squared corrections to the massless modes 
have the same structure as the tachyonic amplitudes. This is because they 
originate from the same set of interactions for the doublet $\zeta(x,\t)$
in the action. The only difference between (\ref{masscorc1}), (\ref{masscorc2}) 
and (\ref{masscorc5}), (\ref{masscorc6})
is that the momentum modes in the external legs of the Feynman diagrams for the 
latter are now the massless modes $\tilde{C}_{w,k}$. 
Upon performing the Matsubara sums, the bosonic amplitudes 
namely (\ref{masscorc5}) and (\ref{masscorc6}) can be recast as  

\beqa
\label{masscorc7}
\Sigma^1_H(0,0,k,k^{'},\beta,q)
&=&\hf \left[\sum_n\f{H_1(k,k^{'},n,n)}{\sqrt{(2n-1)}} \left(\hf
+\f{1}{e^{\beta\sqrt{(2n-1)q}}-1}\right)\right.\nonumber\\
&+& N\left.\sum_m \left(\sum_n\f{\tilde{H}_1(k,k^{'},n,n)}{\o^2_m} 
+ \int \f{dl}{2 \pi \sqrt{q}}\f{H_3(k,k^{'},l-l)}{{\o^2_m}}\right)\right.\nonumber\\
&+& \left.\sum_n\left(\f{7 H_2(k,k^{'},n,n)}{\sqrt{(2n+1)}}\left(\hf
+\f{1}{e^{\beta\sqrt{(2n+1)q}}-1}\right)\right)\right.\nonumber\\
&+& \left.\left(\int \f{dl}{2\pi\sqrt{q}}\f{(7+1/2)\delta_{k,k^{'}}}{(l/\sqrt{q})}
\left(\hf+\f{1}{e^{\beta l}-1}\right)\right)\right]
\eeqa

%\be
%\label{masscorc7}
%\begin{split}
%\Sigma^1_H(k,k^{'},\beta,q)=&\hf [\sum_n\left(\f{H_1(k,k^{'},n,n)}{\sqrt{(2n-1)}} 
%\left(\hf+\f{1}{e^{\beta\sqrt{(2n-1)q}}-1}\right)\right)\\
%&+ \sum_n\left(\f{7 H_2(k,k^{'},n,n)}{\sqrt{(2n+1)}}\left(\hf+\f{1}{e^{\beta\sqrt{(2n+1)q}}-1}
%\right)\right)\\
%&+ \left(\int \f{dl}{(2\pi\sqrt{q})}\f{7 N}{l^2}
%\left((\beta l/2)\coth(\beta l/2)-1\right)\right)+\int \f{dl}{2 \pi \sqrt{q}} 
%\f{H_3(k,k^{'},l,-l)}{\o_m^2}]
%\end{split}
%\ee

\beqa
\label{masscorc8}
&&\Sigma^2_H(k,k^{'},\beta,q)\nonumber\\
&=& -\hf\sum_{n}\left[
\int \f{dl}{2 \pi \sqrt{q}}\f{H_4(k,l,n)H^{*}_4(k^{'},l,n)}{(2n-1)}
\left[\left(\sum_m\f{\sqrt{q}}{\beta\o_m^2} -\f{1}{\sqrt{2n-1}}\left(\f{1}{2} 
+ \f{1}{e^{\sqrt{(2n-1)q}\beta}}-1\right)\right)\right]\right.\nonumber\\ 
&+&\left. qN\int \f{dl}{2 \pi \sqrt{q}}\sum_{m}\f{\tilde{H}_4(k,l,n)
\tilde{H}^{*}_4(k^{'},l,n)}{\o_m^4}\right.\nonumber\\
&+&\left.\int \f{dl}{2 \pi\sqrt{q}}\left[\f{7 H_5(k,l,n) H^{*}_5(k^{'},-l,n)}
{(l/\sqrt{q})^2-(2n+1)} \left(\f{1}{\sqrt{2n+1}}\left(\hf + \f{1}{e^{\sqrt{(2 n+1)q}\beta}-1}
\right)\right.\right.\right.\nonumber\\
&-& \left.\left.\left.\f{1}{(l/\sqrt{q})}\left(\hf 
+ \f{1}{e^{l\beta}-1}\right)\right)\right]\right.\nonumber\\
&+&\left.\int \f{dl}{2 \pi\sqrt{q}}\left[\f{H^{'}_5(k,l,n) H^{'*}_5(k^{'},-l,n)}
{(l/\sqrt{q})^2-(2n-1)} \left(\f{1}{\sqrt{2n-1}}\left(\hf + \f{1}{e^{\sqrt{(2 n-1)q}\beta}-1}
\right)\right.\right.\right.\nonumber\\
&-& \left.\left.\left.\f{1}{(l/\sqrt{q})}
\left(\hf + \f{1}{e^{l\beta}-1}\right)\right)\right]\right.\nonumber\\
&+&\left.\int \f{dl}{2\pi\sqrt{q}}\f{\tilde{H}^{'}_5(k,l,n) 
\tilde{H}^{'*}_5(k^{'},-l,n)}{(l/\sqrt{q})^2}\left(\sum_m\f{1}{\o_{m}^2}
-\f{1}{(l/\sqrt{q})}\left(\hf+\f{1}{e^{l\beta}-1}\right)\right)\right]\nonumber\\
\eeqa
respectively. Similarly using the various vertex functions given in appendix 
(\ref{zerofermions}), the finite-temperature
contribution due to fermions in the loop to the mass-squared corrections for the massless modes 
is given by equation (\ref{fermmasscorc2}). The relevant Feynman 
diagrams are listed in Figure \ref{tfermthree} and Figure \ref{tfermthreecross}.

\begin{figure}[h]\label{tferampfig}
\begin{center}
\begin{psfrags}
\psfrag{c1}[][]{$\tilde{C}_{w,k}$}
\psfrag{c2}[][]{$\tilde{C}_{w,k^{'}}$}
\psfrag{c3}[][]{$R^2_i(m,n)$}
\psfrag{c4}[][]{$L^3_{9-i}(m,l)$}
\psfrag{c5}[][]{$L^2_i(m,n)$}
\psfrag{c6}[][]{$R^3_{9-i}(m,l)$}
\psfrag{c7}[][]{$R^1_i(m,n)$}
\psfrag{c8}[][]{$R^3_{9-i}(m,l)$}
\psfrag{c9}[][]{$L^1_i(m,n)$}
\psfrag{c10}[][]{$L^3_{9-i}(m,l)$}
\psfrag{v1}[][]{$V^R_{H_6}$}
\psfrag{v2}[][]{$V^{R*}_{H_6}$}
\psfrag{v3}[][]{$V^{L}_{H_6}$}
\psfrag{v4}[][]{$V^{L*}_{H_6}$}
\psfrag{v5}[][]{$V^{R}_{H_7}$}
\psfrag{v6}[][]{$V^{R*}_{H_7}$}
\psfrag{v7}[][]{$V^{L}_{H_7}$}
\psfrag{v8}[][]{$V^{L*}_{H_7}$}
\psfrag{a1}[][]{(a)}
\psfrag{a2}[][]{(b)}
\psfrag{a3}[][]{(c)}
\psfrag{a4}[][]{(d)}
%\psfrag{v9}[][]{$\tilde{V}^{'}_5$}
%\psfrag{v10}[][]{$\tilde{V}^{'*}_5$}
\includegraphics[ width= 10cm,angle=0]{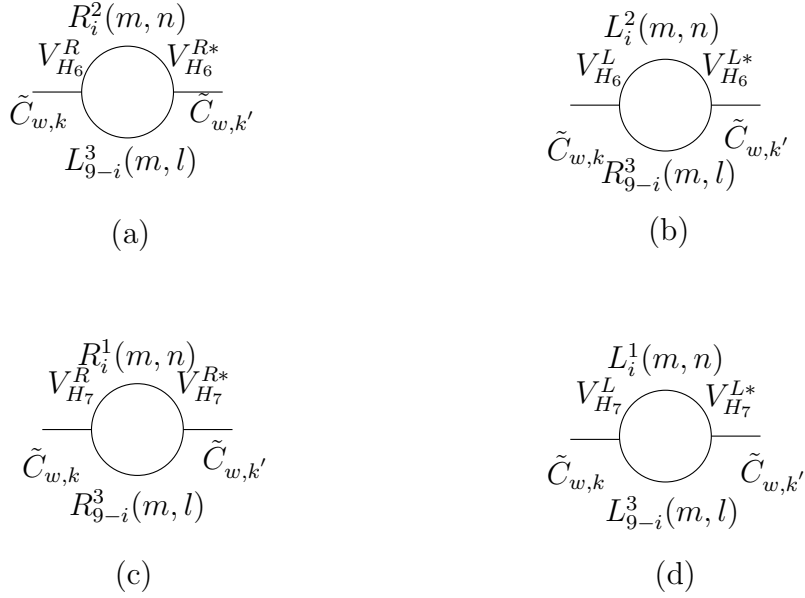}
\end{psfrags}
\caption{Feynman diagrams for the amplitudes with three-point fermionic vertices $V^{R/L}_H$  
and their complex conjugates $V^{*R/L}_H$}
\label{tfermthree}
\end{center}
\end{figure} 

\begin{figure}[h]
\begin{center}
\begin{psfrags}
\psfrag{c1}[][]{$\tilde{C}_{w,k}$}
\psfrag{c2}[][]{$\tilde{C}_{w,k^{'}}$}
\psfrag{c3}[][]{$R^2_i$}
\psfrag{c4}[][]{$L^1_{9-i}$}
\psfrag{c5}[][]{$L^3_{9-i}$}
\psfrag{c6}[][]{$L^3_{9-i}$}
\psfrag{c7}[][]{$L^2_i$}
\psfrag{c8}[][]{$R^1_{9-i}$}
\psfrag{c9}[][]{$R^3_{9-i}$}
\psfrag{c10}[][]{$R^3_{9-i}$}
\psfrag{v1}[][]{$V^R_{H_6}$}
\psfrag{v2}[][]{$V^{L*}_{H_7}$}
\psfrag{v3}[][]{$V^{L}_{H_6}$}
\psfrag{v4}[][]{$V^{R*}_{H_7}$}
\psfrag{a1}[][]{(a)}
\psfrag{a2}[][]{(b)}
%\psfrag{v9}[][]{$\tilde{V}^{'}_5$}
%\psfrag{v10}[][]{$\tilde{V}^{'*}_5$}
\includegraphics[ width= 14cm,angle=0]{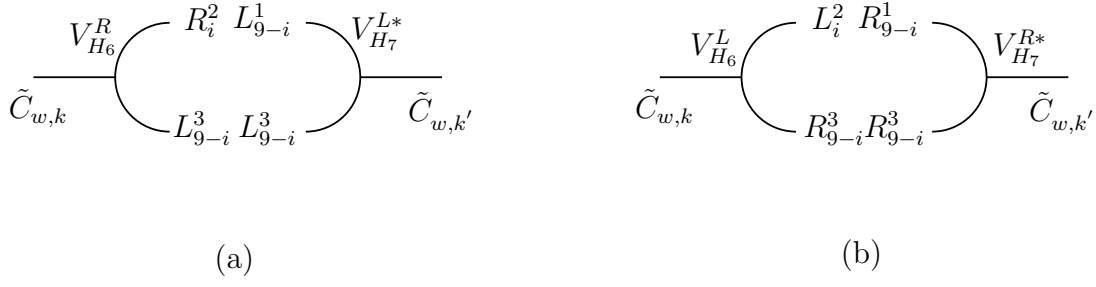}
\end{psfrags}
\caption{Feynman diagrams showing the cross terms in the amplitude with fermions in the loop.}
\label{tfermthreecross}
\end{center}
\end{figure}

\beqa
\label{fermmasscorc2}
\Sigma^3_H(w,w^{'},k,k^{'},\beta,q)&=&- (8 N)\sum_{n,m,m^{'}}\int 
\frac{dl}{2\pi \sqrt{q}}\frac{1}{(i\o_m-\sqrt{\lambda_n^{'}})}\nonumber\\
&\times&\left[\f{H^R_6(k,n,l)H^{R*}_6(k^{'},n,l)}{(i\o_{m^{'}}+l)}
+\f{H^L_6(k,n,l)H^{L*}_6(k^{'},n,l)}{(i\o_{m^{'}}-l)}\right.\nonumber\\
&+ &\left.\f{H^L_7(k,n,l)F^{H*}_7(k^{'},n,l)}{(i\o_{m^{'}}+l)}
+\f{H^R_7(k,n,l)H^{R*}_7(k^{'},n,l)}{(i\o_{m^{'}}-l)}\right.\nonumber\\
&+&\left. \f{H^R_6(k,n,l)H^{L*}_7(k^{'},n,l) + H^{R*}_6(k,n,l)H^L_7(k^{'},n,l)}{(i\o_{m^{'}}+l)}\right.\nonumber\\
&+&\left.\f{H^L_6(k,n,l)H^{R*}_7(k^{'},n,l) 
+ H^{L*}_6(k,n,l)H^R_7(k^{'},n,l)}{(i\o_{m^{'}}-l)}\right]\delta_{w+w^{'}}\non
\eeqa

where $w=m+m^{'}$. After performing the Matsubara sums the 
amplitude (\ref{fermmasscorc2}) assumes the following form.
\beqa
\label{fermmasscorc3}
&&\Sigma^3_H(0,0,k,k^{'},\beta,q)=\nonumber\\
&&(8 N)\sum_{n}\left[\int \frac{dl}{2\pi \sqrt{q}} 
\left(\f{-\beta \tanh \left(\frac{\beta l}{2}\right)
-\beta \tanh \left(\f{1}{2} \beta \sqrt{2 n q}\right)}{2 
\left(l+\sqrt{2 n q}\right)}\right)\right.\nonumber\\
&&\left.[H^R_6(k,n,l)H^{R*}_6(k^{'},n,l) + H^L_7(k,n,l)H^{L*}_7(k^{'},n,l)\right. \nonumber\\ 
&&\left.+ H^R_6(k,n,l)H^{L*}_7(k^{'},n,l) + H^{R*}_6(k,n,l)H^L_7(k^{'},n,l)]\right.\nonumber\\
&+& \left. \int \f{dl}{2 \pi \sqrt{q}}
\left(\f{-\beta \tanh \left(\frac{\beta l}{2}\right)
+\beta \tanh \left(\f{1}{2} \beta \sqrt{2 n q}\right)}
{2 \left(l-\sqrt{2 n q}\right)}\right)\right.\nonumber\\
&&+ \left.[H^L_6(k,n,l)H^{L*}_6(k^{'},n,l) + H^R_7(k,n,l)H^{R*}_7(k^{'},n,l)\right.\nonumber\\ 
&&\left.+ H^L_6(k,n,l)H^{R*}_7(k^{'},n,l) + H^{L*}_6(k,n,l)H^R_7(k^{'},n,l)]\right]
\eeqa

At this point we recall that there is no normalizable eigenfunction 
for the massless modes $\tilde{C}_{w,0}$. Hence the counting starts from $k=1$. 
These massless modes appear as the fluctuations $\tilde{C}_{m,n}$ with $k$ 
replaced as $n$ in the one-loop diagrams for the tree-level 
tachyon (see figures \ref{bosfourfeyn} and \ref{bosthreefeyn}) where  we need to sum over all $n$. 
As mentioned before the two point functions for all the $C_{w,k}$ and 
the $\tilde{C}_{w,k}$ modes are coupled to each other at the one loop level giving rise 
to an infinite dimensional mass-matrix. To get the corrected spectrum 
we must re-diagonalize the mass matrix. Since our approach is to get the final finite values 
of the masses-squared numerically, for simplicity we shall work with a 
finite dimensional matrix for the $\tilde{C}_{w,k}$ modes.
 Like the two point amplitude for the $C_{w,k}$, the two-point functions 
of the massless modes also have contributions from the massless fields 
$\Phi_1^3$, $\Phi_I^3;I \ne 1$ and $A^3_x$ in the loop hence has the problem 
of infrared divergence and will be addressed in the way as mentioned before.
The UV finiteness of the amplitudes for the two-point functions of the fields 
$\tilde{C}_{w,k}$ can be checked using the method used for the tachyonic case.
 
\subsection{Two point function for $\Phi_1^3$}\label{phi1amp}

Using the vertices computed in appendix \ref{vphi13}  we first write down the 
two point amplitude for $\Phi_1^3$.
The Feynman diagrams involving the four-point vertices is depicted in Figure \ref{massless3p11}

\begin{figure}[h]
\begin{center}
\begin{psfrags}
\psfrag{c1}[][]{$\Phi^3_1(w,l)$}
\psfrag{c2}[][]{$\Phi^3_1(w^{'},l^{'})$}
\psfrag{c3}[][]{$\Phi^{(1,2)}_I(m,n)$}
\psfrag{c4}[][]{$C_{m,n}$}
\psfrag{c5}[][]{$\tilde{C}_{m,n}$}
\psfrag{v1}[][]{$V^1_{1}$}
\psfrag{v2}[][]{$V^1_{2}$}
\psfrag{v3}[][]{$\tilde{V}^1_{2}$}
\psfrag{a1}[][]{(a)}
\psfrag{a2}[][]{(b)}
\psfrag{a3}[][]{(c)}
\includegraphics[ width= 12cm,angle=0]{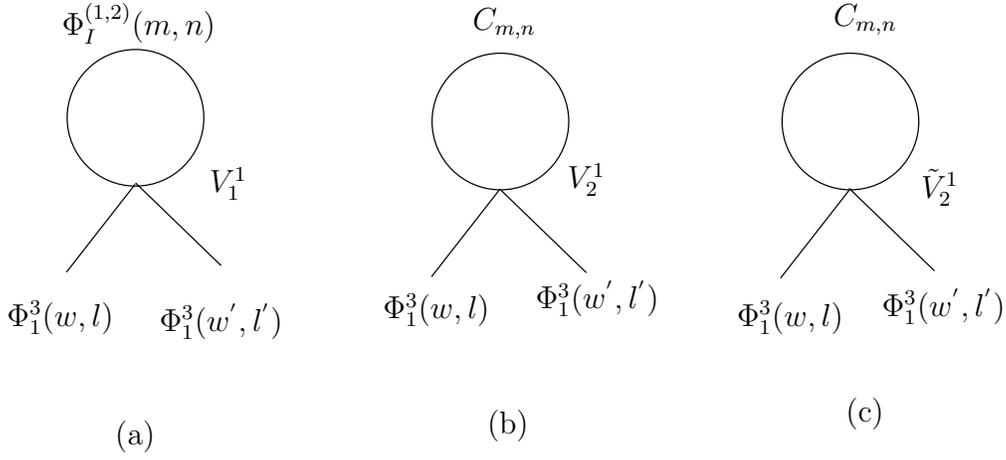}
\end{psfrags}
\caption{{Feynman diagrams for the amplitudes with four-point bosonic vertices 
$V^1_{1},~V^1_2,~\tilde{V}^1_2$}}
\label{massless3p11}
\end{center}
\end{figure} 
The one-loop two-point functions involving the four-point vertices given in section 
(\ref{vphi13}) contributing to the finite-temperature mass-squared corrections to the tree-level 
massless field $\Phi^3_1$ can be collected into 

\be
\label{phi13amp1}
\Sigma^1_{\Phi_1^3-\Phi_1^3}=\hf N \sum_{m,n}\left[(7\times 2) \f{G_1^1(l,l^{'},n,n)}
{(\o_m^2+\gamma_n)}+(2) \f{G_2^1(l,l^{'},n,n)}{(\o_m^2+\lambda_n)} +
(2) \f{\tilde{G}_2^{1}(l,l^{'},n,n)}{(\o_m^2)}  \right]\delta_{w+w^{'}}
\ee
The first term in (\ref{phi13amp1}) has contributions from the massive fields 
$\Phi^{(1,2)}_I$, $I\ne1$ in the loop with four-point vertex and  propagator given in 
(\ref{propphi1I}) and the corresponding Feynman diagram is given by figure 
\ref{massless3p11}(a). The second term bears contributions from the 
fields $C_{m,n}$ with propagator (\ref{propzeta}) and depicted in 
the Feynman diagram in figure \ref{massless3p11}(b). The third term involves the massless fields
$\tilde{C}_{m,n}$ in the loop with propagator (\ref{propzetat}) and corresponding 
Feynman diagram in figure \ref{massless3p11}(c).      
Similarly the three-point bosonic interactions contributing to the finite temperature corrections 
to the massless field $\Phi^3_1(w,l)$ are collected in the 
mass-squared correction (\ref{phi13amp2}). The corresponding Feynman diagrams are given 
in Figure \ref{massless3p12}.

\begin{figure}[h]
\begin{center}
\begin{psfrags}
\psfrag{c1}[][]{$\Phi^3_1(w,l)$}
\psfrag{c2}[][]{$\Phi^3_1(w^{'},l^{'})$}
\psfrag{c3}[][]{$\Phi^{(1,2)}_I(m,n)$}
\psfrag{c4}[][]{$\Phi^{(1,2)}_I(m,n)$}
\psfrag{c5}[][]{$C_{m,n}$}
\psfrag{c6}[][]{$C_{m,n}$}
\psfrag{c7}[][]{$\tilde{C}_{m,n}$}
\psfrag{c8}[][]{$\tilde{C}_{m,n}$}
\psfrag{c9}[][]{$C_{m,n}$}
\psfrag{c10}[][]{$\tilde{C}_{m,n}$}
\psfrag{v1}[][]{$V^1_{1}$}
\psfrag{v2}[][]{$V^{1*}_{1}$}
\psfrag{v3}[][]{${V}^1_{3}$}
\psfrag{v4}[][]{$V^{1*}_{3}$}
\psfrag{v5}[][]{$\tilde{V}^1_{3}$}
\psfrag{v6}[][]{$\tilde{V}^{1*}_{3}$}
\psfrag{v7}[][]{$\tilde{V}^{1'}_{3}$}
\psfrag{v8}[][]{$\tilde{V}^{1'*}_{3}$}
\psfrag{a1}[][]{(a)}
\psfrag{a2}[][]{(b)}
\psfrag{a3}[][]{(c)}
\psfrag{a4}[][]{(d)}
\includegraphics[ width= 12cm,angle=0]{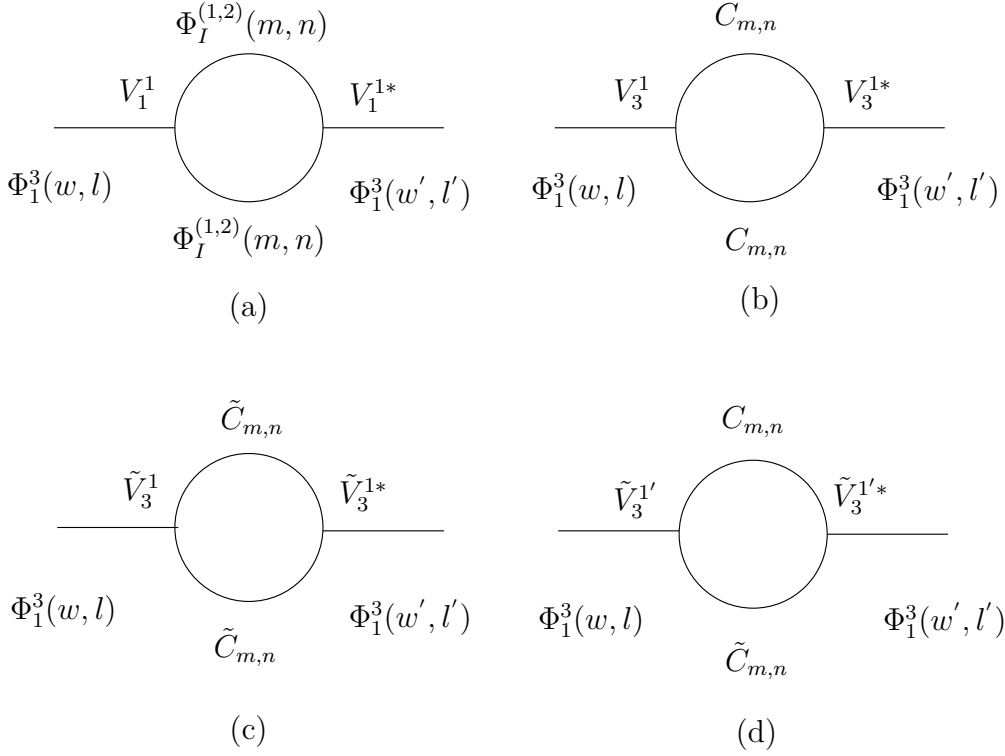}
\end{psfrags}
\caption{{Feynman diagrams with three-point bosonic vertices 
$V^{1'}_{1},~V^1_3,~\tilde{V}^1_3, \tilde{V}^{1'}_3$}}
\label{massless3p12}
\end{center}
\end{figure}

\beqa
\label{phi13amp2}
\Sigma^2_{\Phi_1^3-\Phi_1^3}&=&-\hf qN \sum_{m,n,n^{'}}\left[(7\times 2) 
\f{G_1^{1'}(l,l^{'},n,n)}
{(\o_m^2+\gamma_n)(\o_{m^{'}}^2+\gamma^{'}_{n^{'}})}+ (2)\f{G_3^1(l,n,n^{'})G_3^1(l^{'},n,n^{'})}
{(\o_m^2+\lambda_n)(\o_{m^{'}}^2+\lambda_{n^{'}})}\right.\nonumber\\
&+& \left. (2)\f{\tilde{G}_3^{1}(l,n,n^{'})\tilde{G}_3^{1}(l^{'},n,n^{'})}{(\o_m^2)(\o_{m^{'}}^2)} 
+ (2)\f{\tilde{G}_3^{1'}(l,n,n^{'})\tilde{G}_3^{1'}(l^{'},n,n^{'})}
{\o_m^2(\o_{m^{'}}^2+\lambda_{n^{'}})}\right]\delta_{w+w^{'}}
\eeqa

where $w=m^{'}+m$. The first term in (\ref{phi13amp2}) comprising the 
three-point vertex $G^{1'}_1(l,n,n^{'})$ involves the fields $\Phi^{(1,2)}_I$, $I\ne1$ 
in the loops and is represented by the Feynman diagram in figure \ref{massless3p12}(a). 
Similarly the second term in (\ref{phi13amp2}) involving the vertex
$G^{1}_3(l,n, n^{'})$ comprises of the fields $C_{m,n}$s in the loop. 
The corresponding Feynman diagram is given in figure \ref{massless3p11}(b). 
The third 
term has contributions from $\tilde{C}_{m,n}$s with Feynman diagram in figure 
\ref{massless3p11}(c),while the fourth term represented in figure 
\ref{massless3p11}(d) has contributions from the fields $C_{m,n}$ and $\tilde{C}_{m,n}$.    

\begin{figure}[h]
\begin{center}
\begin{psfrags}
\psfrag{c1}[][]{$\Phi^3_1(w,l)$}
\psfrag{c2}[][]{$\Phi^3_1(w^{'},l^{'})$}
\psfrag{c3}[][]{$\theta_i(m,n)$}
%\psfrag{c4}[][]{$\theta_i(m,n)$}
%\psfrag{c5}[][]{$\theta^{*}_i(m,n)$}
%\psfrag{c6}[][]{$\theta^{*}_i(m,n)$}
%\psfrag{c7}[][]{$\theta^{*}_i(m,n)$}
%\psfrag{c8}[][]{$\theta_i(m,n)$}
%\psfrag{c9}[][]{$\theta_i(m,n)$}
%\psfrag{c10}[][]{$\theta^{*}_i(m,n)$}
\psfrag{v1}[][]{$V^1_{f}$}
\psfrag{v2}[][]{$V^{1*}_{f}$}
\psfrag{v3}[][]{$\tilde{V}^{2}_{f}$}
\psfrag{v4}[][]{$\tilde{V}^{2*}_{f}$}
\psfrag{a1}[][]{(a)}
\psfrag{a2}[][]{(b)}
\includegraphics[ width= 12cm,angle=0]{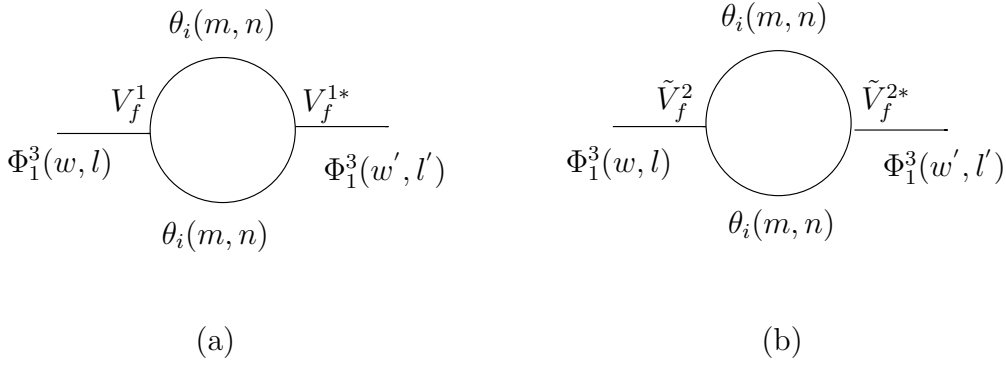}
\end{psfrags}
\caption{{Feynman diagrams involving three-point vertices $V^1_{f},~V^2_f$.}}
\label{masslessp31f2}
\end{center}
\end{figure}

Figure \ref{masslessp31f2} shows the amplitude involving the fermions in the loop. 
The corresponding expression is  

\beqa
\label{phi13amp3}
\Sigma^3_{\Phi_1^3-\Phi_1^3}&=&\hf N \sum_{m,n,n^{'}}\left[ (16\times 2)
\f{G_f^1(l,n,n^{'})G_f^{1*}(l^{'},n,n^{'})}{(i\o_m-\sqrt{\lambda_n^{'}})(i\o_{m^{'}}
-\sqrt{\lambda_{n^{'}}^{'}})}\right.\nonumber\\
&-&\left. (16)\f{G_f^2(l,n,n^{'})G_f^{2*}(l^{'},n,n^{'})}{(i\o_m-\sqrt{\lambda_n^{'}})
(-i\o_{m^{'}}-\sqrt{\lambda_{n^{'}}^{'}})}\right]\delta_{w+w^{'}}
\eeqa

with $w=m^{'}+m$. The first and the second term in (\ref{phi13amp3}) 
are depicted in the Feynman diagrams given in figures 
\ref{masslessp31f2}(a) and \ref{masslessp31f2}(b) respectively.
The various vertices given in section (\ref{vphi13}) that are involved 
in the two-point functions realizing the mass-squared 
corrections to the tree-level massless field $\Phi^3_1$ can be exactly 
computed using the orthogonality relation for Hermite Polynomials. Setting 
the external momenta $l=l^{'}=w=w^{'}=0$, we can write down the various vertices as  

\beqa
\label{g14}
G^1_1[0,0,n,n^{'}] &=& \delta_{n,n^{'}},~~ 
G^1_2[0,0,n,n^{'}]=\hf\delta_{n,n^{'}},~~
\tilde{G}^1_2[0,0,n,n^{'}]=\hf\delta_{n,n^{'}},\\
\label{g11p}
G^{1'}_1[0,n,n^{'}] &=& \sqrt{2n} \delta_{n-1,n^{'}},~~G^1_3[0,n,n^{,}]
=2 \sqrt{\f{2n(n-1)(n-2)}{(2n-1)(2n-3)}} \delta_{n-1,n^{'}},\\
\label{g13t}
\tilde{G}^1_3[0,n,n^{'}]&=& 0,\\
\label{g13tp}
\tilde{G}^{1'}_3[0,n,n^{'}]&=& -\left(\f{\sqrt{2(n-1)}(n+1)}{\sqrt{(2n-1)(2n+1)}} \delta_{n+1,n^{'}} 
- \f{\sqrt{2n}(n-1)}{\sqrt{(2n-1)(2n-3)}} \delta_{n-1,n^{'}}\right)\\
\label{g1f}
G^1_f[0,n,n^{'}]&=& -\f{i}{4}\left(2\delta_{n-1,n^{'}}\right),~~G^2_f[0,n,n^{'}]
=-\f{i}{2}\left(\delta_{n+1,n^{,}} + \delta_{n-1,n^{'}}\right)  
\eeqa

To prove the ultraviolet finiteness of the mass-squared corrections (\ref{phi13amp1}),(\ref{phi13amp2}) 
and (\ref{phi13amp3}), 
we set the external momenta $(l,l^{'},w)=0$ and compute the vertices in the large $n$ limit.
The vertices can be evaluated using the orthogonality condition for Hermite polynomials. 
The various four-point vertices given in (\ref{g14}) in the large $n$ limit 
assume the forms
\beqa
G_1^1(0,0,n,n)= 1\mbox{~~;~~} G_2^1(0,0,n,n)=\tilde{G}_2^1(0,0,n,n)= \hf
\eeqa
where we have used the Kronecker delta's to set $n=n^{'}$. 
The three-point vertices given in (\ref{g11p}) in the large $n$ limit become 
\beqa
G_1^{1'}(0,n,n^{'})\sim \frac{\sqrt{2n}}{2}[2\delta_{n^{'},n-1}] \mbox{~~;~~} 
G_3^1(0,n,n^{'})\sim \frac{\sqrt{2n}}{8}[8\delta_{n^{'},n-1}]
\eeqa
The three-point bosonic vertex $\tilde{G}^1_3(0,n,n^{'})$ is found to be 
identically zero for all values $n$ in (\ref{g13t}). 
Moreover the fermionic three-point vertices are exact for all values $n$. 
Hence the remaining three-point bosonic vertex (\ref{g13tp}) in the 
limit $n \rightarrow \infty$ can be written as

\beqa
\tilde{G}_3^{1'}(0,n,n^{'})\sim  \frac{\sqrt{2n}}{8}[4\delta_{n^{'},n-1}-4\delta_{n^{'},n+1}]
\eeqa

The ultraviolet contribution to the amplitude can now be written down by putting 
these asymptotic values of the vertices 
into (\ref{phi13amp1},\ref{phi13amp2},\ref{phi13amp3}). We get the following from 
the bosonic fields in the loop

\be
\label{uphi31amp1}
\Sigma^1_{\Phi_1^3-\Phi_1^3}\sim\hf N \sum_{m,n}\left[(7\times 2) \f{1}{(\o_m^2+\gamma_n)}
+(2) \f{1/2}{(\o_m^2+\lambda_n)} + (2) \f{1/2}{(\o_m^2)}  \right]
\ee

\beqa
\label{uphi31amp2}
\Sigma^2_{\Phi_1^3-\Phi_1^3}\sim-\hf qN \sum_{m,n}\left[(7\times 2) \f{2n}{(\o_m^2+\gamma_n)^2}
+ (2)\f{2n}{(\o_m^2+\lambda_n)^2}
+ (2)\f{(2n)/2}{\o_m^2(\o_{m}^2+\lambda_{n})}\right]\nonumber\\
\eeqa

Noting that in the large $n$ limit, $\lambda_n=\gamma_n\sim 2nq$,

\beqa
\Sigma^1_{\Phi_1^3-\Phi_1^3}+\Sigma^2_{\Phi_1^3-\Phi_1^3} &\sim& N \sum_{m,n}\left[(8) \f{1}{(\o_m^2+2nq)}
-(8) \f{2nq}{(\o_m^2+2nq)^2}\right]\\
&\sim &\sum_{n}\f{2}{\sqrt{2n}}
\eeqa

In the last line we have done the sum over the Matsubara frequencies $m$ and omitted all the 
finite temperature dependent pieces. 
Similarly the asymptotic form of the fermionic contribution is,

\be
\label{uphi31amp3}
\Sigma^3_{\Phi_1^3-\Phi_1^3}\sim- N \sum_{m,n}\left[ (4)\f{1}{(\o_m^2+\lambda^{'}_n)}-(4)\f{1}{(i\o_m
-\sqrt{\lambda^{'}_n})^2}\right]
\sim -\sum_{n}\f{2}{\sqrt{2n}}
\ee

Thus the one-loop $\Phi_1^3-\Phi_1^3$ amplitude is ultraviolet finite.

One can exactly compute the various amplitudes in (\ref{phi13amp1}),(\ref{phi13amp2}) 
and (\ref{phi13amp3}) using the various vertices presented in (\ref{g14}-\ref{g1f}) 
and their corresponding propagators (see Appendix (\ref{matsubara})). In particular one can   
write down the effective mass-squared for the field $\Phi^3_1$ as a function of $q$ and $\beta$ as

\begin{equation}
\label{mphi}
m^2_{\Phi^3_1}(q,\beta) = m^2_{10} + m^2_1(q,\beta)
\end{equation}
where $ m^2_{10}$ denotes the zero temperature quantum corrections which is made dimensionless
by dividing the physical $m_{10}^2$ by $g^2$ and $m^2_1(q,\beta)$ denotes 
the temperature dependent mass-squared corrections for the
massless field $\Phi^3_1$. The zero temperature quantum corrections for all $n$ can be written as 

%\beqa
%m^2_{10}&=& \f{1}{4\sqrt{2n+1}} \left(7 + \f{(n-1)(8 n^3 + 16n^2 + 1)}{4(2n-1)(2n+1)^2}\right) + \f{1}{\sqrt{2n-1}} \left(\hf - 
%\f{n(28n^2 - 24n - 25)}{4(2n+1)(2n-3)}\right)\nonumber\\
%&+& \f{7}{4}\f{n+1}{\sqrt{2n+3}} - \f{n}{4 (2n-3)^{\f{3}{2}}}\left(\f{(n-1)(n-2)}{(2n-1)}- \f{2n-1}{2n-3}\right) - \f{2}{\sqrt{2n}}
%\eeqa

\beqa
\label{massqp31}
m^2_{10}&=& \left[\sum^\infty_{n=0}\f{7}{2 \sqrt{2n+1}} + \sum^\infty_{n=2}\f{1}{4 \sqrt{2n-1}}
-\f{7}{2}\sum^\infty_{n=1}\left(\f{n}{\sqrt{2n-1}}-\f{n}{\sqrt{2n+1}}\right) 
\right.\nonumber\\
&+&\left.\sum^\infty_{n=2} \left(\f{(n-1)}{(2n-1)^{\f{5}{2}}}\left(\f{(n+1)^2}{(2n+1)} 
+ \f{n(n-1)}{(2n-3)}\right)
-2\f{n(n-1)(n-2)}{(2n-1)(2n-3)}\left(\f{1}{\sqrt{2n-3}} 
- \f{1}{\sqrt{2n-1}}\right)\right)\right]\nonumber\\
&-&\sum^{\infty}_{n=1}\f{4}{\sqrt{2n} +\sqrt{2n-2}} 
\eeqa

The zero temperature quantum correction given by 
(\ref{massqp31}) can be evaluated numerically. The convergent value is given by

\begin{equation}
\label{zetaqp31}
m^2_{10} = 1.579  
\end{equation}
The temperature dependent part in (\ref{mphi}) can be written as

\beqa
\label{masstp31}
m^2_1(q,\beta)&=& \left[\sum^\infty_{n=0} \f{7}{\sqrt{2n+1}}\f{1}{\left(e^{\sqrt{(2n+1)q}\beta}-1\right)}
+ \sum^\infty_{n=2} \f{1}{2\sqrt{2n-1}}\f{1}{\left(e^{\sqrt{(2n-1)q}\beta}-1\right)}\right.\nonumber\\
&-& \left. 7 \sum^\infty_{n=1} \left(\f{n}{\sqrt{2n-1}}\f{1}{\left(e^{\sqrt{(2n-1)q}\beta}-1\right)} - 
\f{n}{\sqrt{2n+1}}\f{1}{\left(e^{\sqrt{(2n+1)q}\beta}-1\right)}\right)\right.\nonumber\\
&-&\left. 4 \sum^\infty_{n=1} \f{n(n-1)(n-2)}{(2n-1)(2n-3)}
\left(\f{1}{\sqrt{2n-3}}\f{1}{\left(e^{\sqrt{(2n-3)q}\beta}-1\right)}- 
\f{1}{\sqrt{2n-1}}\f{1}{\left(e^{\sqrt{(2n-1)q}\beta}-1\right)}\right)\right.\nonumber\\
&+&\left. 2\sum^\infty_{n=2}\left(\f{(n-1)}{(2n-1)^{\f{5}{2}}}\left(\f{(n+1)^2}{(2n+1)} 
+ \f{n(n-1)}{2n-3}\right) \f{1}{\left(e^{\sqrt{(2n-1)q}\beta}-1\right)}\right)
\right.\nonumber\\
&+&\left.\left(\hf-2\sum^\infty_{n=2} \left(\f{(n-1)}{(2n-1)}\left(\f{(n+1)^2}{(2n+1)^2} 
+ \f{n(n-1)}{(2n-3)^2}\right)\right)\right) 
\sum^{\infty}_{m=-\infty} \f{\sqrt{q}\beta}{4\pi^2 m^2}
\right]\nonumber\\      
&+&\left[\sum^{\infty}_{n=1}\left(\left(4\f{1}{(\sqrt{2n}+ \sqrt{2(n-1)})}
+2\f{(\sqrt{2(n+1)} + \sqrt{2(n-1)}-2 \sqrt{2n})}{(\sqrt{2n}- \sqrt{2(n-1)})(\sqrt{2(n+1)}
- \sqrt{2n})}\right)\f{1}{\left(e^{\sqrt{2nq}\beta}+1\right)}\right.\right.\nonumber\\
&+&\left.\left. \left(4\f{1}{(\sqrt{2n}+ \sqrt{2(n-1)})} - \f{2}{\sqrt{2n} 
- \sqrt{2(n-1)}}\right)\f{1}{\left(e^{\sqrt{2(n-1)q}\beta}+1\right)}\right.\right.\nonumber\\
&+&\left.\left.\f{2}{\sqrt{2(n+1)} - \sqrt{2n}}\f{1}{\left(e^{\sqrt{2(n+1)q}\beta}+1\right)}\right)
\right]
\eeqa

%\beqa
%m^2_1(q,\beta)&=& \left(\f{7}{\sqrt{2n+1}}- \f{7}{4} \f{1}{\sqrt{2n+1}} - \f{n(n+1)(n-1)}{(2n-1)(2n+1)^{\f{3}{2}}} 
%+ \f{1}{4}\f{(n-1)(2n-1)}{(2n+1)^\f{5}{2}}\right) \f{1}{(e^{\sqrt{(2n+1)q}\beta}-1)}\nonumber\\ 
%&+& \left(\f{1}{\sqrt{2n-1}} \f{1}{(e^{\sqrt{\lambda_{n}}\beta}-1)}\right)\nonumber\\
%&-& \f{7}{4} \left(\f{1}{2\sqrt{2n+1}} \f{1}{(e^{\sqrt{\gamma_n}\beta}-1)} - \f{n+1}{\sqrt{2n+3}} \f{1}{(e^{\sqrt{\gamma_{n+1}}\beta}-1)}\right.\nonumber\\
%&+&\left. \f{n}{\sqrt{2n-1}} \f{1}{(e^{\sqrt{\gamma_{n-1}}\beta}-1)}\right)\nonumber\\
%&-& 2 \left(\f{n (n-1)}{(2n+1)(2n-3)}\f{1}{\sqrt{2n-1}}\f{1}{(e^{\sqrt{\lambda_n}\beta}-1)} - \f{n(n+1)(n-1)}{(2n-1)(2n+1)^{\f{3}{2}}}
%\f{1}{(e^{\sqrt{\lambda_{n+1}}\beta}-1)} \right.\nonumber\\
%&+&\left.\f{n(n-1)(n-2)}{(2n-1)(2n-3)^{\f{3}{2}}}\f{1}{(e^{\sqrt{\lambda_{n-1}}\beta}-1)}\right)\nonumber\\
%&+& \f{1}{4} \left(\f{(n-1)(2n-1)}{(2n+1)^\f{5}{2}}\f{1}{(e^{\sqrt{\lambda_{n+1}}\beta}-1)} + \f{n(2n-1)}{(2n-3)^\f{5}{2}}\f{1}{(e^{\sqrt{\lambda_{n-1}}\beta}-1)}\right)
%\eeqa   

\subsection{Two point function for $\Phi_I^3$ $(I\ne 1)$}\label{phiIamp}

Let us now write down all the amplitudes that constitute the finite temperature one-loop 
mass-squared corrections to the tree-level massless
fields $\Phi^3_I(m,l)$, where $I\ne1$. The computation of the vertices involving the 
fermions in this section has been done explicitly for $I=2$. 
However the $SO(7)$ invariance of the theory implies that the two point function is the 
same for all $I=2\cdots 8$.
Using the vertices listed in appendix \ref{vphii3} we write below the expressions for 
the two point function. The Feynman diagrams involving 
the four-point bosonic interactions are depicted in the figure \ref{masslessp3I1}.      

\be
\label{phi3iamp1}
\Sigma^1_{\Phi_I^3-\Phi_I^3}=\hf N \sum_{m,n}\left[(6\times 2) \f{G_1^I(l,l^{'},n,n)}
{(\o_m^2+\gamma_n)}+(2) \f{G_2^I(l,l^{'},n,n)}{(\o_m^2+\lambda_n)}
+(2) \f{\tilde{G}_2^I(l,l^{'},n,n)}{\o_m^2}\right]\delta_{w+w^{'}}
\ee

\begin{figure}[h]
\begin{center}
\begin{psfrags}
\psfrag{c1}[][]{$\Phi^3_I(w,l)$}
\psfrag{c2}[][]{$\Phi^3_I(w^{'},l^{'})$}
\psfrag{c3}[][]{$\Phi^{(1,2)}_I(m,n)$}
\psfrag{c4}[][]{$C_{m,n}$}
\psfrag{c5}[][]{$\tilde{C}_{m,n}$}
\psfrag{v1}[][]{$V^I_{1}$}
\psfrag{v2}[][]{$V^I_{2}$}
\psfrag{v3}[][]{$\tilde{V}^I_{2}$}
\psfrag{a1}[][]{(a)}
\psfrag{a2}[][]{(b)}
\psfrag{a3}[][]{(c)}
\includegraphics[ width= 12cm,angle=0]{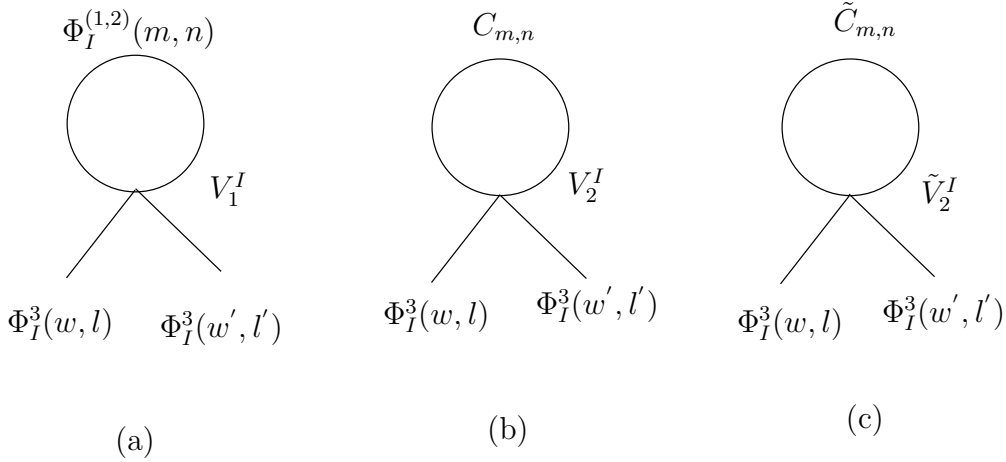}
\end{psfrags}
\caption{{Feynman diagrams with four-point vertices $V^I_{1},~V^I_2,~\tilde{V}^I_2$.}}
\label{masslessp3I1}
\end{center}
\end{figure} 
The three terms in (\ref{phi3iamp1}) are represented by the Feynman diagrams in 
figures \ref{masslessp3I1}(a), \ref{masslessp3I1}(b) and \ref{masslessp3I1}(c)
in the same order. The first term involves the fields $\Phi^{(1,2)}_I$, $I\ne1$, 
the second term involves the fields $C_{m,n}$ and the third term comprises of 
the fields $\tilde{C}_{m,n}$.
Similarly the three-point bosonic interactions of $\Phi^3_I(m,l)$ are represented 
in the Feynman diagrams in Figure \ref{masslessp3I2}.  
\begin{figure}[h]
\begin{center}
\begin{psfrags}
\psfrag{c1}[][]{$\Phi^3_I(w,l)$}
\psfrag{c2}[][]{$\Phi^3_I(w^{'},l^{'})$}
\psfrag{c3}[][]{$\Phi^{(1,2)}_I(m,n)$}
\psfrag{c4}[][]{$C_{m,n}$}
\psfrag{c5}[][]{$\Phi^{(1,2)}_I(m,n)$}
\psfrag{c6}[][]{$\tilde{C}_{m,n}$}
\psfrag{v1}[][]{$V^I_{3}$}
\psfrag{v2}[][]{$V^{I*}_{3}$}
\psfrag{v3}[][]{$\tilde{V}^{I}_{3}$}
\psfrag{v4}[][]{$\tilde{V}^{I*}_{3}$}
\psfrag{a1}[][]{(a)}
\psfrag{a2}[][]{(b)}
\psfrag{a3}[][]{(c)}
\psfrag{a4}[][]{(d)}
\psfrag{a5}[][]{(e)}
\includegraphics[ width= 12cm,angle=0]{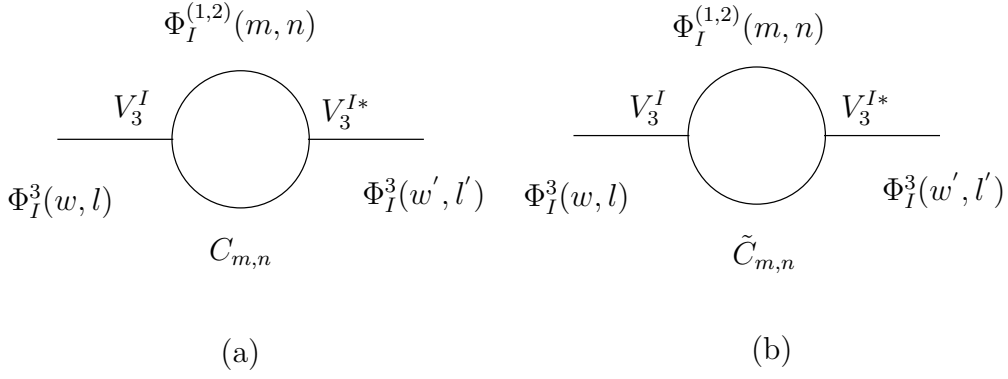}
\end{psfrags}
\caption{{Feynman diagrams with the three-point vertices $V^I_{3},~\tilde{V}^I_2$}}
\label{masslessp3I2}
\end{center}
\end{figure} 
The two-point function involving the bosonic three-point vertices and contributing to the one-loop 
finite temperature mass-corrections for the field $\Phi^3_I$, $I\ne1$ is given by 
\beqa
\label{phi3iamp2}
\Sigma^2_{\Phi_I^3-\Phi_I^3}&=&-\hf qN \sum_{m,n,n^{'}}\left[(2)\f{G_3^I(l,n,n^{'})G_3^I(l^{'},n,n^{'})}
{(\o_m^2+\gamma_n)(\o_{m^{'}}^2+\lambda_{n^{'}})}
+(2)\f{\tilde{G}_3^I(l,n,n^{'})\tilde{G}_3^I(l^{'},n,n^{'})}{(\o_m^2
+\gamma_{n^{'}})\o_{m^{'}}^2}\right]\delta_{w+w^{'}}
\eeqa

where $w=m^{'}+m$. The first term in (\ref{phi3iamp2}) involving the three-point vertex 
$G_3^I(l,n,n^{'})$ involves the fields $\Phi^{(1.2)}_I$, $I\ne1$ and
$C_{m,n}$s. The corresponding Feynman diagram is shown in figure \ref{masslessp3I2}(a). 
Similarly the second term in (\ref{phi3iamp2}) involving the three-point vertex 
$\tilde{G}_3^I(l,n,n^{'})$ involves the fields $\Phi^{(1.2)}_I$, $I\ne1$ and
$\tilde{C}_{m,n}$s. The relevant Feynman diagram is shown in figure \ref{masslessp3I2}(b). 
The Feynman diagram involving the three-point vertex with fermions is drawn in Figure \ref{masslessp3If}.

\begin{figure}[h]
\begin{center}
\begin{psfrags}
\psfrag{c1}[][]{$\Phi^3_I(w,l)$}
\psfrag{c2}[][]{$\Phi^3_I(w^{'},l^{'})$}
\psfrag{c3}[][]{$\theta_i(m,n)$}
%\psfrag{c4}[][]{$\theta_i(m,n)$}
%\psfrag{c5}[][]{$\theta^{*}_i(m^{'},n^{'})$}
%\psfrag{c6}[][]{$\theta^{*}_i(m^{'},n^{'})$}
\psfrag{v1}[][]{$V^I_{f}$}
\psfrag{v2}[][]{$V^{I*}_{f}$}
\psfrag{a1}[][]{(a)}
\psfrag{a2}[][]{(b)}
\psfrag{a3}[][]{(c)}
\psfrag{a4}[][]{(d)}
\psfrag{a5}[][]{(e)}
\includegraphics[ width= 5cm,angle=0]{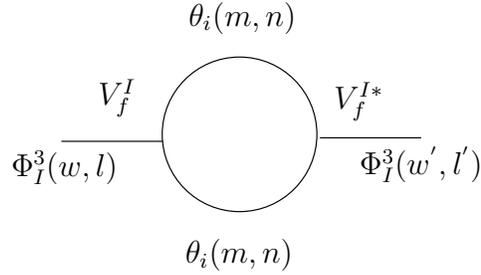}
\end{psfrags}
\caption{{Feynman diagram with three-point vertex $V^I_{f}$ and its complex conjugate.}}
\label{masslessp3If}
\end{center}
\end{figure}

The corresponding amplitude is 
\be
\label{phi3iamp3}
\Sigma^3_{\Phi_I^3-\Phi_I^3}=N \sum_{m,n,n^{'}}\left[ (8)\f{G_f^I(l,n,n^{'})G_f^{I*}(l^{'},n,n^{'})}
{(i \o_m - \sqrt{\lambda^{'}_{n}})
(i \o_{m^{'}}-\sqrt{\lambda^{'}_{n^{'}}})}\right]\delta_{w+w^{'}}
\ee
where $w=m^{'}+m$. As in the case of $\Phi^3_1$, we can use the orthogonality relation 
for Hermite Polynomials to compute exactly, the various 
vertices given in section (\ref{vphii3}) and constituting the mass-squared corrections 
(\ref{phi3iamp1}),(\ref{phi3iamp2}) and (\ref{phi3iamp3}). 
We do these computations after setting the external momenta $l=l^{'}= w=w^{'}=0$. 

\beqa
\label{gI1}
G^I_1[0,0,n,n^{'}] &=& \delta_{n,n^{'}}\\
\label{gI2}
G^I_2[0,0,n,n] &=& \delta_{n,n^{'}}\\
\label{gI2t}
\tilde{G}^I_2[0,0,n,n^{'}] &=& \delta_{n,n^{'}}\\
\label{gI3}
G^I_3[0,n,n^{'}] &=& 0\\
\label{gI3t}
\tilde{G}^I_3[0,n,n^{'}] &=&\sqrt{2n-1} \delta_{n-1,n^{'}}\\
\label{gIf}
G^I_f[0,n,n^{'}]&=& i \delta_{n,n^{'}} 
\eeqa

We now proceed to establish the UV finiteness of finite temperature mass-squared 
corrections to $\Phi^3_I$.
We first analyze the large $n$-behaviour of the various vertices for the one-loop mass-squared 
corrections(\ref{phi3iamp1})
(\ref{phi3iamp2}) and (\ref{phi3iamp3}). The four-point vertices in the amplitudes constituting 
the two-point function (\ref{phi3iamp1}) 
are associated with only one kind of propagator. In the large $n$ limit the four-point vertices 
computed in (\ref{gI1}), (\ref{gI2}) and (\ref{gI2t}) can be written as 
\beqa
G_1^I(0,0,n,n)=G_2^I(0,0,n,n)=\tilde{G}_2^I(0,0,n,n)= 1
\eeqa
where we have used the Kronecker deltas to set $n=n^{'}$. Note that the vertex $G^I_3(0,n,n^{'})$ 
is identically zero for all values of $n$ as shown in (\ref{gI3}). 
Furthermore the fermionic vertex $G^I_f(0,n,n^{'})$ (\ref{gIf}) can be exactly computed 
for all $n$ and remains the same in the UV limit. 
The remaining three-point bosonic vertex $\tilde{G}^I_3(0,n,n^{'})$ (\ref{gI3t}) 
in the UV limit becomes
\beqa
\tilde{G}_3^I(0,n,n^{'})\sim \sqrt{2n}[\delta_{n^{'},n-1}] 
\eeqa

This large $n$ behaviour of the vertices in turn gives rise to the following asymptotic 
forms of the two-point functions for the tree-level massless field $\Phi^3_I$.

\be
\label{uphi3iamp1}
\Sigma^1_{\Phi_I^3-\Phi_I^3}\sim\hf N \sum_{m,n}\left[(6\times 2) \f{1}{(\o_m^2+\gamma_n)}
+(2) \f{1}{(\o_m^2+\lambda_n)} + (2) \f{1}{(\o_m^2)}  \right]
\ee

\beqa
\label{uphi3iamp2}
\Sigma^2_{\Phi_I^3-\Phi_I^3}\sim-\hf qN \sum_{m,n}\left[ \f{(2n)}{\o_m^2(\o_{m}^2+\lambda_{n})}\right]
\eeqa

Thus the total bosonic contribution in the limit $n \rightarrow \infty$ can be written as

\beqa\label{utotbi}
\Sigma^1_{\Phi_I^3-\Phi_I^3}+\Sigma^2_{\Phi_I^3-\Phi_I^3} \sim N \sum_{m,n}\left[(8) \f{1}{(\o_m^2+2nq)}\right]
\sim \sum_{n}\f{4}{\sqrt{2n}}
\eeqa

In this large $n$ limit, the contribution from the fermions coming from $\Sigma^3_{\Phi_I^3-\Phi_I^3}$ 
is same as the right hand side of eqn(\ref{utotbi}) with opposite sign.

Combining the vertices in (\ref{gI1}- \ref{gIf}) with their respective propagators
(see Appendix (\ref{matsubara})), the 
effective mass-squared corrections  for $\Phi^3_I$, $I\ne1$, can be written down as 
a function of $q$ and $\beta$ in the following form;

\begin{equation}
\label{mp3I}
m^2_{\Phi^3_I} (q,\beta) = m^2_{I0} + m^2_{I1}(q,\beta). 
\end{equation}

where $ m^2_{I0}$ and $m^2_{I1}(q,\beta)$ denote the zero temperature quantum corrections 
and the finite temperature corrections respectively 
to the tree-level massless field $\Phi^3_I$, $I\ne1$. The zero temperature 
quantum correction here is made dimensionless in the same way as in the case of $m^2_{10}$ in (\ref{mphi}) and 
can be exactly computed as in the case of $\Phi^3_1$
and found to be 
\begin{equation}
\label{massqp3I}
m^2_{I0} = \sum^{\infty}_{n=0}\f{3}{\sqrt{2n+1}} +  \sum^{\infty}_{n=1}\f{1}{\sqrt{2n-1}} 
- \sum^{\infty}_{n=1}\f{4}{\sqrt{2n}}
\end{equation}
for all $n$, where the first two terms under summation come from the bosonic contributions
and the last term comes from the fermionic contributions.
The various sums in (\ref{massqp3I}) can be reorganized and written in terms of the regularized 
Riemann Zeta function $\zeta\left(\hf\right)$.
The dimensionless zero temperature quantum correction can be evaluated as  
\begin{equation}
\label{zetap3I}
m^2_{I0} = (4(1-\sqrt{2})\zeta\left(\hf\right)-1) = 1.495 
\end{equation}
Similarly the finite temperature part $ m^2_{I1}(q,\beta)$ can be written as,
\beqa
\label{masstp3I}
 m^2_{I1}(q,\beta) &=& \sum^{\infty}_{n=0}\f{6}{\sqrt{2 n +1}}\f{1}{e^{\sqrt{(2n+1)q}\beta}-1}
+\sum^{\infty}_{n=1}\f{2}{\sqrt{2n-1}}\f{1}{e^{\sqrt{(2n-1)q}\beta}-1}\nonumber\\ 
&+& \sum^{\infty}_{n=1}\f{8}{\sqrt{2n}} \f{1}{e^{\sqrt{2 n q}\beta}+1}
\eeqa

\subsection{Two point function for $A^3_x$}\label{A3xamp}

We give below that expression for the two point one loop amplitude for $A^3_x$. The vertices are 
worked out in appendix \ref{va13}. 
The Feynman diagrams comprising the four-point bosonic interactions is given in figure \ref{masslessA31}. 

\begin{figure}[h]
\begin{center}
\begin{psfrags}
\psfrag{c1}[][]{$A^3_x(w,l)$}
\psfrag{c2}[][]{$A^3_x(w^{'},l^{'})$}
\psfrag{c3}[][]{$\Phi^{(1,2)}_I{m,n}$}
\psfrag{c4}[][]{$C_{m,n}$}
\psfrag{c5}[][]{$\tilde{C}_{m,n}$}
\psfrag{v1}[][]{$V^A_{1}$}
\psfrag{v2}[][]{$V^A_{2}$}
\psfrag{v3}[][]{$\tilde{V}^A_{2}$}
\psfrag{a1}[][]{(a)}
\psfrag{a2}[][]{(b)}
\psfrag{a3}[][]{(c)}
\psfrag{a4}[][]{(d)}
\psfrag{a5}[][]{(e)}
\includegraphics[ width= 12cm,angle=0]{massless4point.eps}
\end{psfrags}
\caption{{Feynman diagrams with four-point vertices $V^A_{1},~V^A_2,~\tilde{V}^A_2$.}}
\label{masslessA31}
\end{center}
\end{figure} 

The amplitudes that are represented by the Feynman diagrams in Figure \ref{masslessA31} are 
collected together into the two-point 
finite temperature mass-squared corrections to the tree-level massless field $A^3_x$ in the 
following equation, namely   

\beqa
\label{A3xamp1}
\Sigma^1_{A^3_x-A^3_x}=\hf N \sum_{m,n}\left[(7\times 2) \f{G_1^A(n,n,l,l^{'})}
{(\o_m^2+\gamma_n)}+(2) \f{G_2^A(n,n,l,l^{'})}{(\o_m^2+\lambda_n)} 
+ (2) \f{\tilde{G}_2^A(n,n,l,l^{'})}{\o_m^2}\right]\delta_{w+w^{'}}\nonumber\\
\eeqa

The first second and third terms in (\ref{A3xamp1}) are represented by the Feynman diagrams 
in figures \ref{masslessA31}(a), \ref{masslessA31}(b) and 
\ref{masslessA31}(c) respectively. The fields involved in four-point vertices in the first, 
second and third terms  are $\Phi^{(1,2)}_I$, $I\ne1$, $C_{m,n}$ and  
$\tilde{C}_{m,n}$ respectively. The Feynman diagrams depicting the various three-point bosonic 
interactions including  $A^3_x$ are presented 
in figure \ref{masslessA32}.

\beqa
\label{A3xamp2}
\Sigma^2_{A^3_x-A^3_x}=&-&\hf qN \sum_{m,n,n^{'}}\left[ \f{G_3^A(n,n^{'},l)G_3^A(n,n^{'}, 
l^{'})}{(\o_m^2+\lambda_n)(\o_{m^{'}}^2+\lambda_{n^{'}})} 
+ \f{\tilde{G}_3^{A}(n,n^{'},l)\tilde{G}_3^{A}(n,n^{'},l^{'})}{(\o_m^2)(\o_{m^{'}}^2)}\right.\nonumber\\ 
&+& \left.(2)\f{\tilde{G}_3^{A'}(n,n^{'},l)\tilde{G}_3^{A'}(n,n^{'},l^{'})}{(\o_m^2+\lambda_n)
(\o_{m^{'}}^2)} + (7)\f{G_4^A(n,n^{'},l)G_4^A(n,n^{'},l^{'})}
{(\o_m^2+\gamma_n)(\o_{m^{'}}^2+\gamma_{n^{'}})}\right]\delta_{w+w^{'}}\nonumber\\
\eeqa

where $w=m^{'}+m$. The first term in (\ref{A3xamp2}) comprising the three-point 
vertex $G^{A}_3(l,n,n^{'})$ involves the fields $C_{m,n}$ and $C^{'}_{m,n}$. 
in the loops and is represented by the Feynman diagram in figure \ref{masslessA32}(a). 
The second term in (\ref{A3xamp2}) involving the vertex
$\tilde{G}^{A}_3(l,n, n^{'})$ comprises of the fields $\tilde{C}_{m,n}$s and 
$\tilde{C}^{'}_{m,n}$ in the loop. The corresponding Feynman diagram is given 
in figure \ref{masslessA32}(b). The third term has contributions from 
$\tilde{C}_{m,n}$s and $\tilde{C}_{m,n}$ in the loop with 
Feynman diagram shown in figure \ref{masslessA32}(c). Lastly the fourth 
term depicted in figure \ref{masslessA32}(d) has contributions 
from the fields $\Phi^{(1,2)}_I$, $I\ne1$.   

\begin{figure}[h]
\begin{center}
\begin{psfrags}
\psfrag{c1}[][]{$A^3_x(w,l)$}
\psfrag{c2}[][]{$A^3_x(w^{'},l^{'})$}
\psfrag{c3}[][]{$C_{m,n}$}
\psfrag{c4}[][]{$C_{m,n}$}
\psfrag{c5}[][]{$\tilde{C}_{m,n}$}
\psfrag{c6}[][]{$\tilde{C}_{m,n}$}
\psfrag{c7}[][]{$C_{m,n}$}
\psfrag{c8}[][]{$\tilde{C}_{m,n}$}
\psfrag{c9}[][]{$\Phi^{(1,2)}_I{m,n}$}
\psfrag{c10}[][]{$\Phi^{(1,2)}_I{m,n}$}
\psfrag{v1}[][]{$V^A_{3}$}
\psfrag{v2}[][]{$V^{A*}_{3}$}
\psfrag{v3}[][]{$\tilde{V}^A_{3}$}
\psfrag{v4}[][]{$\tilde{V}^{A*}_{3}$}
\psfrag{v5}[][]{$\tilde{V}^{A'}_{3}$}
\psfrag{v6}[][]{$\tilde{V}^{A'*}_{3}$}
\psfrag{v7}[][]{$V^{A}_{4}$}
\psfrag{v8}[][]{$V^{A*}_{4}$}
\psfrag{a1}[][]{(a)}
\psfrag{a2}[][]{(b)}
\psfrag{a3}[][]{(c)}
\psfrag{a4}[][]{(d)}
\psfrag{a5}[][]{(e)}
\includegraphics[ width= 12cm,angle=0]{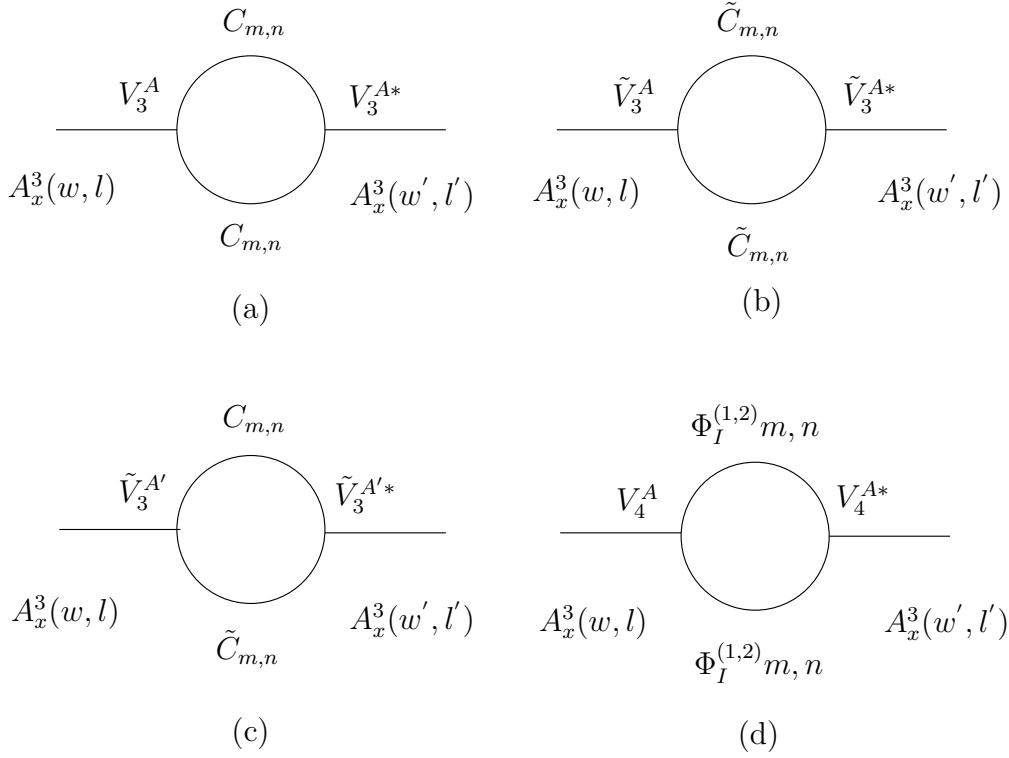}
\end{psfrags}
\caption{{Feynman diagrams with three-point vertices 
$V^{A}_{3},~\tilde{V}^A_3,~\tilde{V}^{A'}_3$ and $V^4_3$.}}
\label{masslessA32}
\end{center}
\end{figure} 

Similarly the amplitude involving fermions in the loop is (\ref{A3xamp3}).
The corresponding Feynman diagrams are presented in figure \ref{masslessA3f} 
.
\beqa
\label{A3xamp3}
\Sigma^3_{A^3_x-A^3_x}=&&\hf q N\sum_{m,n,n^{'}}\left[(16)\f{G_f^{A1}(n,n^{'},l)G_f^{A1*}
(n,n^{'},l^{'})}{(i \o_m - \sqrt{\lambda^{'}_n})(i \o_{m^{'}} - \sqrt{\lambda^{'}_{n^{'}}})})\right.\nonumber\\
&&\left.-(16)\f{G_f^{A2}(n,n^{'},l)G_f^{A2*}
(n,n^{'},l^{'})}{(i \o_m - \sqrt{\lambda^{'}_n})(-i \o_{m^{'}} - \sqrt{\lambda^{'}_{n^{'}}})}) \right]\delta_{w+w^{'}}
\eeqa

where $w=m^{'}+m$. The fermionic three-point vertex $G^{A1}_f(n,n^{'},l)$ constituting 
the first term in the two-point function (\ref{A3xamp3}) has contributions 
from the fermionic fields $\theta_i(m,n)$. The amplitude is represented in the 
Feynman diagram presented in figure \ref{masslessA3f}(a). The second term 
in (\ref{A3xamp3}) on the other hand involves  the fields $\theta_i(m,n)$ and their 
complex conjugate $\theta^{*}_i(m,n)$ in the vertex $G^{A2}_f(n,n^{'},l)$. 
The corresponding Feynman diagram is given in figure \ref{masslessA3f}(b). 

\begin{figure}[h]
\begin{center}
\begin{psfrags}
\psfrag{c1}[][]{$A^3_x(w,l)$}
\psfrag{c2}[][]{$A^3_x(w^{'},l^{'})$}
\psfrag{c3}[][]{$\theta_i(m,n)$}
%\psfrag{c4}[][]{$\theta_i(m,n)$}
%\psfrag{c5}[][]{$\theta^{*}_i(m,n)$}
%\psfrag{c6}[][]{$\theta^{*}_i(m,n)$}
%\psfrag{c7}[][]{$\theta^{*}_i(m,n)$}
%\psfrag{c8}[][]{$\theta_i(m,n)$}
%\psfrag{c9}[][]{$\theta_i(m,n)$}
%\psfrag{c10}[][]{$\theta^{*}_i(m,n)$}
\psfrag{v1}[][]{$V^{A1}_{f}$}
\psfrag{v2}[][]{$V^{A1*}_{f}$}
\psfrag{v3}[][]{$\tilde{V}^{A2}_{f}$}
\psfrag{v4}[][]{$\tilde{V}^{A2*}_{f}$}
\psfrag{a1}[][]{(a)}
\psfrag{a2}[][]{(b)}
\psfrag{a3}[][]{(c)}
\psfrag{a4}[][]{(d)}
\psfrag{a5}[][]{(e)}
\includegraphics[ width= 14cm,angle=0]{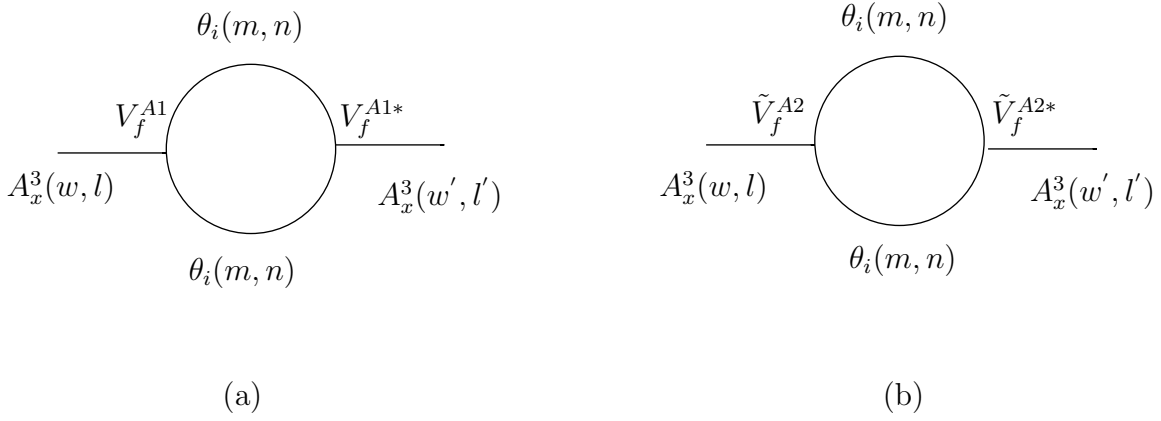}
\end{psfrags}
\caption{{Feynman diagrams involving  three-point vertices $V^{A1}_{f},~V^{A2}_f$.}}
\label{masslessA3f}
\end{center}
\end{figure}

The vertices given in section (\ref{va13}) and participating in the two-point functions 
that produce the mass-squared 
corrections to the tree-level massless field $A^3_x$ can be exactly computed following 
the same procedure as discussed for $\Phi^3_1$ and $\Phi^3_I$, $I\ne1$.   

\beqa
\label{gA1}
G^A_1[n,n^{'},0,0] &=& \delta_{n,n^{'}},~~ 
\label{gA2}
G^A_2[n,n^{'},0,0]=\hf \delta_{n,n^{'}},~~\tilde{G}^A_2[n,n^{'},0,0]=\hf \delta_{n,n^{'}},\\
\label{gA3}
G^{A}_3[n,n^{'},0] &=& 2\left(\sqrt{\f{2n(n+1)(n-1)}{(2n-1)(2n+1)}} \delta_{n+1,n^{'}} - 
\sqrt{\f{2n(n-1)(n-2)}{(2n-1)(2n-3)}} \delta_{n-1,n^{'}}\right),\\
\tilde{G}^A_3[n,n^{,},0] &=& 0, \\
\label{gA3tp}
\tilde{G}^{A'}_3[n,n^{'},0]&=&-\sqrt{2}\left(\f{(n+1)\sqrt{n-1}}{\sqrt{(2n-1)(2n+1)}} 
\delta_{n+1,n^{'}} + \f{\sqrt{n}(n-1)}{\sqrt{(2n-1)(2n-3)}}\delta_{n-1,n^{'}}\right),\\
\label{gA4}
G^A_4[n,n^{'},0]&=& \left(\sqrt{2(n+1)}\delta_{n+1,n^{'}} - \sqrt{2n}\delta_{n-1,n^{'}}\right),\\
\label{gAf1}
G^{A1}_f[n,n^{'},0]&=& -\f{i}{2}\left(\delta_{n+1,n^{,}}+\delta_{n-1,n^{'}}\right),\\
\label{gAf2}
G^{A2}_f[n,n^{'},0]&=&-\f{i}{2}\left(2\delta_{n+1,n^{,}}-2\delta_{n-1,n^{'}}\right).  
\eeqa

In the same spirit as for $\Phi^3_1$ and $\Phi^3_I$ we now proceed to establish the 
UV finiteness for the one-loop two-point functions for the field $A^3_x$.
In the large $n$ limit the various vertices  in eqns (\ref{gA2}-\ref{gAf2}) reduce to

\beqa
G_1^A(n,n,0)= 1\mbox{~~;~~} G_2^A(n,n,0)=\tilde{G}_2^A(n,n,0,0)= \hf
\eeqa

\beqa
G_3^{A}(n,n^{'},0)\sim \frac{\sqrt{2n}}{8}[8\delta_{n^{'},n-1}-8\delta_{n^{'},n+1}] 
\mbox{~~;~~} \tilde{G}_3^{A^{'}}(n,n^{'},0)\sim 
-\frac{\sqrt{2n}}{8}[4\delta_{n^{'},n-1}+4\delta_{n^{'},n+1}]
\eeqa

\beqa
\tilde{G}_3^A(n,n^{'},0)= 0 \mbox{~~;~~}\tilde{G}_4^{A}(n,n^{'},0)\sim  
\sqrt{2n}[\delta_{n^{'},n+1}-\delta_{n^{'},n-1}]
\eeqa

%\beqa
%G_f^{A1}(n,n^{'},0)\sim \frac{i}{2}[\delta_{n^{'},n-1}+\delta_{n^{'},n+1}] \mbox{~~;~~}G_f^{A2}(n,n^{'},0)\sim  \frac{i}{2}[2\delta_{n^%%{'},n-1}-2\delta_{n^{'},n+1}]
%\eeqa

With these, the amplitudes in the ultraviolet limit is same as the right hand side of the equations 
(\ref{uphi31amp1}), (\ref{uphi31amp1}) and (\ref{uphi31amp3}), 
thus showing that the one-loop $A^3_x-A^3_x$ amplitude is ultraviolet finite.

Once again we write down the effective mass-squared for the field $A^3_x$ as a 
function of $q$ and $\beta$ as

\begin{equation}
\label{mA}
m^2_{A^3_x}(q,\beta) = m^2_{x0} + m^2_{x1}(q,\beta)
\end{equation}
where $ m^2_{x0}$ denotes the dimensionless (same as $m^2_{10}$ and $m^2{I0}$) zero temperature 
quantum corrections and $m^2_{x1}(q,\beta)$ 
denotes the temperature dependent mass-squared corrections for the
tree-level massless field $A^3_x$. The zero temperature quantum corrections can be written as 

%\beqa
%m^2_{10}&=& \f{1}{4\sqrt{2n+1}} \left(7 + \f{(n-1)(8 n^3 + 16n^2 + 1)}{4(2n-1)(2n+1)^2}\right) + \f{1}{\sqrt{2n-1}} \left(\hf - 
%\f{n(28n^2 - 24n - 25)}{4(2n+1)(2n-3)}\right)\nonumber\\
%&+& \f{7}{4}\f{n+1}{\sqrt{2n+3}} - \f{n}{4 (2n-3)^{\f{3}{2}}}\left(\f{(n-1)(n-2)}{(2n-1)}- \f{2n-1}{2n-3}\right) - \f{2}{\sqrt{2n}}
%\eeqa

\beqa
\label{massqA3x}
m^2_{x0}&=& \left[\sum^\infty_{n=0}\f{7}{2 \sqrt{2n+1}} + \sum^\infty_{n=2}\f{1}{4 \sqrt{2n-1}}
-\f{7}{4}\sum^\infty_{n=0}\left(\f{1}{\sqrt{2n+1}}- \f{n+1}{\sqrt{2n+3}}\right)
-\f{7}{4}\sum^\infty_{n=1}\f{n}{\sqrt{2n-1}} 
\right.\nonumber\\
&-&\left.\sum^\infty_{n=2} \left(\f{n(n-1)}{(2n+1)(2n-3)}\f{1}{\sqrt{2n-1}} 
- \f{n(n+1)(n-1)}{(2n-1)(2n+1)}\f{1}{\sqrt{2n+1}}
+ \f{n(n-1)(n-2)}{(2n-1)(2n-3)}\f{1}{\sqrt{2n-3}}\right)\right.\nonumber\\
&+&\left.\sum^{\infty}_{n=2}\left(\f{(n-1)(n+1)^2}{(2n-1)^{\f{5}{2}}(2n+1)} 
+ \f{n(n-1)^2}{(2n-1)^{\f{5}{2}}(2n-3)}\right)\right]\nonumber\\
&-&\sum^{\infty}_{n=1}\f{2}{\sqrt{2n}+\sqrt{2(n+1)}}+ \f{2}{\sqrt{2n}+\sqrt{2(n-1)}}
\eeqa

We compute the dimensionless zero temperature quantum corrections given by (\ref{massqA3x}) 
numerically. The convergent value is, 

\begin{equation}
\label{zetaqA3x}
m^2_{x0} = 1.514
\end{equation}
The temperature dependent part in (\ref{mA}) can be written as

\beqa
\label{masstA3x}
m^2_{x1}(q,\beta)&=& \left[\sum^\infty_{n=0} \f{7}{\sqrt{2n+1}}\f{1}{\left(e^{\sqrt{(2n+1)q}\beta}-1\right)}
+ \sum^\infty_{n=2} \f{1}{2\sqrt{2n-1}}\f{1}{\left(e^{\sqrt{(2n-1)q}\beta}-1\right)}\right.\nonumber\\
&-& \left. \f{7}{2} \sum^\infty_{n=1} \left(\f{1}{\sqrt{2n+1}}\f{1}{\left(e^{\sqrt{(2n+1)q}\beta}-1\right)}
-\f{n+1}{\sqrt{2n+3}}\f{1}{\left(e^{\sqrt{(2n+3)q}\beta}-1\right)}\right)\right.\nonumber\\
&-&\left. \f{7}{2}\sum^\infty_{n=1}\f{n}{\sqrt{2n-1}}\f{1}{\left(e^{\sqrt{(2n-1)q}\beta}-1\right)}
-2 \sum^\infty_{n=2} \left( \f{n(n-1)}{(2n+1)(2n-3)}\f{1}{\sqrt{2n-1}}
\f{1}{\left(e^{\sqrt{(2n-1)q}\beta}-1\right)}\right.\right.\nonumber\\
&-&\left.\left. \f{n(n+1)(n-1)}{(2n-1)(2n+1)^{\f{3}{2}}}\f{1}{\left(e^{\sqrt{(2n+1)q}\beta}-1\right)}
+ \f{n(n-1)(n-2)}{(2n-1)(2n-3)^{\f{3}{2}}}\f{1}{\left(e^{\sqrt{(2n-3)q}\beta}-1\right)} 
\right)\right.\nonumber\\
&+&\left. 2\sum^\infty_{n=2}\left(\f{(n-1)(n+1)^2}{(2n-1)^{\f{5}{2}}(2n+1)} 
+ \f{n(n-1)^2}{(2n-1)^{\f{5}{2}}(2n-3)}\right)\f{1}{\left(e^{\sqrt{(2n-1)q}\beta}-1\right)}
\right.\nonumber\\
&+&\left.\left(\hf-2 \sum^{\infty}_{n=2}\left(\f{(n-1)(n+1)^2}{(2n-1)(2n+1)^{2}} 
+ \f{n(n-1)^2}{(2n-1)(2n-3)^2}\right)\right) 
\sum^{\infty}_{m=-\infty} \f{\sqrt{q}\beta}{4\pi^2 m^2}
\right]\nonumber\\      
&+&\left[\sum^{\infty}_{n=1}\left(\left(2\f{(2 \sqrt{2n} + \sqrt{2(n-1)} 
+ \sqrt{2(n+1)})}{(\sqrt{2n}+ \sqrt{2(n-1)})(\sqrt{2n}
+ \sqrt{2(n+1)})}\right.\right.\right.\nonumber\\
&+&\left.\left. \left.8\f{(\sqrt{2(n+1)} + \sqrt{2(n-1)}-2 \sqrt{2n})}{(\sqrt{2n}
- \sqrt{2(n-1)})(\sqrt{2(n+1)}
- \sqrt{2n})}\right)\f{1}{\left(e^{\sqrt{2nq}\beta}+1\right)}\right.\right.\nonumber\\
&+&\left.\left. \left(\f{2}{\sqrt{2(n-1)} + \sqrt{2n}} - \f{8}{\sqrt{2n} 
- \sqrt{2(n-1)}}\right)\f{1}{\left(e^{\sqrt{2(n-1)q}\beta}+1\right)}\right.\right.\nonumber\\
&+&\left.\left.\left(\f{2}{\sqrt{2(n+1)} + \sqrt{2n}} + \f{8}{\sqrt{2(n+1)} 
- \sqrt{2n}}\right)\f{1}{\left(e^{\sqrt{2(n+1)q}\beta}+1\right)}\right)
\right]
\eeqa

In the end the effective masses-squared for the massless fields namely 
$m^2_{\Phi^3_1}$, $m^2_{\Phi^3_I}$ and $m^2_{A^3_x}$ depend only on the parameter $q$ and temperature.
Later in section (\ref{numerics2}), we present the behaviour of the effective masses-squared 
with temperature for different values of $q$.

\section{Finite part of effective Tachyon mass}\label{finitetachmass}

Having computed the temperature corrected one-loop mass-squared for the various massless 
fields, we can now proceed to compute the mass-squared corrections for the 
tree-level tachyons. We have already established the UV finiteness of the tachyonic 
amplitudes by demonstrating the cancellation of leading order divergences from
the zero temperature bosonic and fermionic quantum corrections to the tree-level 
tachyon mass-squared. However given the fairly complicated mathematical form of the
various corrections given in (\ref{masscorc1}),(\ref{masscorc2}) and (\ref{masscorc3}), 
extracting the finite part of the amplitudes appears to be very difficult.
Hence we are unable to give analytical expressions for the finite part of the 
zero temperature quantum corrections to the tree-level tachyon mass-squared. 
This complications also prevents us from computing the transition temperature 
analytically. Given these handicaps we are compelled to resort to numerical means.
In the following section we present a numerical computation of the transition temperature.

\begin{figure}[htb]
\begin{center}
\includegraphics[width= 8cm,angle=0]{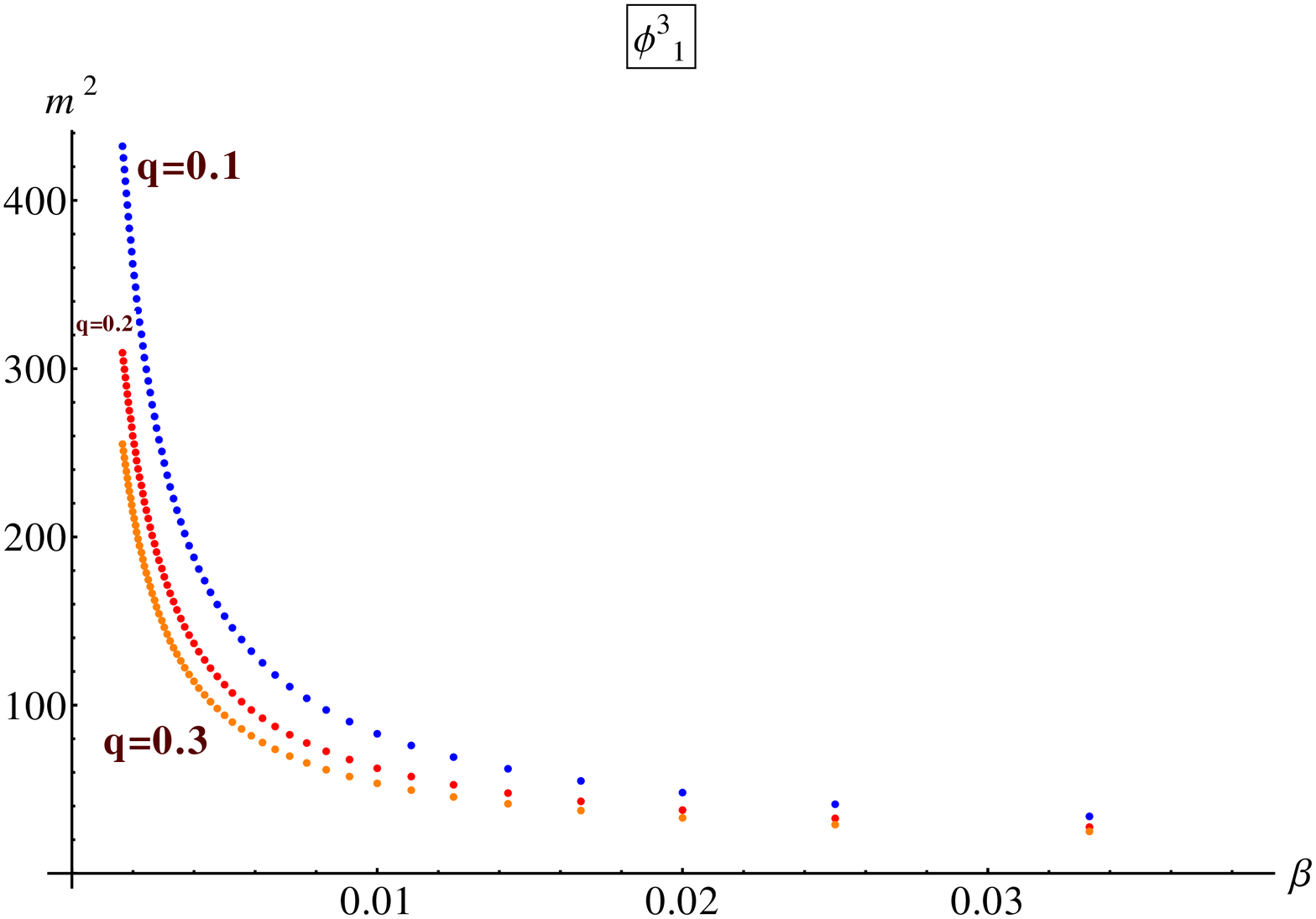}
\caption{{\small{Plots of the mass-squared correction to the massless field 
$\phi^3_1$ against $\beta = \f{1}{T}$ for 
$g^2=0.01$ and $q=0.1$,$0.2$,$0.3$.}}} 
\label{plot1a}
\end{center}
\end{figure}

\begin{figure}[htb]
\begin{center}
\includegraphics[width= 8cm,angle=0]{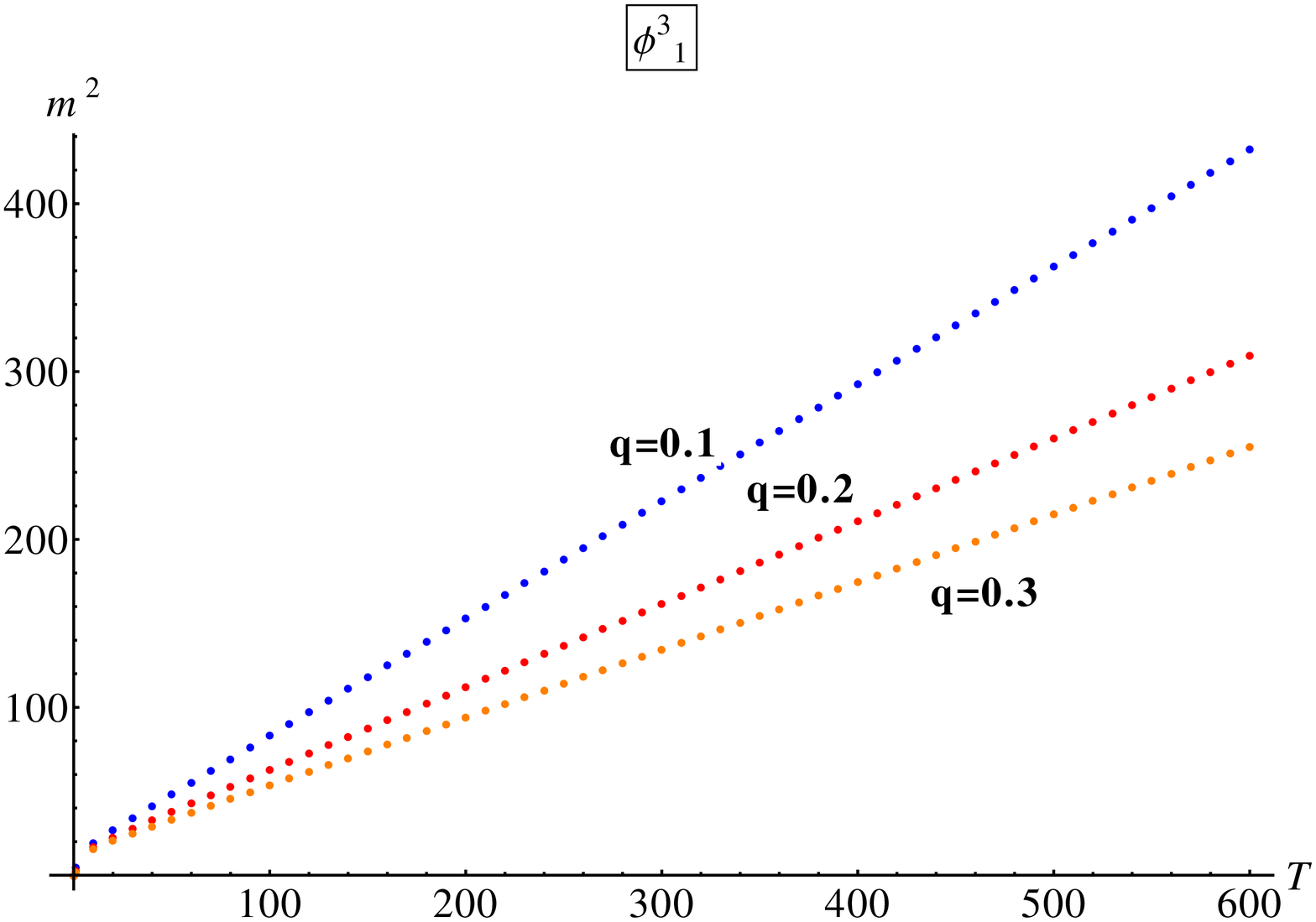}
\caption{{\small{Plot of the mass-squared correction to the massless field $\phi^3_1$ against 
$T$ for 
$g^2=0.01$ and $q=0.1$,$0.2$,$0.3$.}}}
\label{plot1b}
\end{center} 
\end{figure}

\begin{figure}[htb]
\begin{center}
\includegraphics[width= 8cm,angle=0]{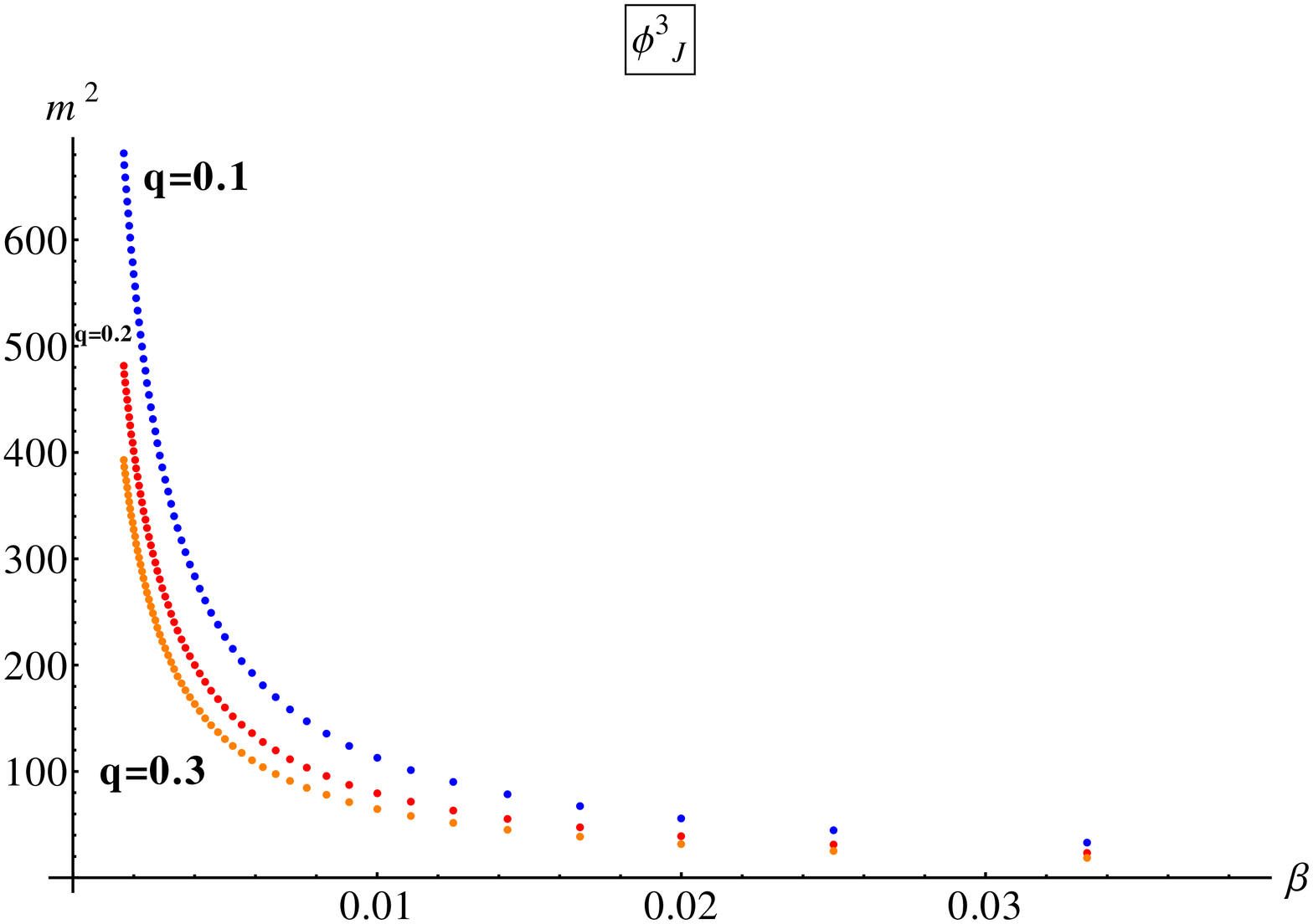}
\caption{{\small{Plot of the mass-squared correction to the massless field $\phi^3_I$ 
against $\beta=\f{1}{T}$ for 
$g^2=0.01$ and $q=0.1$,$0.2$,$0.3$.}}}
\label{plot2a}
\end{center}
\end{figure}

\begin{figure}[htb]
\begin{center}
\includegraphics[width= 8cm,angle=0]{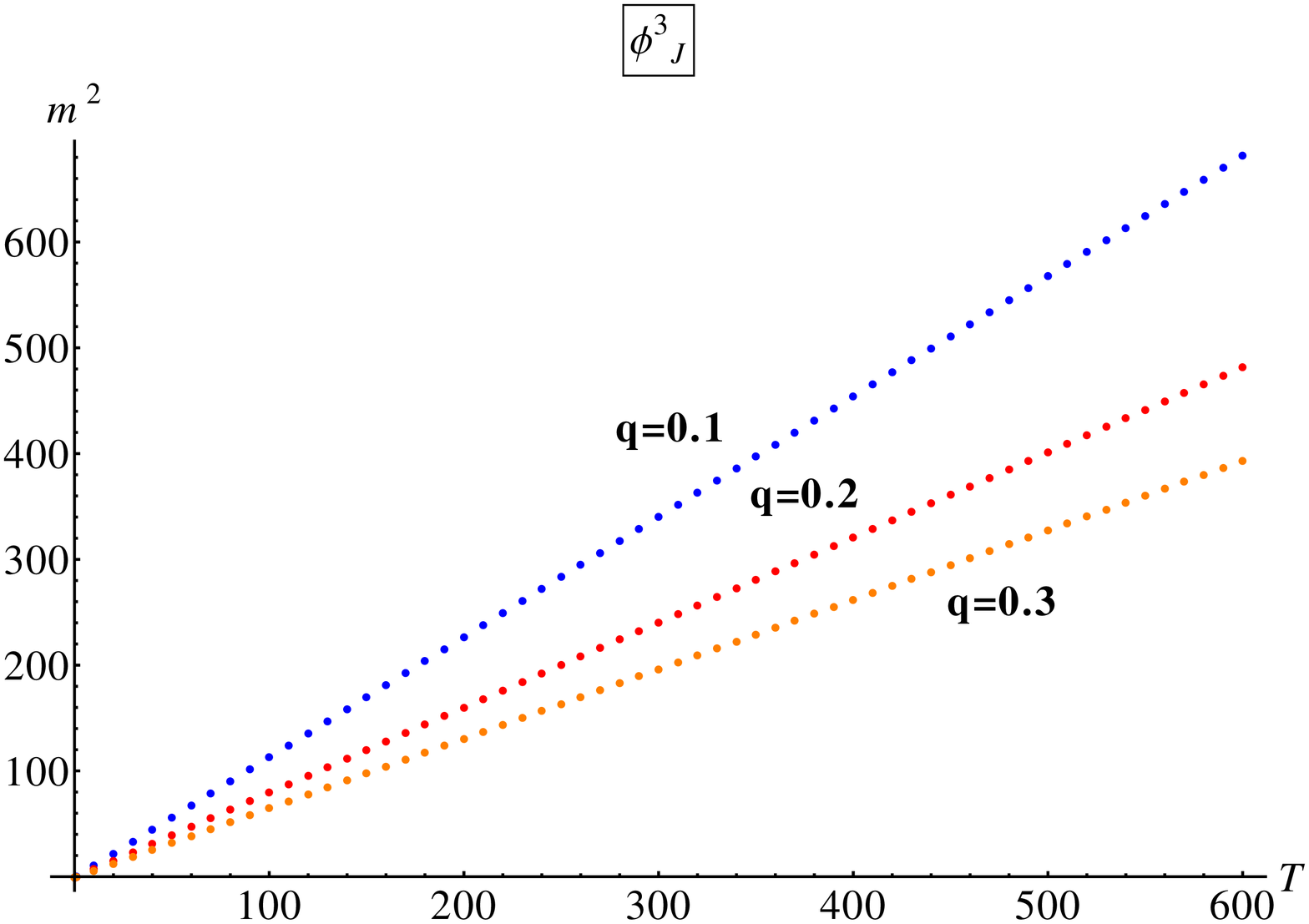}
\caption{{\small{Plot of the mass-squared correction to the massless field $\phi^3_I$ against $T$
for $g^2=0.01$ and $q=0.1$,$0.2$,$0.3$.}}}
\label{plot2b}
\end{center}
\end{figure}

\begin{figure}[htb]
\begin{center}
\includegraphics[width= 8cm,angle=0]{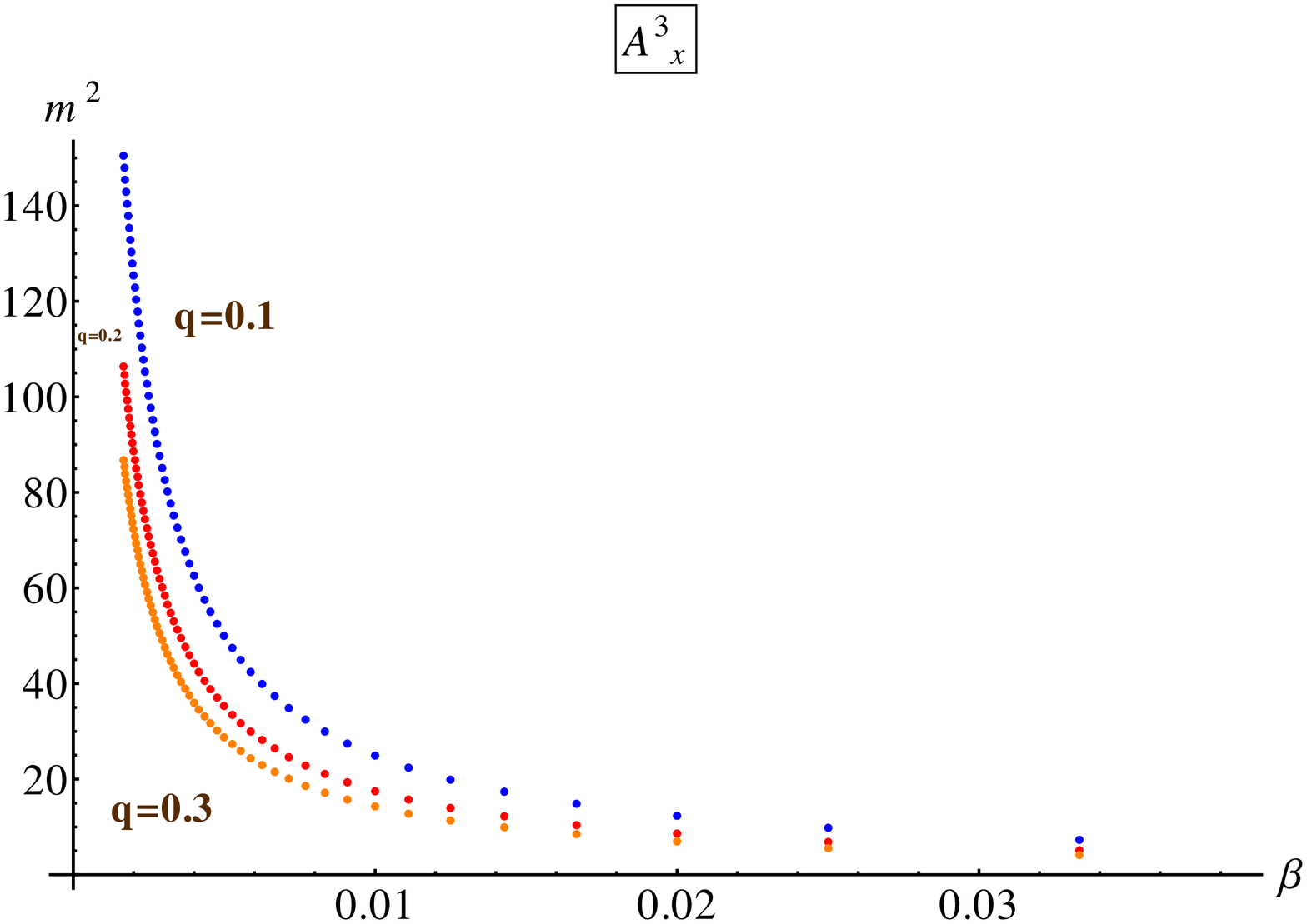}
\caption{{\small{Plot of the mass-squared correction to the massless field $A^3_x$ 
against $\beta = \f{1}{T}$ for 
$g^2=0.01$ and $q=0.1$,$0.2$,$0.3$.}}}
\label{plot3a}
\end{center}
\end{figure}

\begin{figure}[htb]
\begin{center}
\includegraphics[width= 8cm,angle=0]{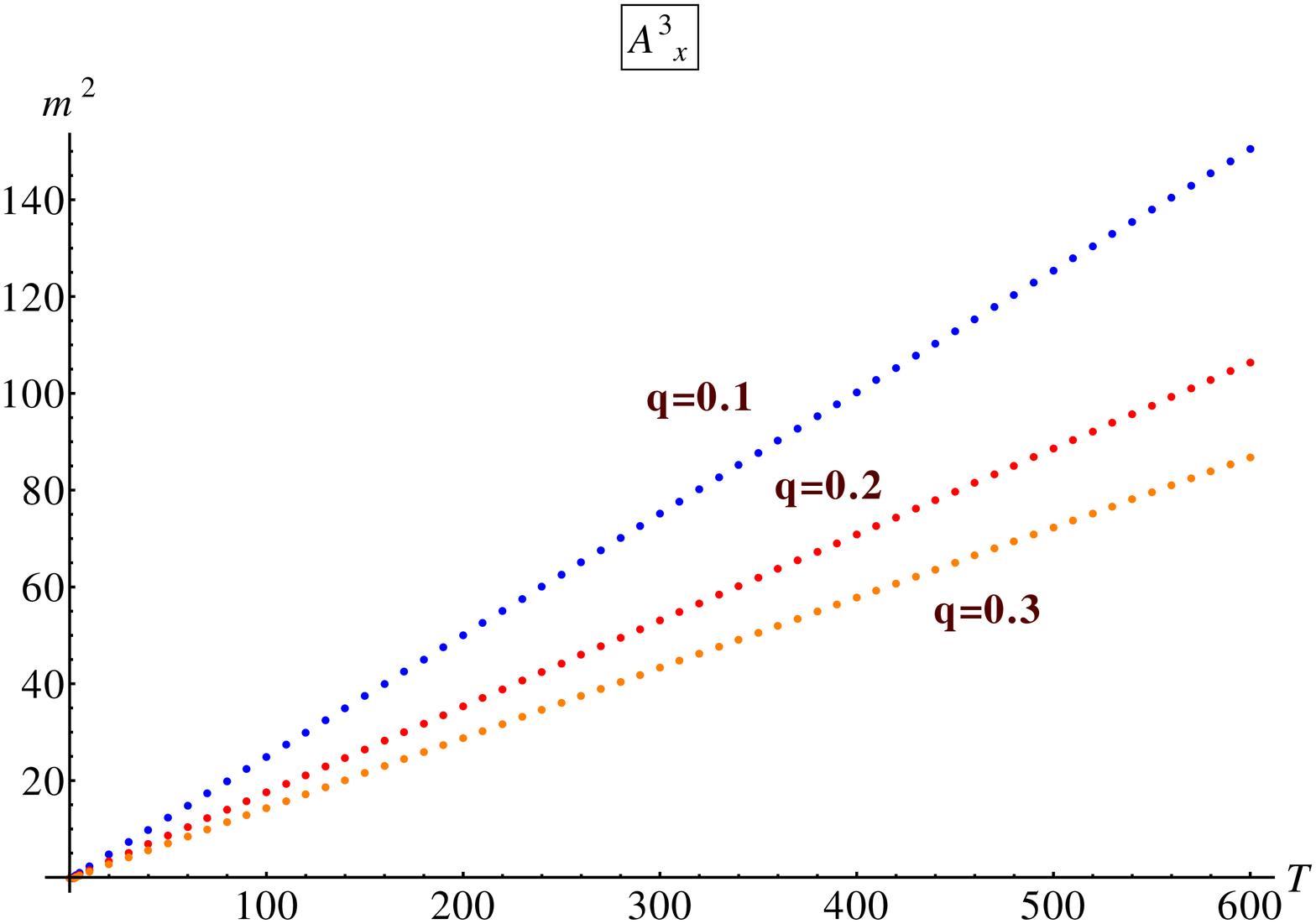}
\caption{{\small{Plot of the mass-squared correction to the massless field $A^3_x$ against $T$ for 
$g^2=0.01$ and $q=0.1$,$0.2$,$0.3$.}}}
\label{plot3b}
\end{center}
\end{figure}

\begin{figure}[htb]
\begin{center}
\includegraphics[width= 8cm,angle=0]{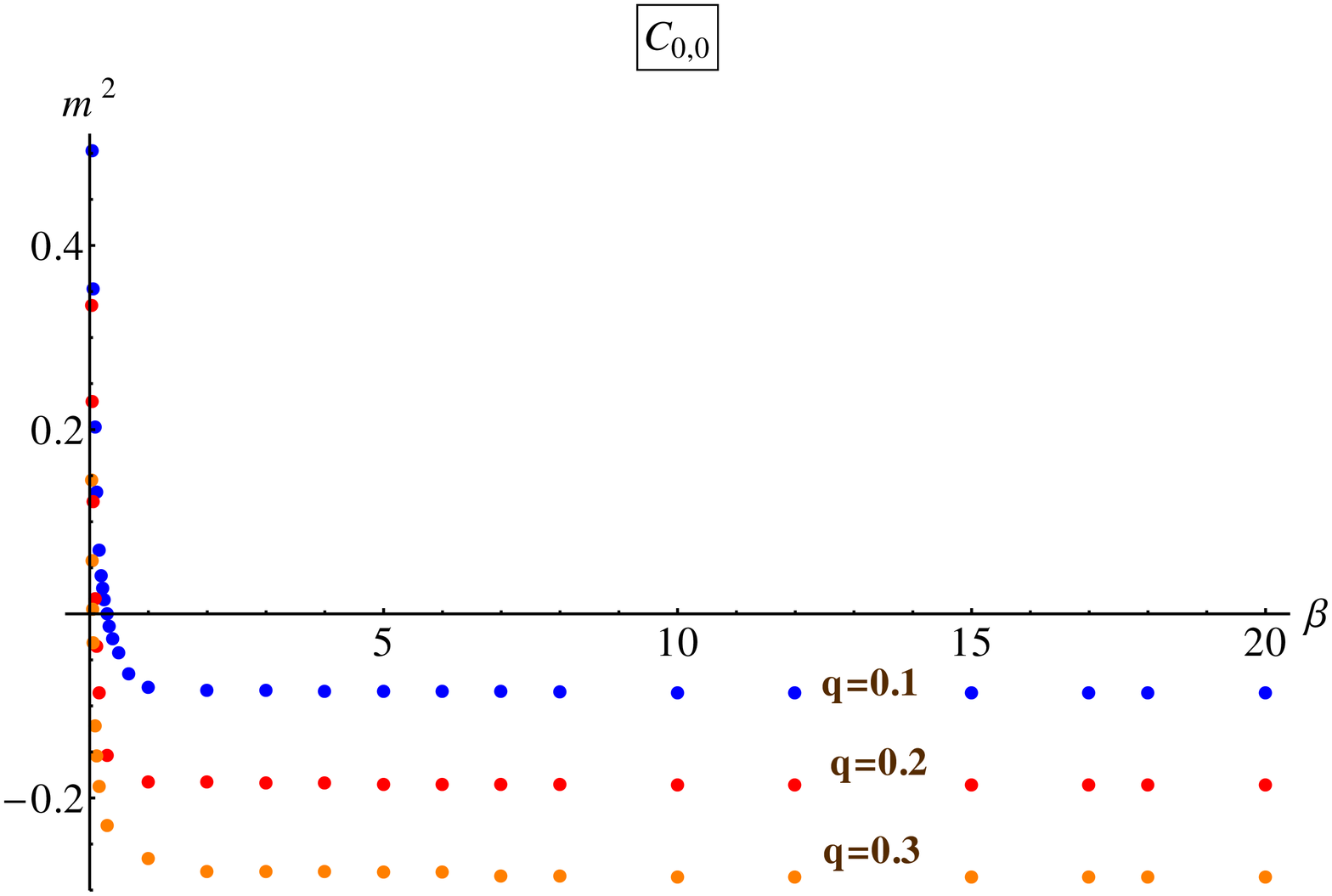}
\caption{{\small{Plot of the mass-squared correction to the tree-level tachyon 
against $\beta = \f{1}{T}$ for $g^2=0.01$, 
$q=0.1$, $0.2$ and $0.3$.}}}
\label{plot4a}
\end{center}
\end{figure}

\begin{figure}[htb]
\begin{center}
\includegraphics[width= 8cm,angle=0]{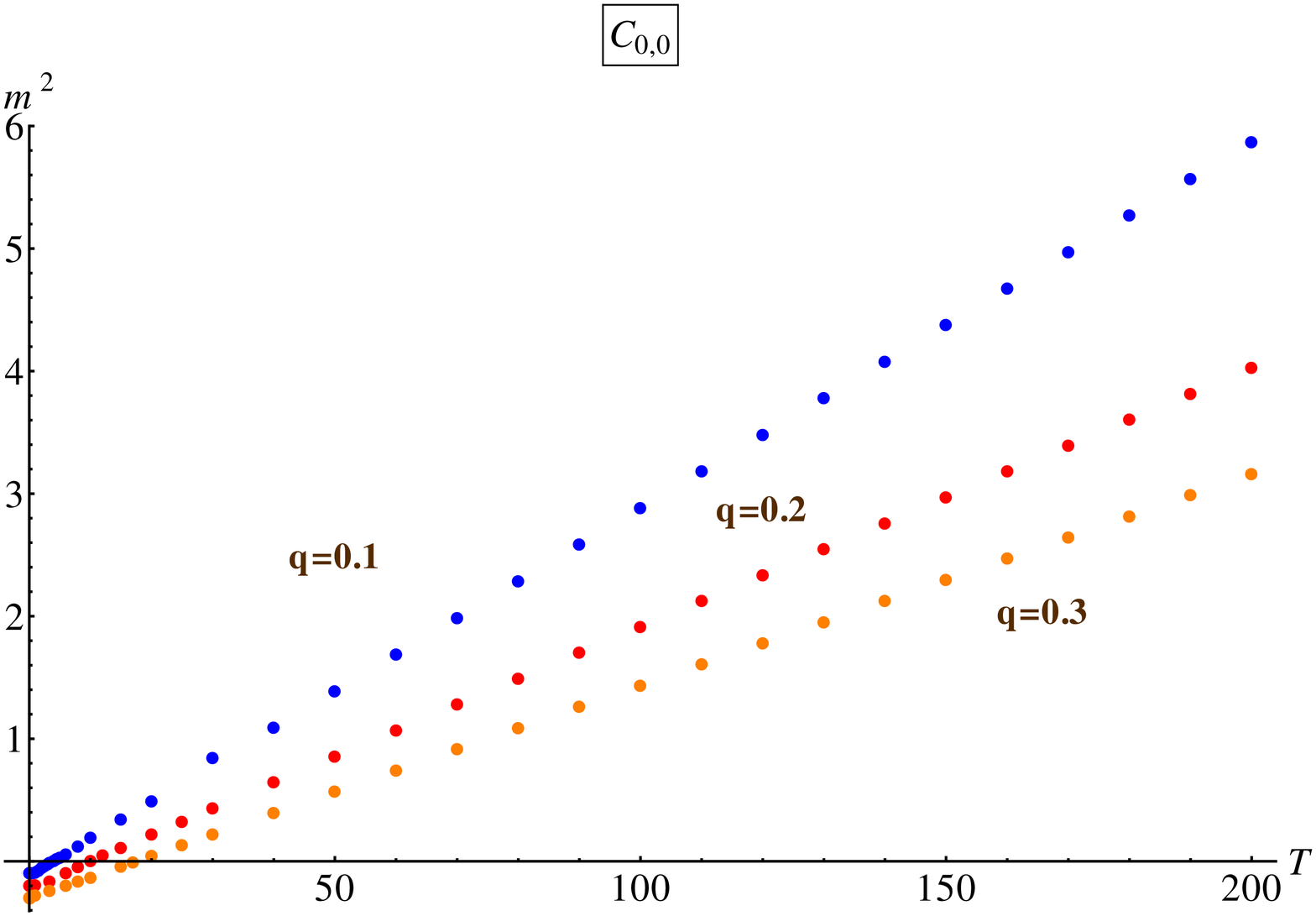}
\caption{{\small{Plot of the mass-squared correction to the tree-level tachyons 
against $T$ for 
$g^2=0.01$, $q=0.1$, $0.2$ and $0.3$.}}}
\label{plot4b}
\end{center}
\end{figure}

\begin{figure}[htb]
\begin{center}
\includegraphics[width= 8cm,angle=0]{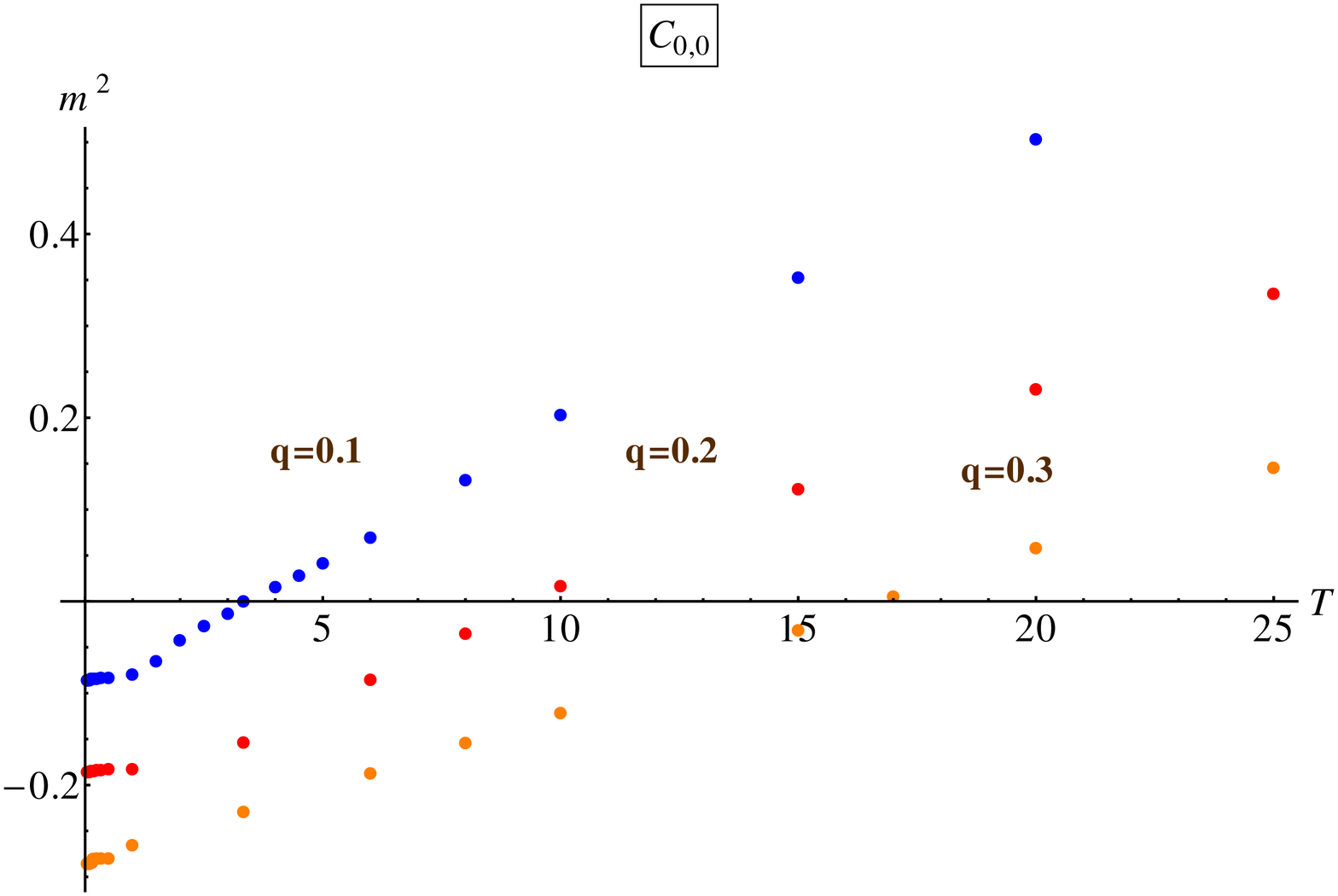}
\caption{{\small{Plot of the mass-squared correction to the tree-level tachyons against $T$ for 
$g^2=0.01$, $q=0.1$, $0.2$ and $0.3$ }}}
\label{plot4c}
\end{center}
\end{figure}

\subsection{Numerical results}\label{numerics2}
The computation of the one-loop finite temperature mass-squared for all the tree-level 
massless degrees of freedom is crucial because these
temperature dependent masses modify the corresponding propagators of the massless fields 
thereby ensuring infra-red finiteness of the one-loop effective masses-squared
of the tachyons. All the finite temperature corrections to the tree-level masses-squared 
are now shown to be UV finite. The tachyonic instability in the bulk is 
proposed to give rise to BCS Cooper-pairing instability in the boundary theory 
{\cite{KalyanaRama:2011ny}}. In this section we demonstrate that the instability 
is removed by finite temperature effects. The tree-level tachyon mass-squared is $-\f{q}{g^2}$, 
where $g$ is the dimensionfull Yang-Mills coupling 
in $(1+1)$-dimensions. The finite temperature one-loop correction including the zero temperature 
quantum corrections is $\mathcal{O}(1)$. The temperature-dependent 
mass-squared corrections is always increasing and there exists a critical temperature where the 
effective mass-squared of the tachyonic fields become zero.
Beyond the critical temperature the effective mass-squared of the tachyon is found to be positive 
and increasing. This bears hallmark of a phase-transition 
from the unstable phase to the stable phase. In the boundary theory this is proposed in 
{\cite{KalyanaRama:2011ny}} to correspond to a superconducting phase-transition.             
As mentioned in sections \ref{intro} and \ref{finitetachmass}, the critical temperature 
of phase transition cannot be computed analytically. 
We therefore tread a different path. We demonstrate numerically the behaviour of the 
masses-squared with varying $\beta$ as well as $T$ due to zero temperature quantum 
corrections + the finite temperature effects without computing them separately. In all 
the mass-squared corrections, the UV divergent pieces in the zero temperature corrections from 
the bosonic side cancel with that from the fermionic side. At large values of the momenta $n$ and $l$ 
the finite part of the quantum corrections fall off very sharply and eventually only the finite 
temperature corrections dominate. 
The parameter $q$ provides a scale for supersymmetry breaking in the present 
brane-configuration under study and has the dimension of mass-squared. 
It is also related to the angle between the branes as
\begin{equation}
\label{q}
q= \f{1}{\pi \alpha^{'}} \tan\left(\f{\theta}{2}\right) 
\end{equation}
The two-point functions for the tachyons have infra-red problem due to the presence of massless 
fields in the loops. As mentioned in section \ref{ir} we need to
modify the propagators of the tree-level massless fields by introducing a mass-squared shift 
provided by the finite temperature corrections to their tree-level masses-squared.
The leading order behaviour of the temperature dependent part of the mass-squared corrections 
in $1+1$-dimensions is linear with increasing temperature at high temperatures.
The finite temperature effective mass-squared of the tree-level tachyon in $(1+1)$-dimensions 
can thus be estimated within perturbation theory to be (in dimensionless variables)

\begin{equation}
\label{massest}
m^2_{\text{eff}}(C_{0,0}) = -\f{q}{g^2} +  \left[m_0^2
+ \f{T}{\sqrt{q}}\underbrace{\left(\sum_n \f{1}{\lambda_n} + \cdots\right)}_{=x}\right] 
+ \mathcal{O}(\f{g^2}{q}). 
\end{equation}
The physical mass is thus $m^2_{eff}g^2$ and $m_0^2$ represents the dimensionless zero temperature 
quantum corrections independent of $q$ and $g^2$ and collected from all the amplitudes in equations 
(\ref{masscorc3}), (\ref{masscorc4}) and (\ref{fermmasscorc1}). In equation (\ref{massest}), $x$ is a 
dimensionless number (independent of  $g^2,$ and $q$) specifying the temperature dependent contribution.
In equation (\ref{massest}), $\mathcal{O}(\f{g^2}{q})$ represents the next higher order 
in quantum corrections given by two-loop Feynman diagrams. 
The dimensionless zero temperature 
quantum correction to the tree-level tachyon mass-squared and is found numerically to be 
approximately equal to $m^2_0 = 1.6$.
The behaviour of the finite temperature masses-squared of the massless 
fields as well as the tachyonic fields are depicted pictorially by plotting the masses-squared 
against temperature $T$ and $\beta=\f{1}{T}$.      
We proceed to present the plots. 
 
In all the plots we have displayed the one-loop effective masses-squared as multiplied by $g^2$. 
The figures (\ref{plot1a}) (\ref{plot2a}) and (\ref{plot3a}) depict the behaviour of the 
masses-squared namely $m^2_{\phi^3_1}$, $m^2_{\phi^3_I}$ and $m^2_{A^3_x}$
with varying $\beta$. The mass-squared decreases with increasing $\beta$ as expected. 
In the figures (\ref{plot1b}), (\ref{plot2b}) and (\ref{plot3b}), $m^2_{\phi^3_1}$, $m^2_{\phi^3_I}$ 
and $m^2_{A^3_x}$ are shown to increase almost linearly with increasing temperature. This behaviour is 
expected from finite temperature field theory as finite temperature corrections are always 
known to be positive and increasing.

As discussed in section (\ref{zeroamp}), we also calculate the mass-matrix for the 
massless modes $\tilde{C}_{w,k}$ numerically and diagonalize the matrix. 
Using the temperature dependent masses-squared of the various massless fields discussed 
above the effective mass-squared for the fields $C_{0,0}$ 
can be evaluated numerically as a function of $\beta$ (or $T$). The effective mass-squared  
for the tree-level tachyon are plotted against $\beta$ in the Figure \ref{plot4a}. 
for three values of $q$, namely $q=0.1$ and $q=0.2$ and $q=0.3$ and $g^2=0.01$. The plots of the  
mass-squared against temperature $T$ for $q=0.1$ and $q=0.2$ and $q=0.3$ are given in 
Figure \ref{plot4b} and Figure \ref{plot4c}. 
As expected the finite temperature corrections dominate with increasing temperature 
which is clear from the $m^2_{eff}(C_{0,0})$ {\it vs} $T$ plots in Figure \ref{plot4b} where the behaviour 
of the mass-squared appears to be almost linear at higher temperatures. The effective 
mass-squared  $m^2_{eff}(C_{0,0})$ is equal to zero at a value of temperature (the transition 
temperature $T_c$) where 
the graphs intersect the $\beta$ and $T$-axes in the plots given in Figure \ref{plot4a}, 
Figure \ref{plot4b} and Figure \ref{plot4c}. 
The plots in Figure \ref{plot4c} are drawn for smaller range of $T$ in order to show the 
critical points ($m^2_{\text{eff}}(C_{0,0})=0$) more clearly.   
Putting $m^2_{\text{eff}}(C_{0,0})=0$ in equation (\ref{massest}) we get
\begin{equation}
\label{criticalT}
T_c = {1\over x}[q^{\f{1}{2}}(\f{q}{g^2}- m_0^2)]
\end{equation}
where $T_c$ is the dimensionfull transition temperature (having dimension of mass). 
One can also define a dimensionless transition temperature by $T_c=\tilde{T}_c q^{\f{3}{2}}/g^2$. 
One should note that the dimensionless tree-level 
mass-squared of the tachyon for $q=0.1$, $0.2$ and $0.3$ are $q/g^2= 10$, $20$ and $30$ respectively for $g^2=0.01$,
whereas the quantum correction $m_0^2$ is much smaller.  Now if $m_0^2=0$, $\tilde T_c = {1\over x}$ and 
is thus independent of $q, g^2$. Thus  to a good approximation one expects that $\tilde T_c$ will be 
independent of $q,g^2$. One has to note that this scaling relation for the transition temperature is 
the dominant term at the level of one-loop. At the level of higher loops the dependence of the effective mass-squared
on temperature will be much more complicated. However at least in weak coupling, the quantum corrections 
at successive higher loops will be smaller and smaller, and the dominant term in $\tilde{T}_c$ will still 
be given by this one loop relation. From the plots given in Figure \ref{plot4a} and Figure \ref{plot4c}, 
the numerical values of $T_c$ 
for $g^2=0.01$ and $q=0.1$, $q=0.2$ and $q=0.3$ are 
$T_c=3.34$, $T_c=9.48$ and $T_c=16.73$ respectively. This gives  
$\tilde{T}_c =1.0562$, $1.0599$ and $1.0182$ for 
$q=0.1$, $0.2$ and $0.3$ and $g^2=0.01$ respectively, confirming our expectation that when $m_0^2$ is small $\tilde T_c$
is approximately independent of $g,q$.

\section{Discussion and Outlook}
We have computed the one-loop finite temperature corrections to the tree-level tachyon mass-squared 
in intersecting $D1$-branes in a self consistent manner.  We have shown the UV finiteness (at one loop).
We have seen that at high temperatures the tachyonic field becomes massive as expected and 
we have computed this critical temperature in a one loop approximation - improved by 
incorporating mass-squared corrections to the massless fields in the spirit of the RG.
This takes care of the IR divergences.
Thus our calculation has no UV or IR divergences.

This model resembles the holographic BCS superconductor model discussed in 
\cite{KalyanaRama:2011ny}. We expect that with the
techniques developed in this paper it should be possible to tackle the model in 
\cite{KalyanaRama:2011ny} involving higher branes. These techniques should also 
be useful in other contexts where D brane constructions are used and supersymmetry spontaneously broken.

The entire computation is done in temporal gauge, $A^a_0=0$. This gauge choice helps 
to avoid ghosts in the theory.  
The original theory describing the world-volume of two $D1$-branes is a 
$(1+1)$-dimensional supersymmetric $SU(2)$ Yang-Mills theory which is a UV finite 
theory. The choice of the background $\langle \phi^3_B \rangle = qx$ breaks 
supersymmetry without tampering with the other degrees of freedom. Hence we find that  
the UV behaviour of the amplitudes in a broken supersymmetry scenario remains 
the same as in the supersymmetric case. In order to establish the UV finiteness of the 
one-loop corrections we rely on asymptotic expansion of the vertices and find that 
the leading order divergent pieces from the bosonic and fermionic loops
cancel among themselves. The effective mass-squared of the tachyon is found to grow 
linearly with temperature in accordance to the expected behaviour in $(1+1)$-dimensions.
The kinetic terms for the bosons are scale independent in $(1+1)$-dimensions. 
Hence the zero-temperature one loop quantum corrections are found to be independent
of the supersymmetry breaking scale. The crossing of the $m^2_{\text{eff}}(C_{0,0})$ $\it{vs}$ $T$
curves from negative to positive values indicates two distinct phases. This bears the signature of a phase transition.
We also find that the dimensionless critical
temperature ($T_c g^2/q^{3/2}$) for this phase transition has very closely placed values namely 
$\tilde{T}_c=1.0562$, $1.0599$ and $1.0182$.
In order to do the complete stability analysis 
of the intersecting $D1$-branes at finite temperature
one has to compute the full tachyon effective action at finite temperature which rely 
on higher loop calculations.
The results demonstrated in this paper can be generalized to higher dimensional branes 
without much difficulty. In particular for two intersecting $Dp$-branes 
the effective mass-squared for the tree-level tachyons is expected to grow as $T^{p-1}$.        
\\

\noindent
{\large{\textbf{Acknowledgments:}}}
The authors would like to acknowledge the Institute of Mathematical Sciences, Chennai for providing 
a free and vibrant academic atmosphere and 
the Department of Atomic Energy for providing funds for this project. S.P.C. would like to 
thank colleagues and compatriots Swastik Bhattacharya and 
Shankhadeep Chakrabortty, Saurabh Gupta and Akhilesh Nautiyal and T Geetha for fruitful as 
well as refreshing lighthearted discussions.  
S.S. would like to thank Institute of Mathematical Sciences for the kind hospitality during 
various stages of this work. 
The work of S.S. is partially supported by the Research and Development Grant (2013-2014), 
University of Delhi.

\newpage
\appendix

\section{Dimensional reduction of $D=10$, ${\cal N}=1$, $SU(2)$ SYM to $D=2$}\label{dimr}

We first write down the action for $D=10$, ${\cal N}=1$ SYM \footnote{We will use the metric $\mbox{diagonal}(+1, -1, \cdots, -1)$.},

\beqa
S_{9+1}=\f{1}{g^2}\mbox{tr}\int d^{10}x \left[-\f{1}{2}F_{MN}F^{MN}+i\bar{\Psi}\Gamma^{M}D_{M}\Psi\right]
\eeqa

\beqa
F_{MN}&=&\partial_{M}A_{N}-\partial_{N}A_{M}-i\left[A_{M},A_{N}\right]\\
D_{M}\Psi&=&\partial_{M}\Psi-i\left[A_{M},\Psi\right]
\eeqa

where $M,N=0, \cdot\cdot\cdot 9$ with $A_{M}=\f{\sigma^a}{2}A^a_{M}$, $\Psi=\f{\sigma^a}{2}\Psi^a_{M}$ and,

\beqa
\left[\f{\sigma^a}{2},\f{\sigma^b}{2}\right]=i\epsilon^{abc}\f{\sigma^c}{2} \mbox{~~;~~}
\f{1}{2} \mbox{tr} \left(\sigma^a\sigma^b\right)=\delta^{ab}
\eeqa

$\Gamma^M$ are $32\times32$ imaginary matrices and. In $(9+1)$-dimensions the gamma-matrix which anti commutes with all other gamma matrices is  

\beqa
\label{gamma11}
\gamma^{11}= \left(\begin{array}{cc}
\mathbb{I}_{16 \times 16} & 0\\
0 & -\mathbb{I}_{16 \times 16}
\end{array}\right)
\eeqa
The chiral projection operator in $(9+1)$-dimensions giving rise to left and right-moving chiral fermions is given by
\begin{equation}
\label{chiral}
\mathcal{P} = \f{1\pm \gamma^{11}}{2}
\end{equation}
Under the chiral projection (\ref{chiral}), the $32$-component Dirac fermions become   
\beqa\label{16psi}
\Psi=\left(
\begin{array}{c}
\Psi_L \\ 
0
\end{array}\right)
\eeqa
$\Psi$ is a $32$ component Majorana-Weyl spinor with
$16$ non-zero components.

The $32$ dimensional $\Gamma$ matrices satisfy the Dirac algebra 
$\{\Gamma^M, \Gamma^N\}=2\eta^{MN}$. Under the decomposition $SO(9,1)\rightarrow SO(1,1)\times SO(8)$, they have 
to be written in terms of the $16$ dimensional $spin(8)$ matrices. For $M$ corresponding to the
$SO(8)$ directions (that we label by $I$) we call the  $16$ dimensional $spin(8)$ matrices as
$\gamma^I$, $(I=1,...,8)$. The $\gamma^I$'s thus satisfy the  $spin(8)$ algebra $\{\gamma^I,\gamma^J\}=2\delta^{IJ}$. They
are however reducible and can be written in terms of the $8$ dimensional representations $\alpha^{I}$ as,

\beqa
\gamma^{I}= \left(\begin{array}{cc}
0 & \alpha^I\\
\alpha^{IT} & 0
\end{array}\right)
\eeqa

where the $\alpha^I$'s now satisfy $\{\alpha^I,\alpha^J\}=2\delta^{IJ}$. A representation of the $\alpha^I$'s can be written follows \cite{GSW},

\beqa\label{gamma}
\alpha^1&=&\tau\otimes\tau\otimes\tau \mbox{~~~~~~~~}\alpha^2=1\otimes\sigma^1\otimes\tau\\
\alpha^3&=&1\otimes\sigma^3\otimes\tau \mbox{~~~~~~~~}\alpha^4=\sigma^1\otimes\tau\otimes 1\non
\alpha^5&=&\sigma^3\otimes\tau\otimes 1 \mbox{~~~~~~~~}\alpha^6=\tau\otimes 1 \otimes\sigma^1\non
\alpha^7&=&\tau\otimes 1\otimes\sigma^3 \mbox{~~~~~~~~}\alpha^8= 1\otimes 1\otimes 1\nonumber
\eeqa

where $\tau=i\sigma^2$. We can construct one more $\gamma$ matrix that anti commutes with all the other $\gamma^I$'s. 
It is given by $\gamma^9=\gamma^1\gamma^2\cdot\cdot\cdot\gamma^8$. In matrix form,

\beqa
\gamma^{9}= \left(\begin{array}{cc}
1_{8}&0\\
0&-1_{8}
\end{array}\right)
\eeqa

The nine $32$ dimensional $\Gamma^M$ $(M=1,...9)$ matrices can thus be formed out of the nine $\gamma$ matrices. 
Since there is no tenth $\gamma$ matrix, we need to construct a tenth $\Gamma$ matrix that anti commutes with all the others. Ultimately we can write,

\begin{eqnarray}\label{Gamma}
\Gamma^0&=&\sigma^2\otimes 1_{16}\\
\Gamma^{I}&=&i\sigma^1\otimes\gamma^{I}\non
\Gamma^9&=&i\sigma^1\otimes\gamma^{9}\nonumber
\end{eqnarray}

With this, the sixteen dimensional $\Psi_L$ in equation (\ref{16psi}) can further be written as,

\beqa
\Psi_L=\left(\begin{array}{c}
\psi_L \\ \psi_R
\end{array}\right)
\eeqa

where $\psi_L$ and $\psi_R$ are now $8$ component fermions. In other words, the sixteen component $\Psi_L$ decomposes as 
$16= (1,8)+(\bar{1},\bar{8})$. Thus in $1+1$ dimensions we have $8$ one-component left-moving plus
$8$ one-component right-moving fermions.

We can now write down the dimensionally reduced action,

\beqa\label{2action}
S_{1+1}&=&S^1_{1+1}+S^2_{1+1}\\
S^1_{1+1}&=&\f{1}{g^2}\mbox{tr} \int d^2x \left[-\f{1}{2}F_{\mu\nu}F^{\mu\nu}+ D_{\mu}\Phi_I D^{\mu}\Phi_I+\f{1}{2}\left[\Phi_I,\Phi_J\right]^2\right]\\
S^2_{1+1}&=&\f{i}{g^2}\mbox{tr} \int d^2x \left[\psi^T_L D_0 \psi_L+\psi^T_R D_0 \psi_R+\psi^T_L D_1 \psi_L-\psi^T_R D_1 \psi_R \right.\\
&+& \left. 2i\psi^T_R\alpha^T_I\left[\Phi_I,\psi_L\right]\right]\nonumber
\eeqa

where, $D_{\mu}=\partial_{\mu}-i\left[A_{\mu},\star\right]$ and all the fermions, $\psi_L$ and $\psi_R$ are anti commuting.

\section{Tables of fields, eigenfunctions and normalizations.}\label{table}

\begin{table}[h]

\begin{center}
\begin{tabular}{|c|c|}
\hline
\hline
\multicolumn{2}{ |c| }{{\bf Dimensionfull constants}}\\
\hline
$q$&Slope of intersecting brane configuration\\
\hline
$\beta$&Inverse of temperature $T$\\
\hline
$g^2$&Yang-Mills coupling constant\\
\hline
\hline
\multicolumn{2}{ |c| }{{\bf Normalizations}}\\
\hline
$N$&$\sqrt{q}/\beta$\\
\hline
${\cal N}(n)$&$\f{1}{\sqrt{\sqrt{\pi} 2^n (4n^2-2)(n-2)!}}$\\
\hline
$\tilde{{\cal N}}(n)$&$\f{1}{\sqrt{\sqrt{\pi} 2^n (4n-2)(n-1)!}}$\\
\hline
$\mathcal{N}^{'}(n)$&$\f{1}{\sqrt{\sqrt{\pi}2^n n!}}$\\
\hline
${\cal N}_F(n)$&$\f{1}{\sqrt{\sqrt{\pi} 2^{n+1} (n-1)!}}$\\
\hline
\hline
\end{tabular}
\end{center}
\caption{Dimensionfull constants and normalizations.}
\label{t1}
\end{table}
The various fields together with their eigenfunctions, momenta and momentum modes are tabulated below. 

\begin{table}

\begin{center}
\scalebox{0.75}{
\begin{tabular}{|c|c|c|c|}
\hline
\hline
{\bf Field}& {\bf Eigenfunction}& {\bf Tree-level mass} & {\bf Momentum} \\
&&&{\bf mode}\\
\hline
\multicolumn{4}{ |c| }{{\bf Bosons}} \\
\hline\\
$
\left(\begin{array}{c}
A_x^2\\
\phi_1^1
\end{array}\right)
$&$\left(\begin{array}{c}
A_n(x)\\
\phi_n(x)
\end{array}\right)=\left(\begin{array}{c}
{\cal N}(n)e^{- q x^2/2}  \left(H_n (\sqrt{q} x) + 2 n H_{n-2} (\sqrt{q} x)\right)\\
{\cal N}(n)e^{- q x^2/2} \left(H_n (\sqrt{q} x) - 2 n H_{n-2} (\sqrt{q} x) \right)
\end{array}\right)$&$\f{\lambda_n}{g^2}=\f{(2n-1)q}{g^2}$& $C_{w,n}$\\
\hline\\
$
\left(\begin{array}{c}
A_x^2\\
\phi_1^1
\end{array}\right)
$&$\left(\begin{array}{c}
\tilde{A}_n(x)\\
\tilde{\phi}_n(x)
\end{array}\right)=\left(\begin{array}{c}
\tilde{{\cal N}}(n)e^{- q x^2/2}  \left(H_n (\sqrt{q} x) - 2 (n-1) H_{n-2} (\sqrt{q} x)\right)\\
\tilde{{\cal N}}(n)e^{- q x^2/2}  \left(H_n (\sqrt{q} x) + 2 (n-1) H_{n-2} (\sqrt{q} x) \right)
\end{array}\right)$&$\f{\tilde{\lambda}_n}{g^2}=0$& $\tilde{C}_{w,n}$\\
\hline\\
$
\left(\begin{array}{c}
A_x^1\\
\phi_1^2
\end{array}\right)
$&$\left(\begin{array}{c}
-A_n(x)\\
\phi_n(x)
\end{array}\right)=\left(\begin{array}{c}
-{\cal N}(n)e^{- q x^2/2}  \left(H_n (\sqrt{q} x) + 2 n H_{n-2} (\sqrt{q} x)\right)\\
{\cal N}(n)e^{- q x^2/2}  \left(H_n (\sqrt{q} x) - 2 n H_{n-2} (\sqrt{q} x) \right)
\end{array}\right)$&$\f{\lambda_n}{g^2}=\f{(2n-1)q}{g^2}$& $C^{'}_{w,n}$\\
\hline\\
$
\left(\begin{array}{c}
A_x^1\\
\phi_1^2
\end{array}\right)
$&$\left(\begin{array}{c}
-\tilde{A}_n(x)\\
\tilde{\phi}_n(x)
\end{array}\right)=\left(\begin{array}{c}
-\tilde{{\cal N}}(n)e^{- q x^2/2}  \left(H_n (\sqrt{q} x) - 2 (n-1) H_{n-2} (\sqrt{q} x)\right)\\
\tilde{{\cal N}}(n)e^{- q x^2/2}  \left(H_n (\sqrt{q} x) + 2 (n-1) H_{n-2} (\sqrt{q} x) \right)
\end{array}\right)$&$\f{\tilde{\lambda}_n}{g^2}=0$& $\tilde{C}^{'}_{w,n}$\\
\hline\\
$\Phi_I^a, (a=1,2) $&${\cal N}^{'}(n)e^{- q x^2/2}  \left(H_n (\sqrt{q} x)\right)$&$\f{\gamma_n}{g^2}=\f{(2n+1)q}{g^2}$&$\Phi_I^a (w,n)$\\
$(I=2\cdots 8)$&&&\\
\hline\\
$\Phi_I^3 $&$ e^{ilx} $&$0$&$\Phi_I^3 (w,l)$\\
$(I=1\cdots 8)$&&&\\
\hline\\
$A_x^3 $&$ e^{ilx} $&$0$&$A_x^3 (w,l)$\\
\hline\\
\multicolumn{4}{ |c| }{{\bf Fermions}} \\
\hline\\
$
\left(\begin{array}{c}
L_1^1\\
R_8^2
\end{array}\right)
$&$\left(\begin{array}{c}
L_n(x)\\
R_n(x)
\end{array}\right)=\left(\begin{array}{c}
{\cal N}_F(n)e^{- q x^2/2}  \left(-\f{i}{\sqrt{2n}} H_n (\sqrt{q} x) + H_{n-1} (\sqrt{q} x)\right)\\
{\cal N}_F(n)e^{- q x^2/2} \left(-\f{i}{\sqrt{2n}}H_n (\sqrt{q} x) -  H_{n-1} (\sqrt{q} x) \right)
\end{array}\right)$&$\f{\lambda^{'}_n}{g^2}=\f{2nq}{g^2}$& $\theta_i(w,n)$\\
and for $8$ other sets&&&\\
\hline\\
$
\left(\begin{array}{c}
L_1^1\\
R_8^2
\end{array}\right)
$&$\left(\begin{array}{c}
L^{*}_n(x)\\
R^{*}_n(x)
\end{array}\right)=\left(\begin{array}{c}
{\cal N}_F(n)e^{- q x^2/2}  \left(\f{i}{\sqrt{2n}} H_n (\sqrt{q} x) + H_{n-1} (\sqrt{q} x)\right)\\
{\cal N}_F(n)e^{- q x^2/2} \left(\f{i}{\sqrt{2n}}H_n (\sqrt{q} x) -  H_{n-1} (\sqrt{q} x) \right)
\end{array}\right)$&$\f{\lambda^{'}_n}{g^2}=\f{2nq}{g^2}$& $\theta^{*}_i(w,n)$\\
and for $8$ other sets&&&\\
\hline\\
$
\left(\begin{array}{c}
L_1^2\\
R_8^1
\end{array}\right)
$&$\left(\begin{array}{c}
L_n(x)\\
-R_n(x)
\end{array}\right)=\left(\begin{array}{c}
{\cal N}_F(n)e^{- q x^2/2}  \left(-\f{i}{\sqrt{2n}} H_n (\sqrt{q} x) + H_{n-1} (\sqrt{q} x)\right)\\
-{\cal N}_F(n)e^{- q x^2/2} \left(-\f{i}{\sqrt{2n}}H_n (\sqrt{q} x) -  H_{n-1} (\sqrt{q} x) \right)
\end{array}\right)$&$\f{\lambda^{'}_n}{g^2}=\f{2nq}{g^2}$& $\theta_j(w,n)$\\
and for $8$ other sets&&&\\
\hline\\
$
\left(\begin{array}{c}
L_1^2\\
R_8^1
\end{array}\right)
$&$\left(\begin{array}{c}
L^{*}_n(x)\\
-R^{*}_n(x)
\end{array}\right)=\left(\begin{array}{c}
{\cal N}_F(n)e^{- q x^2/2}  \left(\f{i}{\sqrt{2n}} H_n (\sqrt{q} x) + H_{n-1} (\sqrt{q} x)\right)\\
-{\cal N}_F(n)e^{- q x^2/2} \left(\f{i}{\sqrt{2n}}H_n (\sqrt{q} x) -  H_{n-1} (\sqrt{q} x) \right)
\end{array}\right)$&$\f{\lambda^{'}_n}{g^2}=\f{2nq}{g^2}$& $\theta^{*}_j(w,n)$\\
and for $8$ other sets&&&\\
\hline\\
$L_i^3$&$e^{ilx}$&$0$&$L_i^3(w,l)$\\
$(i=1\cdots8)$&&&\\
\hline\\
$R_i^3$&$e^{ilx}$&$0$&$R_i^3(w,l)$\\
$(i=1\cdots8)$&&&\\
\hline
\hline
\end{tabular}}
\end{center}
\caption{Eigenfunctions, Tree-level masses and momentum modes.}
\label{t2}
\end{table}

\newpage
\section{Propagators and vertices for computation of two-point $C_{w,k}$ amplitudes}\label{apv}

\subsection{Bosons}\label{bosons}

Propagator for $C_{m,n}$
\beqa
\label{propzeta}
\expect{C_{m,n}C_{m^{'},n^{'}}}=g^2\f{\delta_{m,-m^{'}}\delta_{n,n^{'}}}{\o_m^2+\lambda_n}
\eeqa

Propagator for $\tilde{C}_{m,n}$
\beqa
\label{propzetat}
\expect{\tilde{C}_{m,n}\tilde{C}_{m^{'},n^{'}}}=g^2\f{\delta_{m,-m^{'}}\delta_{n,n^{'}}}{\o_m^2}
\eeqa

Propagator for the $\Phi_I^1$ and  $\Phi_I^2$ fluctuations ($I\ne1$)are same and is given by
\beqa
\label{propphi1I}
\expect{\Phi_I^1(m,n)\Phi_I^1(m^{'},n^{'})}=g^2\f{\delta_{m,-m^{'}}\delta_{n,n^{'}}}{\o_m^2+\gamma_n}
\eeqa

Propagator for the $\Phi_I^3$ fluctuations for all $I$ is given by
\beqa
\label{propphi3I}
\expect{\Phi_I^3(m,l)\Phi_I^3(m^{'},l^{'})}=g^2 \f{\delta_{m,-m^{'}}2\pi\delta(l+l^{'})}{\o_m^2+l^2}
\eeqa

Propagator for the $A^3_x$ fluctuations assumes the form because there is no term in the Lagrangian (\ref{Lagrangian3}) with spatial derivatives on $A^3_x$. 
\beqa
\label{propA31}
\expect{A^3_x(m,l) A^3_x(m^{'},l^{'})}=g^2 \f{\delta_{m,-m^{'}}2\pi\delta(l+l^{'})}{\o_m^2}
\eeqa

\begin{figure}[h]
\begin{center}
\begin{psfrags}
\psfrag{c1}[][]{$C_{w,k}$}
\psfrag{c2}[][]{$C_{w^{'},k^{'}}$}
\psfrag{d1}[][]{$C_{m,n}$}
\psfrag{d2}[][]{$C_{m^{'},n^{'}}$}
\psfrag{a11}[][]{$\Phi_I^{1,2}(m,n)$}
\psfrag{a12}[][]{$\Phi_I^{1,2}(m^{'},n^{'})$}
\psfrag{v1}[][]{$V_1$}
\psfrag{v2}[][]{$V_2$}
\includegraphics[width= 10cm,angle=0]{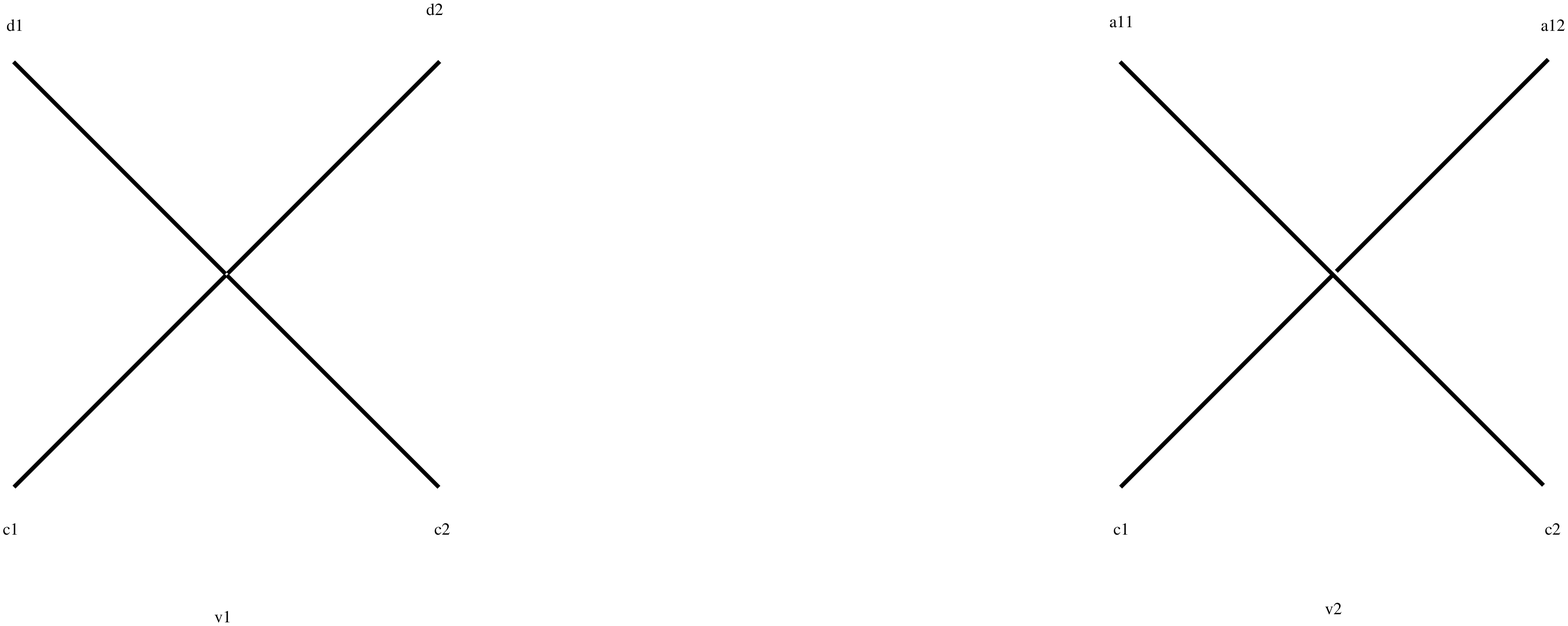}
\caption{$V_1$ , $V_2$ vertices}
\label{v1v2}
\end{psfrags}
\end{center}
\end{figure}

\beqa
\label{V1}
V_1=-\f{N}{2g^2} F_1(k,k^{'},n,n^{'}) \delta_{\o+\o^{'}+m+m^{'}} \mbox{~~~~~~~(Figure \ref{v1v2})}
\eeqa

\beqa
\label{F1}
F_1(k,k^{'},n,n^{'})&=&\sqrt{q}\int dx \left[\phi_k(x)\phi_{k^{'}}(x)A_n(x) A_{n^{'}}(x) + 2 A_k(x)\phi_{k^{'}}(x)\phi_n(x) A_{n^{'}}(x)\right.\nonumber\\
&+& \left. 2 \phi_k(x) A_{k^{'}}(x)\phi_n(x) A_{n^{'}}(x) + A_k(x) A_{k^{'}}(x) \phi_n(x) \phi_{n^{'}}(x)\right] 
\eeqa

\beqa
\label{V2}
V_2=-\f{N}{2g^2}\left[F_2(k,k^{'},n,n^{'})\right]\delta_{\o+\o^{'}+m+m^{'}} \mbox{~~~~~~~(Figure \ref{v1v2})}
\eeqa

\beqa
\label{F2}
F_2(k,k^{'},n,n^{'})=\sqrt{q}\int dx e^{-qx^2}\left [A_k(x) A_{k^{'}}(x)+\phi_k(x)\phi_{k^{'}}(x)\right] \left[H_n(x)  H_{n^{'}}(x)\right]
\eeqa

\begin{figure}[h]
\begin{center}
\begin{psfrags}
\psfrag{c1}[][]{$C_{w,k}$}
\psfrag{c2}[][]{$C_{w^{'},k^{'}}$}
\psfrag{d1}[][]{$\tilde{C}_{m,n}$}
\psfrag{d2}[][]{$\tilde{C}_{m^{'},n^{'}}$}
\psfrag{a11}[][]{$\Phi_I^{3}(m,l)$}
\psfrag{a12}[][]{$\Phi_I^{3}(m^{'},l^{'})$}
\psfrag{v1}[][]{$\tilde{V}_1$}
\psfrag{v2}[][]{$V^{'}_2$}
\includegraphics[width= 10cm,angle=0]{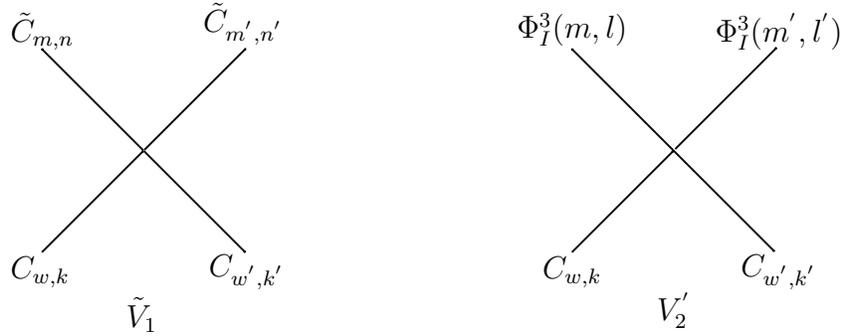}
\caption{$\tilde{V}_1$ , $V^{'}_2$ vertices}
\label{tv1v2}
\end{psfrags}
\end{center}
\end{figure}

\beqa
\label{V1t}
\tilde{V}_1=-\f{N}{2g^2} \tilde{F}_1(k,k^{'},n,n^{'}) \delta_{\o+\o^{'}+m+m^{'}}\mbox{~~~~~~~(Figure \ref{tv1v2})}
\eeqa

\beqa
\label{F1t}
\tilde{F}_1(k,k^{'},n,n^{'})&=&\sqrt{q}\int dx \left[\phi_k(x)\phi_{k^{'}}(x)\tilde{A}_n(x) \tilde{A}_{n^{'}}(x) + 
2 A_k\phi_{k^{'}}(x)\tilde{\phi}_n(x) \tilde{A}_{n^{'}}(x)\right.\nonumber\\
&+& \left. 2 \phi_k(x) A_{k^{'}}(x)\tilde{\phi}_n(x) \tilde{A}_{n^{'}}(x) + A_k(x) A_{k^{'}}(x) \tilde{\phi}_n(x) \tilde{\phi}_{n^{'}}(x)\right] 
\eeqa

\beqa
\label{V2p}
V^{'}_2=-\f{N}{2g^2}\left[F^{'}_2(k,k^{'},l,l^{'})\right]\delta_{\o+\o^{'}+m+m^{'}}\mbox{~~~~~~~(Figure \ref{tv1v2})}
\eeqa

\beqa
\label{F2p}
F^{'}_2(k,k^{'},l,l^{'})=\sqrt{q}\int dx \left[A_k(x) A_{k^{'}}(x)+\phi_k(x)\phi_{k^{'}}(x)\right] \left[e^{ilx}e^{il^{'}x}\right]
\eeqa

\begin{figure}[h]
\begin{center}
\begin{psfrags}
\psfrag{c1}[][]{$C_{w,k}$}
\psfrag{c2}[][]{$C_{w^{'},k^{'}}$}
\psfrag{c}[][]{$C_{w,k}$}
%\psfrag{a1}[][]{$D_{m^{'},n}$}
%\psfrag{a3}[][]{$A^3_x(m,l)$}
\psfrag{a31}[][]{$A^3_x(m,l)$}
\psfrag{a32}[][]{$A^3_x(m^{'},l^{'})$}
\psfrag{p13}[][]{$\Phi_1^3(m,l)$}
\psfrag{p23}[][]{$\Phi_1^3(m^{'},l^{'})$}
\psfrag{v3}[][]{$V_3$}
\psfrag{v3p}[][]{$V^{'}_3$}
\includegraphics[width=10cm,angle=0]{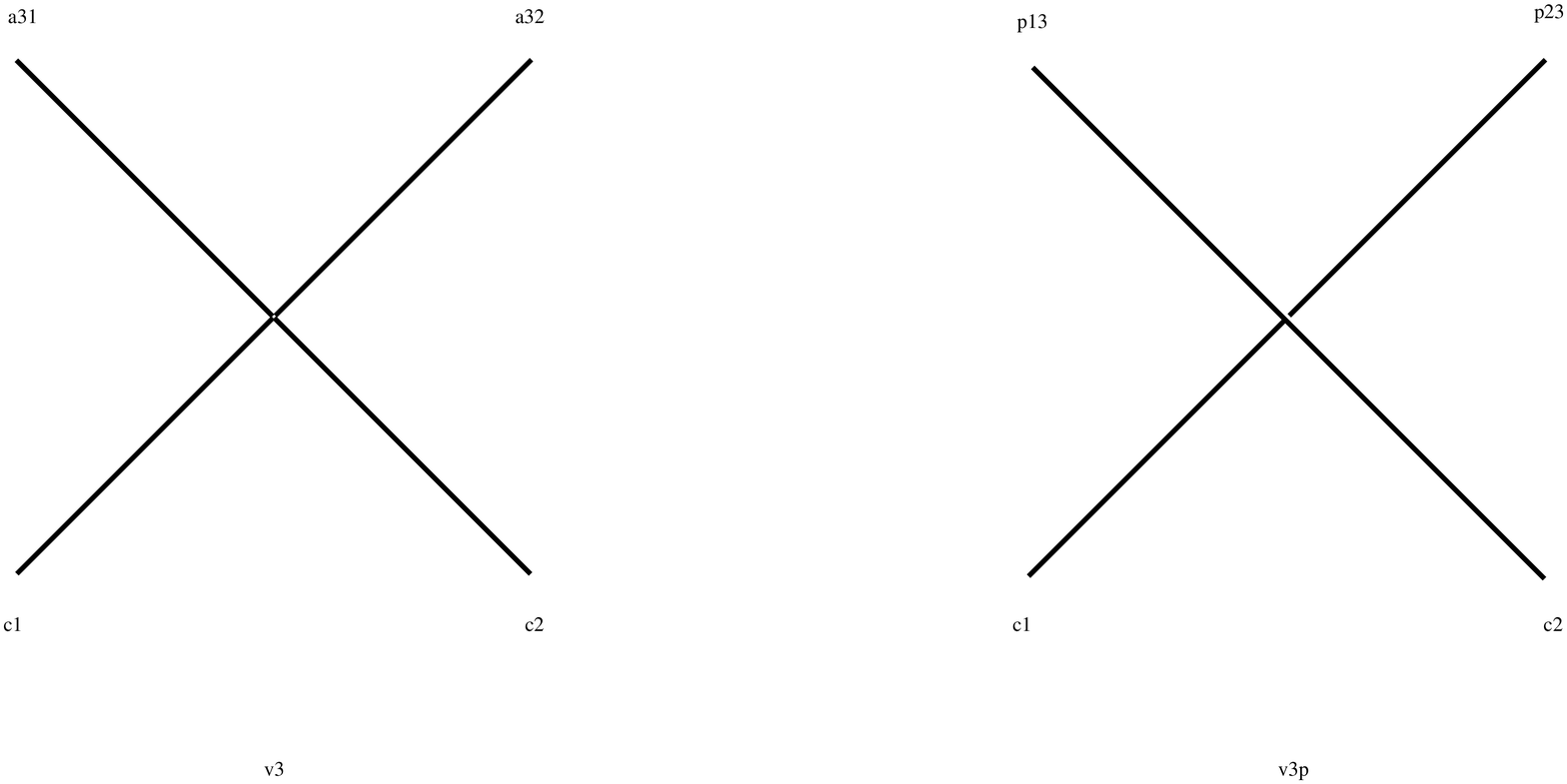} 
\caption{$V_3$ and $V^{'}_3$ vertices}
\label{v3v3p}
\end{psfrags}
\end{center}
\end{figure}

\beqa
\label{V3}
V_3=-\f{N}{2g^2}F_3(k,k^{'},l,l^{'})\delta_{\o+\o^{'}+m+m^{'}}\mbox{~~~~~~~(Figure \ref{v3v3p})}
\eeqa

\beqa
\label{F3}
F_3(k,k^{'},l,l^{'})=\sqrt{q}\int dx \left[\phi_k(x) \phi_{k^{'}(x)}\right]e^{i (l+l^{'}) x}
\eeqa

\beqa
\label{V3p}
V^{'}_3=-\f{N}{2g^2}F^{'}_3(k,k^{'},l,l^{'})\delta_{\o+\o^{'}+m+m^{'}}\mbox{~~~~~~~(Figure \ref{v3v3p})}
\eeqa

\beqa
\label{F3p}
F^{'}_3(k,k^{'},l,l^{'})=\sqrt{q}\int dx \left[A_k(x) A_{k^{'}(x)}e^{ilx}e^{il^{'}x}\right]
\eeqa

\begin{figure}[h]
\begin{center}
\begin{psfrags}
\psfrag{c}[][]{$C_{w,k}$}
\psfrag{a1}[][]{$C^{'}_{n,m}$}
\psfrag{a3}[][]{$A_x^3(l,m^{'})$}
\psfrag{v5}[][]{$V_4$}
\psfrag{c1}[][]{$A_x^3(l,m^{'})$}
\psfrag{c2}[][]{$\tilde{C}^{'}_{m,n}$}
\psfrag{v5p}[][]{$\tilde{V}_4$}
\includegraphics[width= 12cm,angle=0]{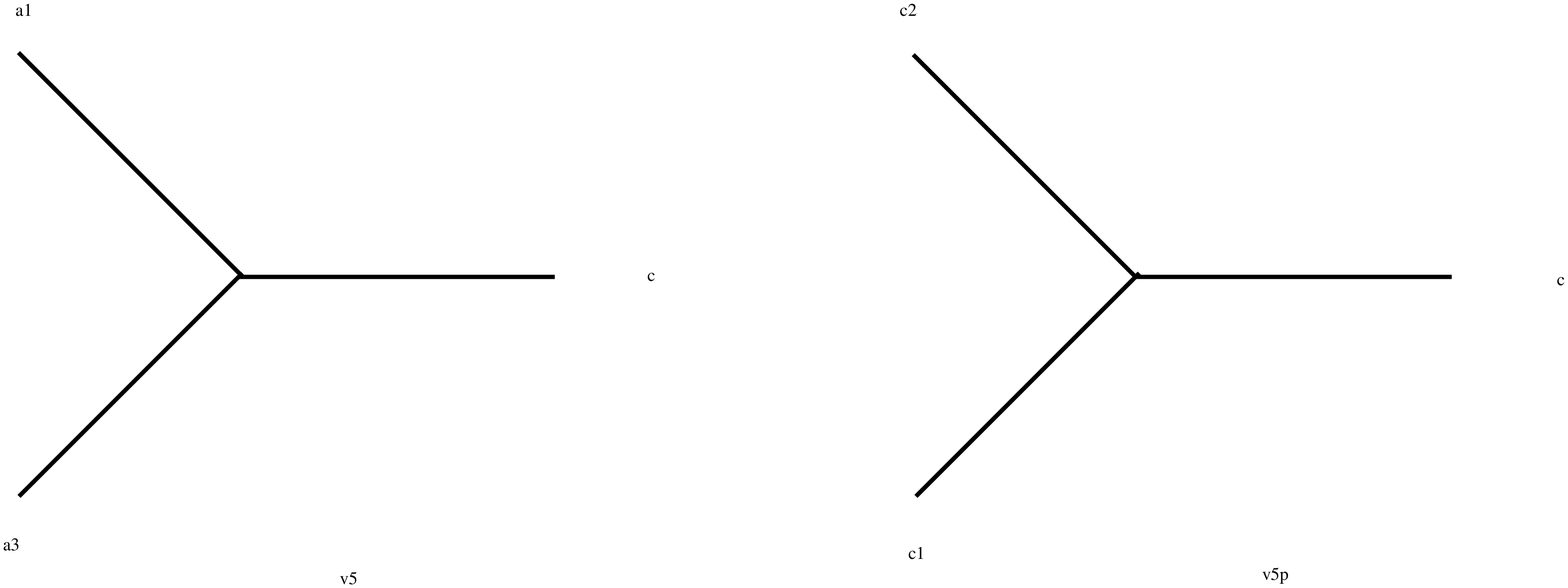} 
\caption{$V_4$ and $\tilde{V}_4$ vertices}
\label{v4v4t}
\end{psfrags}
\end{center}
\end{figure}

\beqa
\label{V4}
V_4&=&-\f{N^{3/2}}{g^2} F_4(k,l,n) \beta\delta_{\o+m+m^{'}}\mbox{~~~~~~~(Figure \ref{v4v4t})}
\eeqa

\beqa
\label{F4}
F_4(k,l,n)&=&\int dx \left[-\phi_n(x)\partial_x \phi_k(x)+\phi_k(x)\partial_x
\phi_n(x)-\phi_n(x)A_k(x) (qx)\right.\nonumber\\
&+& \left. A_n(x)\phi_k(x) (qx) \right]e^{ilx}
\eeqa

\beqa
\label{V4t}
\tilde{V}_4&=&-\f{N^{3/2}}{g^2} \tilde{F}_4(k,l,n) \beta\delta_{\o+m+m^{'}}\mbox{~~~~~~~(Figure \ref{v4v4t})}
\eeqa

\beqa
\label{F4t}
\tilde{F}_4(k,l,n)&=&\int dx \left[-\tilde{\phi}_n(x)\partial_x
\phi_k(x)+\phi_k(x)\partial_x \tilde{\phi}_n(x)-\tilde{\phi}_n(x)A_k(x)
(qx)\right.\nonumber\\
&+& \left.\tilde{A}_n(x)\phi_k(x) (qx)\right]e^{ilx}
\eeqa

\begin{figure}[h]
\begin{center}
\begin{psfrags}
\psfrag{c1}[][]{$C_{w,k}$}
\psfrag{c2}[][]{$C_{w^{'},k^{'}}$}
\psfrag{c}[][]{$C_{w,k}$}
\psfrag{a1}[][]{$\Phi_I^{1,2}(n,m)$}
\psfrag{a3}[][]{$\Phi_I^3(l,m^{'})$}
\psfrag{v4}[][]{$V_5$}
\includegraphics[width= 6 cm,angle=0]{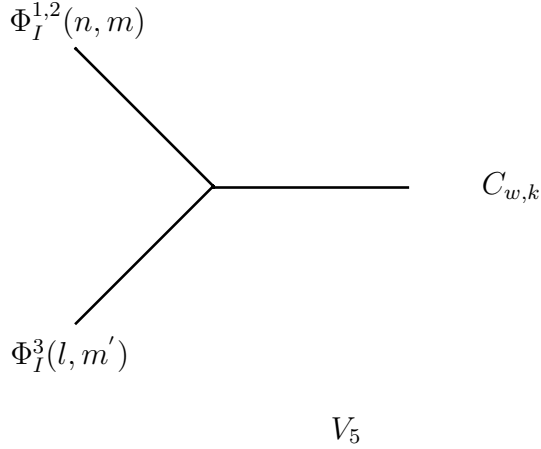} 
\caption{$V_5$ vertex.}
\label{v5}
\end{psfrags}
\end{center}
\end{figure}

\beqa
\label{V5}
V_5=-\f{N^{3/2}}{2g^2}F_5(k,l,n)\beta\delta_{\o+m+m^{'}} \mbox{~~~~~~~(Figure \ref{v5})}
\eeqa

\beqa
\label{F}
F_5(k,l,n)&=&\int dx  e^{-qx^2/2} \left[e^{ilx}A_k(x)\partial_x H_n(x)-(qx)e^{ilx}A_k(x) H_n(x)\right.\nonumber\\
&-& \left.(il)e^{ilx}A_k(x) H_n(x)+
(qx) e^{ilx}\phi_k(x) H_n(x)\right]
\eeqa

\begin{figure}[h]
\begin{center}
\begin{psfrags}
\psfrag{c}[][]{$C_{w,k}$}
\psfrag{a1}[][]{$\Phi_1^3(n,m^{'})$}
\psfrag{a3}[][]{$C^{'}_{n,m}$}
\psfrag{v5}[][]{$V^{'}_5$}
\psfrag{c1}[][]{$\tilde{C}^{'}_{n,m}$}
\psfrag{c2}[][]{$\Phi_1^3(l,m^{'})$}
\psfrag{v5p}[][]{$\tilde{V}^{'}_5$}
\includegraphics[width= 12cm,angle=0]{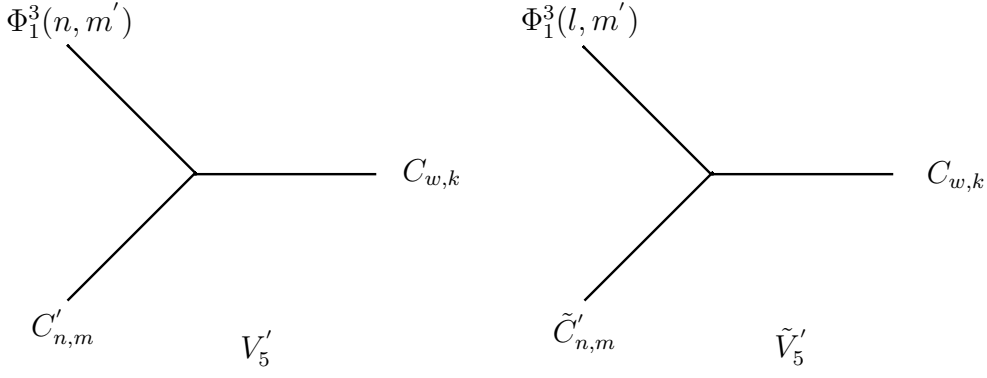} 
\caption{$V^{'}_5$ and $\tilde{V}^{'}_5$ vertices}
\label{v5ptv5p}
\end{psfrags}
\end{center}
\end{figure}

\beqa
\label{V5p}
V^{'}_5=-\f{N^{3/2}}{g^2}F^{'}_5(k,l,n)\beta\delta_{\o+m+m^{'}} \mbox{~~~~~~~(Figure \ref{v5ptv5p})}
\eeqa

\beqa
\label{F5p}
F^{'}_5(k,l,n)&=&\int dx \left[(il)e^{ilx}\phi_k(x)A_n(x)+(il)e^{ilx}A_k(x)\phi_n(x)\right.\nonumber\\
&-&\left.e^{ilx}\partial_x\phi_k(x) A_n(x)-e^{ilx}\partial_x\phi_n(x) A_k(x)\right]
\eeqa

\beqa
\label{V5tp}
\tilde{V}^{'}_5=-\f{N^{3/2}}{g^2}\tilde{F}^{'}_5(k,l,n)\beta\delta_{\o+m+m^{'}} \mbox{~~~~~~~(Figure \ref{v5ptv5p})}
\eeqa

\beqa
\label{F5tp}
\tilde{F}^{'}_5(k,l,n)&=&\int dx \left[(il)e^{ilx}\phi_k(x)\tilde{A}_n(x)+(il)e^{ilx}A_k(x)\tilde{\phi}_n(x)\right.\nonumber\\
&&\left. -e^{ilx}\partial_x\phi_k(x) \tilde{A}_n(x)-e^{ilx}\partial_x\tilde{\phi}_n(x) A_k(x)\right]
\eeqa

\subsection{Fermions}\label{fermions}

Propagator for the $L^a_i$ and $R^a_i$ modes $(a=1,2)$,

\beqa
\label{propferm1}
\expect{\theta_j(m,n)\theta_k^*(m^{'},n^{'})}=\frac{g^2}{N^{1/2}}\f{\delta_{jk}\delta_{m,m^{'}}\delta_{n'n^{'}}}{i\o_m+\sqrt{\lambda^{'}_n}}
\eeqa

$\lambda^{'}_n=2nq$.

Propagator for the $L^3_i$ and $R^3_i$ modes.
\beqa
\label{propferm2}
\expect{L^3_i(m,l)L^{3}_k(m^{'},l^{'})}&=&\frac{g^2}{N^{1/2}}\f{\delta_{ik}\delta_{m,-m^{'}}2\pi\delta(l+l^{'})}{i\o_m+l}\nonumber\\
\expect{R^3_i(m,l)R^{3}_k(m^{'},l^{'})}&=&\frac{g^2}{N^{1/2}}\f{\delta_{ik}\delta_{m,-m^{'}}2\pi\delta(l+l^{'})}{i\o_m-l}
\eeqa

\begin{figure}[h]
\begin{center}
\begin{psfrags}
\psfrag{c}[][]{$C_{w,k}$}
\psfrag{a1}[][]{$R_i^2(n,m)$}
\psfrag{a3}[][]{$L_{(9-i)}^3(l,m^{'})$}
\psfrag{v5}[][]{$V_6^R$}
\psfrag{c1}[][]{$R_{(9-i)}^3(l,m^{'})$}
\psfrag{c2}[][]{$L_i^2(n,m)$}
\psfrag{v5p}[][]{$V^L_6$}
\includegraphics[width= 12cm,angle=0]{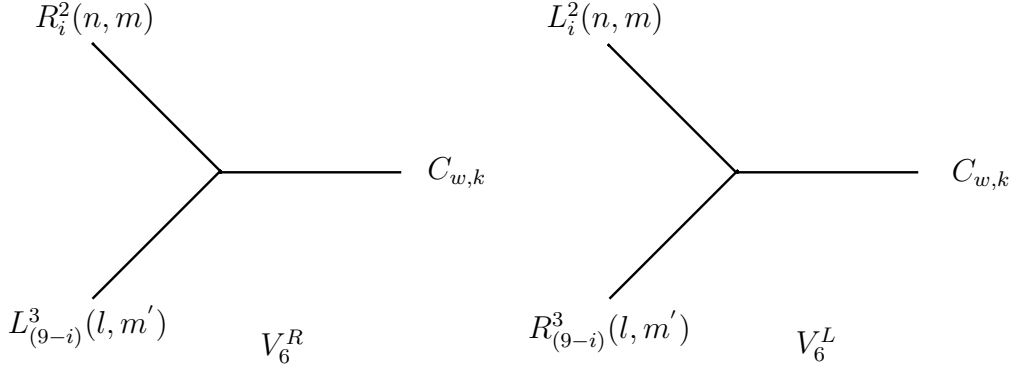} 
\caption{$V_6^R$ and $V^L_6$ vertices}
\label{v6rl}
\end{psfrags}
\end{center}
\end{figure}

\beqa
\label{VRL6}
V^{R/L}_{6}=i\f{N}{g^2}F^{R/L}_6(k,n,l)\delta_{w+m+m^{'}}\mbox{~~~~~~~(Figure \ref{v6rl})}
\eeqa

Where,

\beqa
\label{FR6}
F^R_6(k,n,l)=\sqrt{q}\int dx \phi_k(x)R_n(x)e^{ilx} 
\eeqa

There are eight such vertices involving $R_i^2$ and $L_{(9-i)}^3$ ($i=1 \cdots 8$).  We also have eight 
vertices involving $L_i^2$ and $R_{(9-i)}^3$ ($i=1 \cdots 8$). The corresponding vertices would be given by replacing $R_n(x)$
with $L_n(x)$. This vertex is thus,

\beqa
\label{FL6}
F^L_6(k,n,l)=\mp\sqrt{q}\int dx \phi_k(x)L_n(x)e^{ilx} 
\eeqa

The $\mp$ is due to the fact that half 
of the above vertices come with sign opposite to that of the other half in the Lagrangian.
There is no $\mp$ in ({\ref{FR6}) as the minus sign coming from the vertices in the Lagrangian is compensated by the ones from the 
eigenfunctions.

Similarly the other eight vertices involving both $R^3_i$ and $R^a_i$ with $(a=1,2)$ (or the $L$ legs) have the same structure.

\begin{figure}[h]
\begin{center}
\begin{psfrags}
\psfrag{c}[][]{$C_{w,k}$}
\psfrag{a1}[][]{$R_i^1(n,m)$}
\psfrag{a3}[][]{$R_{i}^3(l,m^{'})$}
\psfrag{v5}[][]{$V_7^R$}
\psfrag{c1}[][]{$L_{i}^3(l,m^{'})$}
\psfrag{c2}[][]{$L_i^1(n,m)$}
\psfrag{v5p}[][]{$V^L_7$}
\includegraphics[width= 12cm,angle=0]{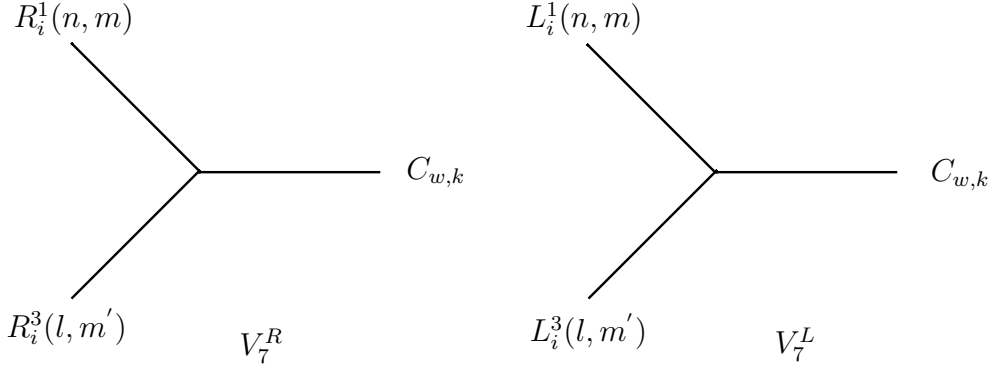} 
\caption{$V_7^R$ and $V^L_7$ vertices}
\label{v7rl}
\end{psfrags}
\end{center}
\end{figure}

\beqa
\label{VRL7}
V^{R/L}_{7}=i\f{N}{g^2}F^{R/L}_7(k,n,l)\delta_{w+m+m^{'}} \mbox{~~~~~~~(Figure \ref{v7rl})}
\eeqa

\beqa
\label{FRL7}
F^R_7(k,n,l)&=&\mp\sqrt{q}\int dx A_k(x)R_n(x)e^{ilx}\nonumber\\ 
F^L_7(k,n,l)&=&\sqrt{q}\int dx A_k(x)L_n(x)e^{ilx} 
\eeqa

Here the $\mp$ in the expression for $F^R_7(k,n,l)$ is due to the fact that half of the eigenfunctions come with a sign opposite to the other
half and in the Lagrangian all the terms come with the same sign.

\section{Vertices for computation of two-point $\tilde{C}_{w,k}$ amplitudes } \label{zeromassvertex}
Since the zero-eigenfunctions are also massless fields we compute their two-point finite temperature amplitudes. The amplitudes are given in section (\ref{zeroamp}).
Here we present the various four-point and three-point vertices that occur in the calculation for the amplitudes of the zero-eigenfunctions.

\subsection{Bosonic vertices}\label{zerobosvertex}

\begin{figure}[h]
\begin{center}
\begin{psfrags}
\psfrag{c1}[][]{$\tilde{C}_{w,k}$}
\psfrag{c2}[][]{$\tilde{C}_{w^{'},k^{'}}$}
\psfrag{d1}[][]{$C_{m,n}$}
\psfrag{d2}[][]{$C_{m^{'},n^{'}}$}
\psfrag{a11}[][]{$\Phi_I^{1,2}(m,n)$}
\psfrag{a12}[][]{$\Phi_I^{1,2}(m^{'},n^{'})$}
\psfrag{v1}[][]{$V_{H_1}$}
\psfrag{v2}[][]{$V_{H_2}$}
\includegraphics[width= 10cm,angle=0]{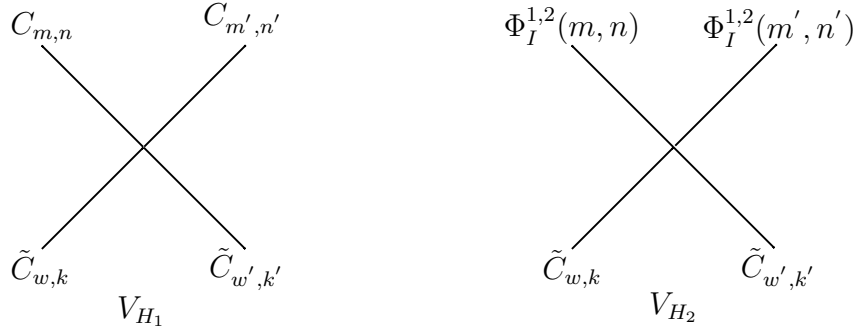}
\caption{$V_{H_1}$ , $V_{H_2}$ vertices}
\label{hv1v2}
\end{psfrags}
\end{center}
\end{figure}

\beqa
\label{VH1}
V_{H_1}=-\f{N}{2g^2} H_1(k,k^{'},n,n^{'}) \delta_{\o+\o^{'}+m+m^{'}} \mbox{~~~~~~~(Figure \ref{hv1v2})}
\eeqa

\beqa
H_1(k,k^{'},n,n^{'})&=&\sqrt{q}\int dx \left[\tilde{\phi}_k(x)\tilde{\phi}_{k^{'}}(x){A}_n(x) {A}_{n^{'}}(x) + 2 \tilde{A}_k(x)\tilde{\phi}_{k^{'}}(x){\phi}_n(x)  
{A}_{n^{'}}(x)\right.\nonumber\\
&+& \left. 2 \tilde{\phi}_k(x) \tilde{A}_{k^{'}}(x){\phi}_n(x) {A}_{n^{'}}(x) + \tilde{A}_k(x)\tilde{A}_{k^{'}}(x){\phi}_n(x){\phi}_{n^{'}}(x)\right] 
\eeqa

\beqa
\label{VH2}
V_{H_2}=-\f{N}{2g^2}\left[H_2(k,k^{'},n,n^{'})\right]\delta_{\o+\o^{'}+m+m^{'}} \mbox{~~~~~~~(Figure \ref{hv1v2})}
\eeqa

\beqa
H_2(k,k^{'},n,n^{'})=&&\sqrt{q}\int dx e^{-qx^2}\left [\tilde{A}_k(x) \tilde{A}_{k^{'}}(x)+\tilde{\phi}_k(x)\tilde{\phi}_{k^{'}}(x)\right]\nonumber\\
&& \times \left[H_n(\sqrt{q}x) H_{n^{'}}(\sqrt{q}x)\right]
\eeqa

\begin{figure}[h]
\begin{center}
\begin{psfrags}
\psfrag{c1}[][]{$\tilde{C}_{w,k}$}
\psfrag{c2}[][]{$\tilde{C}_{w^{'},k^{'}}$}
\psfrag{d1}[][]{$\tilde{C}_{m,n}$}
\psfrag{d2}[][]{$\tilde{C}_{m^{'},n^{'}}$}
\psfrag{a11}[][]{$\Phi_I^{3}(m,l)$}
\psfrag{a12}[][]{$\Phi_I^{3}(m^{'},l^{'})$}
\psfrag{v1}[][]{$\tilde{V}_{H_1}$}
\psfrag{v2}[][]{$V^{'}_{H_2}$}
\includegraphics[width= 10cm,angle=0]{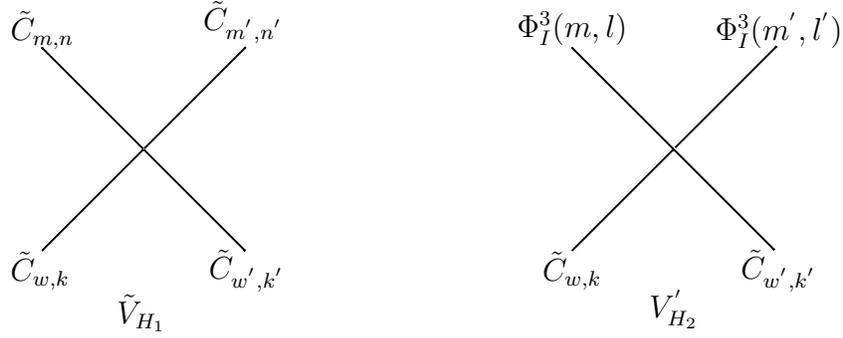}
\caption{$\tilde{V}_{H_1}$ , $V^{'}_{H_2}$ vertices}
\label{htv1v2}
\end{psfrags}
\end{center}
\end{figure}

\beqa
\label{VH1t}
\tilde{V}_{H_1}=-\f{N}{2g^2} \tilde{H}_1(k,k^{'},n,n^{'}) \delta_{\o+\o^{'}+m+m^{'}} \mbox{~~~~~~~(Figure \ref{htv1v2})}
\eeqa

\beqa
\tilde{H}_1(k,k^{'},n,n^{'})&=&\sqrt{q}\int dx \left[\tilde{\phi}_k(x)\tilde{\phi}_{k^{'}}(x)\tilde{A}_n(x) \tilde{A}_{n^{'}}(x) 
+ 2 \tilde{A}_k(x)\tilde{\phi}_{k^{'}}(x)\tilde{\phi}_n(x)  \tilde{A}_{n^{'}}(x)\right.\nonumber\\
&+& \left. 2 \tilde{\phi}_k(x) \tilde{A}_{k^{'}}(x)\tilde{\phi}_n(x) \tilde{A}_{n^{'}}(x) 
+ \tilde{A}_k(x)\tilde{A}_{k^{'}}(x)\tilde{\phi}_n(x) \tilde{\phi}_{n^{'}}(x)\right]
\eeqa

\beqa
\label{VH2p}
V^{'}_{H_2}=-\f{N}{2g^2}\left[H^{'}_2(k,k^{'},l,l^{'})\right]\delta_{\o+\o^{'}+m+m^{'}} \mbox{~~~~~~~(Figure \ref{htv1v2})}
\eeqa

\beqa
H^{'}_2(k,k^{'},l,l^{'})=\sqrt{q}\int dx \left[\tilde{A}_k(x) \tilde{A}_{k^{'}}(x)+\tilde{\phi}_k(x)\tilde{\phi}_{k^{'}}(x)\right] \left[e^{ilx}e^{il^{'}x}\right]
\eeqa

\begin{figure}[h]
\begin{center}
\begin{psfrags}
\psfrag{c1}[][]{$\tilde{C}_{w,k}$}
\psfrag{c2}[][]{$\tilde{C}_{w^{'},k^{'}}$}
\psfrag{c}[][]{$C_{w,k}$}
%\psfrag{a1}[][]{$D_{m^{'},n}$}
%\psfrag{a3}[][]{$A^3_x(m,l)$}
\psfrag{a31}[][]{$A^3_x(m,l)$}
\psfrag{a32}[][]{$A^3_x(m^{'},l^{'})$}
\psfrag{p13}[][]{$\Phi_1^3(m,l)$}
\psfrag{p23}[][]{$\Phi_1^3(m^{'},l^{'})$}
\psfrag{v3}[][]{$V_{H_3}$}
\psfrag{v3p}[][]{$V^{'}_{H_3}$}
\includegraphics[width=10cm,angle=0]{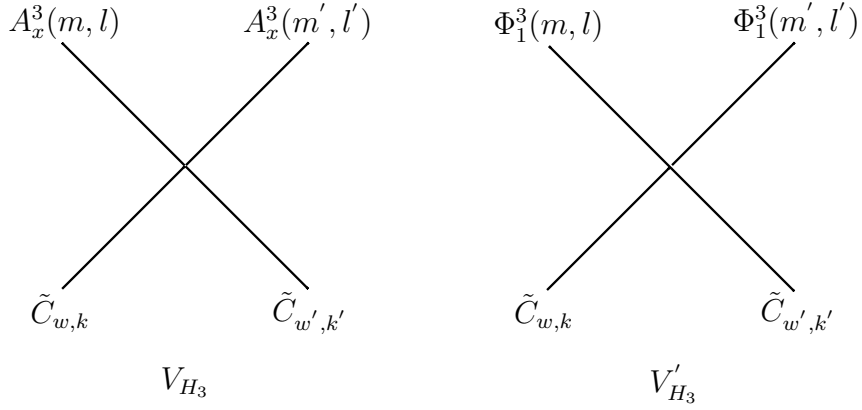} 
\caption{$V_{H_3}$ and $V^{'}_{H_3}$ vertices}
\label{hv3v3p}
\end{psfrags}
\end{center}
\end{figure}

\beqa
\label{VH3}
V_{H_3}=-\f{N}{2g^2}H_3(k,k^{'})\delta_{\o+\o^{'}+m+m^{'}} \mbox{~~~~~~~(Figure \ref{hv3v3p})}
\eeqa

\beqa
H_3(k,k^{'}l,l^{'})=\sqrt{q}\int dx \left[\tilde{\phi}_k(x) \tilde{\phi}_{k^{'}(x)}\right]e^{ilx}e^{il^{'}x}
\eeqa

\beqa
\label{VH3p}
V^{'}_{H_3}=-\f{N}{2g^2}H^{'}_3(k,k^{'},l,l^{'})\delta_{\o+\o^{'}+m+m^{'}} \mbox{~~~~~~~(Figure \ref{hv3v3p})}
\eeqa

\beqa
H^{'}_3(k,k^{'},l,l^{'})=\sqrt{q}\int dx \left[\tilde{A}_k(x) \tilde{A}_{k^{'}(x)}e^{ilx}e^{il^{'}x}\right]
\eeqa

\begin{figure}[h]
\begin{center}
\begin{psfrags}
\psfrag{c}[][]{$\tilde{C}_{w,k}$}
\psfrag{a1}[][]{$C^{'}_{n,m}$}
\psfrag{a3}[][]{$A_x^3(l,m^{'})$}
\psfrag{v5}[][]{$V_{H_4}$}
\psfrag{c1}[][]{$A_x^3(l,m^{'})$}
\psfrag{c2}[][]{$\tilde{C}^{'}_{m,n}$}
\psfrag{v5p}[][]{$\tilde{V}_{H_4}$}
\includegraphics[width= 12cm,angle=0]{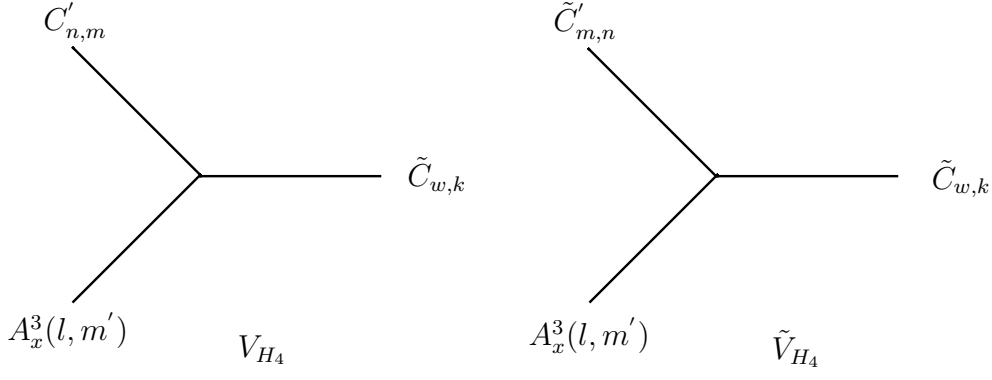} 
\caption{$V_{H_4}$ and $\tilde{V}_{H_4}$ vertices}
\label{hv4v4t}
\end{psfrags}
\end{center}
\end{figure}

\beqa
\label{VH4}
V_{H_4}&=&-\f{N^{3/2}}{g^2} H_4(k,l,n) \beta\delta_{\o+m+m^{'}}\mbox{~~~~~~~(Figure \ref{hv4v4t})}
\eeqa

\beqa
H_4(k,l,n)&=&\int dx \left[-\phi_n(x)\partial_x \tilde{\phi}_k(x)+\tilde{\phi}_k(x)\partial_x \phi_n(x)-\phi_n(x)\tilde{A}_k(x) (qx)\right.\nonumber\\
&+&\left.\tilde{\phi}_k(x)A_n(x) (qx)\right]e^{ilx}
\eeqa

\beqa
\label{VH4t}
\tilde{V}_{H_4}&=&-\f{N^{3/2}}{g^2} \tilde{H}_4(k,l,n) \beta\delta_{\o+m+m^{'}} \mbox{~~~~~~~(Figure \ref{hv4v4t})}
\eeqa

\beqa
\tilde{H}_4(k,l,n)&=&\int dx \left[-\tilde{\phi}_n(x)\partial_x \tilde{\phi}_k(x)+\tilde{\phi}_k(x)\partial_x 
\tilde{\phi}_n(x)-\tilde{\phi}_n(x)\tilde{A}_k(x) (qx)\right.\nonumber\\ 
&+&\left. \tilde{\phi}_k(x)\tilde{A}_n(x) (qx) \right]e^{ilx}
\eeqa

\begin{figure}[h]
\begin{center}
\begin{psfrags}
%\psfrag{c1}[][]{$\tilde{C}_{w,k}$}
%\psfrag{c2}[][]{$\tilde{C}_{w^{'},k^{'}}$}
\psfrag{c}[][]{$\tilde{C}_{w,k}$}
\psfrag{a1}[][]{$\Phi_I^{1,2}(n,m)$}
\psfrag{a3}[][]{$\Phi_I^3(l,m^{'})$}
\psfrag{v4}[][]{$V_{H_5}$}
\includegraphics[width= 6 cm,angle=0]{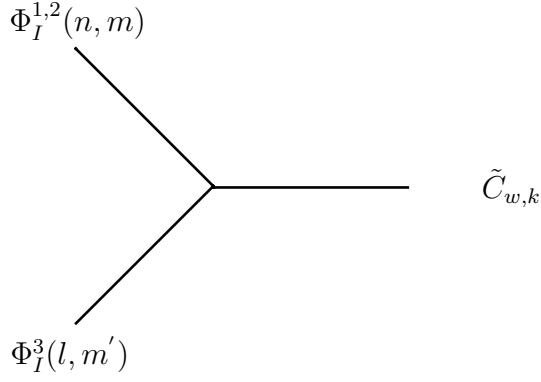} 
\caption{$V_{H_5}$ vertex.}
\label{hv5}
\end{psfrags}
\end{center}
\end{figure}

\beqa
\label{VH5}
V_{H_5}=-\f{N^{3/2}}{2g^2}H_5(k,l,n)\beta\delta_{\o+m+m^{'}}\mbox{~~~~~~~(Figure \ref{hv5})}
\eeqa

\beqa
H_5(k,l,n)&=&\int dx  e^{-qx^2/2} \left[e^{ilx}\tilde{A}_k(x)\partial_x H_n(\sqrt{q}x)-(qx)e^{ilx}\tilde{A}_k(x) H_n(\sqrt{q}x)\right.\nonumber\\
&& \left. -(il)e^{ilx}\tilde{A}_k(x) H_n(\sqrt{q}x)+
(qx) e^{ilx}\tilde{\phi}_k(x) H_n(\sqrt{q}x)\right]
\eeqa

\begin{figure}[h]
\begin{center}
\begin{psfrags}
\psfrag{c}[][]{$\tilde{C}_{w,k}$}
\psfrag{a1}[][]{$\Phi_1^3(n,m^{'})$}
\psfrag{a3}[][]{$C^{'}_{n,m}$}
\psfrag{v5}[][]{$V^{'}_5$}
\psfrag{c1}[][]{$\tilde{C}^{'}_{n,m}$}
\psfrag{c2}[][]{$\Phi_1^3(l,m^{'})$}
\psfrag{v5p}[][]{$\tilde{V}^{'}_{H_5}$}
\includegraphics[width= 12cm,angle=0]{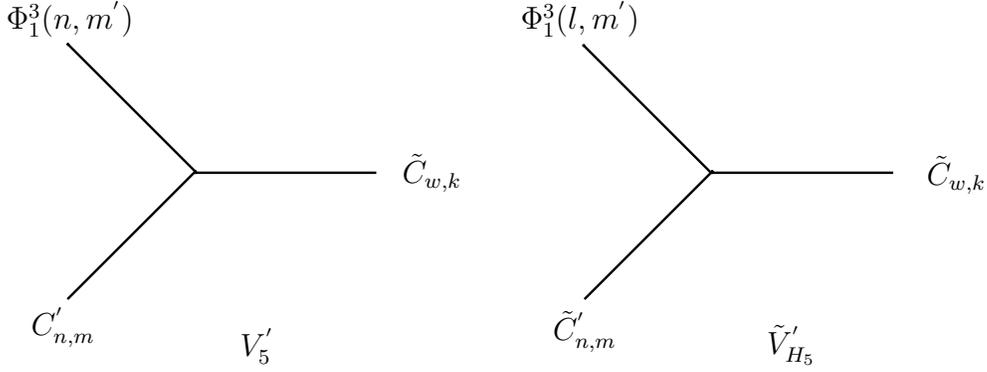} 
\caption{$V^{'}_{H_5}$ and $\tilde{V}^{'}_{H_5}$ vertices}
\label{hv5ptv5p}
\end{psfrags}
\end{center}
\end{figure}

\beqa
\label{VH5p}
V^{'}_{H_5}=-\f{N^{3/2}}{g^2}H^{'}_5(k,l,n)\beta\delta_{\o+m+m^{'}}\mbox{~~~~~~~(Figure \ref{hv5ptv5p})}
\eeqa

\beqa
H^{'}_5(k,l,n)&=&\int dx \left[(il)e^{ilx}\tilde{\phi}_k(x)A_n(x)+(il)e^{ilx}\tilde{A}_k(x)\phi_n(x)\right.\nonumber\\
&&\left. -e^{ilx}\partial_x\tilde{\phi}_k(x) A_n(x)-e^{ilx}\partial_x\phi_n(x) \tilde{A}_k(x)\right]
\eeqa

\beqa
\label{VH5tp}
\tilde{V_{H_5}}^{'}=-\f{N^{3/2}}{g^2}\tilde{H}^{'}_5(k,l,n)\beta\delta_{\o+m+m^{'}}\mbox{~~~~~~~(Figure \ref{hv5ptv5p})}
\eeqa

\beqa
\tilde{H}^{'}_5(k,l,n)&=&\int dx \left[(il)e^{ilx}\tilde{\phi}_k(x)\tilde{A}_n(x)+(il)e^{ilx}\tilde{A}_k(x)\tilde{\phi}_n(x)\right.\nonumber\\
&-&\left.e^{ilx}\partial_x \tilde{\phi}_k(x) \tilde{A}_n(x)-e^{ilx}\partial_x\tilde{\phi}_n(x) \tilde{A}_k(x)\right]
\eeqa

\subsection{Fermionic vertices}\label{zerofermions}
The fermionic three-point vertices for the $\tilde{C}_{w,k}-\tilde{C}_{w^{'},k^{'}}$ two-point functions at finite temperature participate in the cancellation
of UV divergence as in the case of the tachyonic amplitudes. The propagators for the fermionic loops are given in appendix (\ref{fermions}). 
The various three-point vertices are given below.

\begin{figure}[h]
\begin{center}
\begin{psfrags}
\psfrag{c}[][]{$\tilde{C}_{w,k}$}
\psfrag{a1}[][]{$R_i^2(n,m)$}
\psfrag{a3}[][]{$L_{(9-i)}^3(l,m^{'})$}
\psfrag{v5}[][]{$V_{H_6}^R$}
\psfrag{c1}[][]{$R_{(9-i)}^3(l,m^{'})$}
\psfrag{c2}[][]{$L_i^2(n,m)$}
\psfrag{v5p}[][]{$V^L_{H_6}$}
\includegraphics[width= 12cm,angle=0]{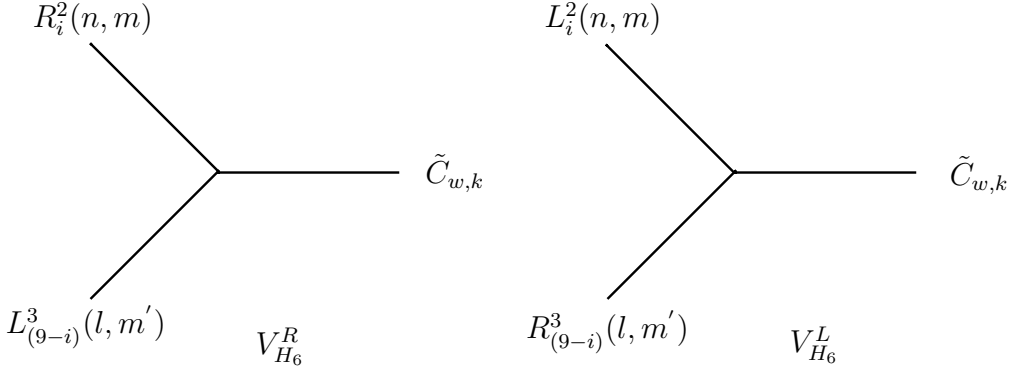} 
\caption{$V_{H_6}^R$ and $V^L_{H_6}$ vertices}
\label{hv6rl}
\end{psfrags}
\end{center}
\end{figure}

\beqa
\label{VHRL6}
\tilde{V}^{R/L}_{H_6}=i\f{N}{g^2}\tilde{H}^{R/L}_6(k,n,l)\delta_{w+m+m^{'}}\mbox{~~~~~~~(Figure \ref{hv6rl})}
\eeqa

\beqa\label{HR6}
\tilde{H}^R_6(k,n,l)=\sqrt{q}\int dx \tilde{\phi}_k(x)R_n(x)e^{ilx} 
\eeqa

\beqa
\label{HL6}
\tilde{H}^L_6(k,n,l)=\mp\sqrt{q}\int dx \tilde{\phi}_k(x)L_n(x)e^{ilx} 
\eeqa

\begin{figure}[h]
\begin{center}
\begin{psfrags}
\psfrag{c}[][]{$\tilde{C}_{w,k}$}
\psfrag{a1}[][]{$R_i^1(n,m)$}
\psfrag{a3}[][]{$R_{i}^3(l,m^{'})$}
\psfrag{v5}[][]{$V_7^R$}
\psfrag{c1}[][]{$L_{i}^3(l,m^{'})$}
\psfrag{c2}[][]{$L_i^1(n,m)$}
\psfrag{v5p}[][]{$V^L_7$}
\includegraphics[width= 12cm,angle=0]{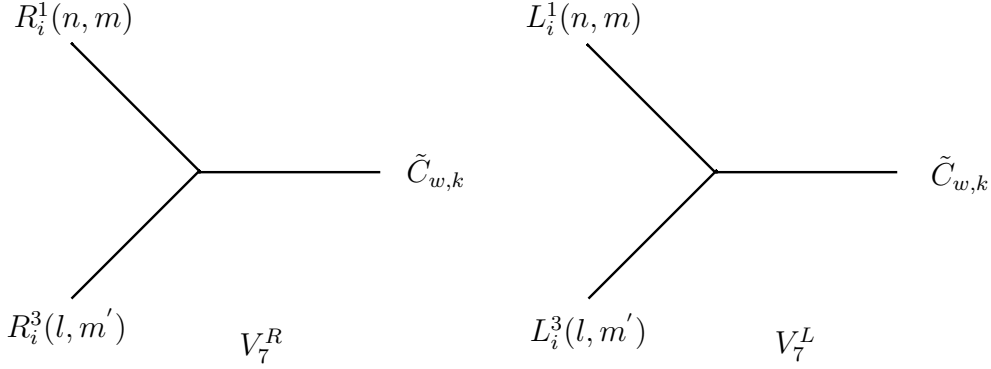} 
\caption{$V_{H_7}^R$ and $V^L_{H_7}$ vertices}
\label{hv7rl}
\end{psfrags}
\end{center}
\end{figure}

\beqa
\label{VHRL7}
\tilde{V}^{R/L}_{H_7}=i\f{N}{g^2}\tilde{H}^{R/L}_7(k,n,l)\delta_{w+m+m^{'}} \mbox{~~~~~~~(Figure \ref{hv7rl})}
\eeqa

\beqa
\label{HRL7}
\tilde{H}^R_7(k,n,l)&=&\mp\sqrt{q}\int dx \tilde{A}_k(x)R_n(x)e^{ilx}\nonumber\\ 
\tilde{H}^L_7(k,n,l)&=&\sqrt{q}\int dx \tilde{A}_k(x)L_n(x)e^{ilx} 
\eeqa

The origin of the $\mp$ sign is explained in Appendix \ref{fermions}.

\section{Vertices for computation of two point amplitudes for $\Phi_I^3$ and $A^3_x$}

\subsection{$\Phi_1^3$ vertices}\label{vphi13}
We list here  all the four-point and three-point bosonic and three-point fermionic vertices that constitute the two-point functions for $\Phi^3_1-\Phi^3_1$. 
\begin{figure}[h]
\begin{center}
\begin{psfrags}
\psfrag{c1}[][]{$\Phi^3_1(m,l)$}
\psfrag{c2}[][]{$\Phi^3_1(m^{'},l^{'})$}
%\psfrag{c}[][]{$\tilde{C}_{w,k}$}
\psfrag{a1}[][]{$\Phi^1_I{m^{'},n}, \Phi^2_I{m^{'},n}$}
\psfrag{a3}[][]{$A^3_x(m)$}
\psfrag{a31}[][]{$\Phi^{(1,2)}_I{m,n}$}
\psfrag{a32}[][]{$\Phi^{(1,2)}_I{m^{'},n^{'}}$}
\psfrag{p13}[][]{$C_{m,n}$}
\psfrag{p23}[][]{$C_{m^{'},n^{'}}$}
\psfrag{v3}[][]{$V^1_1$}
\psfrag{v3p}[][]{$V^1_2$}
\includegraphics[width=10cm,angle=0]{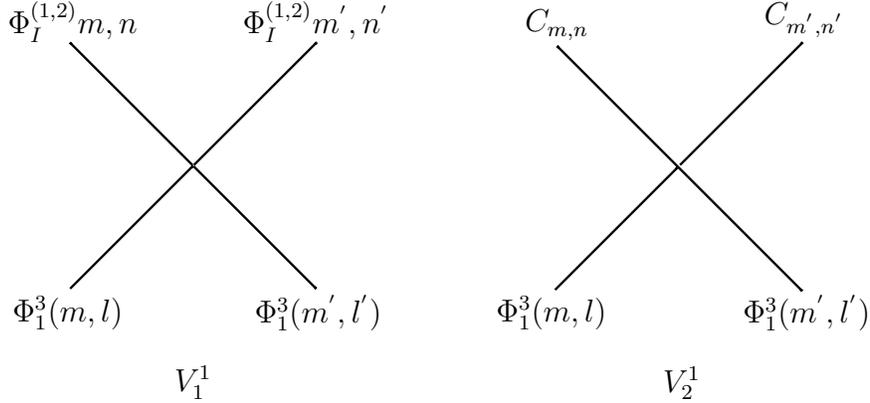} 
\caption{$V^1_1$ and $V^1_2$ vertices}
\label{v11v12}
\end{psfrags}
\end{center}
\end{figure}

\beqa
\label{V11}
V_1^1=-\f{N}{2g^2} G_1^1(l,l^{'},n,n^{'}) \delta_{w+w^{'}+m+m^{'}} \mbox{~~~~~~~(Figure \ref{v11v12})}
\eeqa

\beqa
\label{G11}
G_1^1(l,l^{'},n,n^{'})=\sqrt{q}\int dx e^{-qx^2}H_n(x)H_{n^{'}}(x)e^{ilx}e^{il^{'}x}
\eeqa

\beqa
\label{V21}
V_2^1=-\f{N}{2g^2} G_2^1(l,l^{'},n,n^{'}) \delta_{w+w^{'}+m+m^{'}} \mbox{~~~~~~~(Figure \ref{v11v12})}
\eeqa

\beqa
\label{G21}
G_2^1(l,l^{'},n,n^{'})=\sqrt{q}\int dx A_n(x)A_{n^{'}}(x)e^{ilx}e^{il^{'}x}
\eeqa

\begin{figure}[h]
\label{phi31fourpoint2}
\begin{center}
\begin{psfrags}
\psfrag{c1}[][]{$\Phi^3_1(m,l)$}
\psfrag{c2}[][]{$\Phi^3_1(m^{'},l^{'})$}
\psfrag{a11}[][]{$\tilde{C}_{m,n}$}
\psfrag{a12}[][]{$\tilde{C}_{m^{'},n^{'}}$}
\psfrag{v2}[][]{$\tilde{V}_2^1$}
\includegraphics[width= 4cm,angle=0]{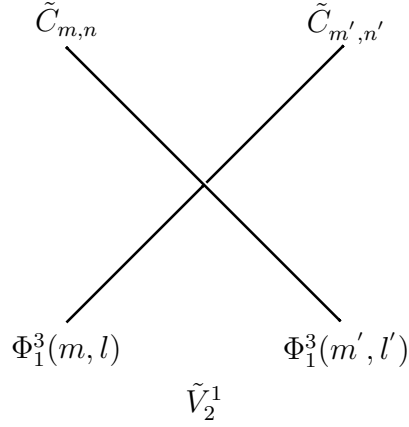}
\caption{$\tilde{V}_2^1$ vertex}
\label{tv12}
\end{psfrags}
\end{center}
\end{figure}

\beqa
\label{V21t}
\tilde{V}_2^1=-\f{N}{2g^2} \tilde{G}_2^1(l,l^{'},n,n^{'}) \delta_{w+w^{'}+m+m^{'}}\mbox{~~~~~~~(Figure \ref{tv12})}
\eeqa

\beqa
\label{G21t}
\tilde{G}_2^1(l,l^{'},n,n^{'})=\sqrt{q}\int dx \tilde{A}_n(x)\tilde{A}_{n^{'}}(x)e^{ilx}e^{il^{'}x}
\eeqa

\begin{figure}[h]
\begin{center}
\begin{psfrags}
\psfrag{c}[][]{$\Phi^3_1(m,l)$}
\psfrag{a1}[][]{$\Phi^{(1,2)}_I(m,n)$}
\psfrag{a3}[][]{$\Phi^{(1,2)}_I(m^{'},n^{'})$}
\psfrag{v5}[][]{$V^{1'}_1$}
\psfrag{c1}[][]{$C_{n,m}$}
\psfrag{c2}[][]{$C_{m^{'},n^{'}}$}
\psfrag{v5p}[][]{$V^1_3$}
\includegraphics[width= 12cm,angle=0]{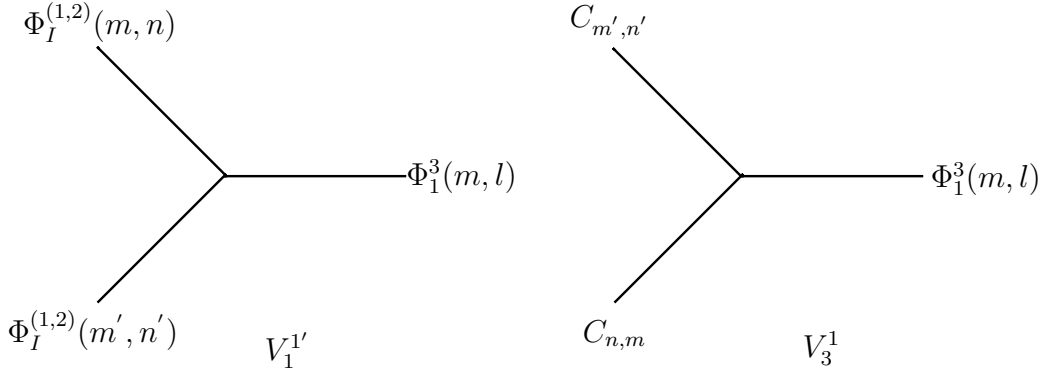} 
\caption{$V^{1'}_1$ and $V^{1}_3$ vertices}
\label{v11pv13}
\end{psfrags}
\end{center}
\end{figure}

\beqa
\label{V11p}
V_1^{1'}=-\f{N}{2g^2} G_1^{1'}(l,n,n^{'}) \delta_{w+m+m^{'}}\mbox{~~~~~~~(Figure \ref{v11pv13})}
\eeqa

\beqa
\label{G11p}
G_1^{1'}(l,n,n^{'})=\sqrt{q}\int dx~\left[q x e^{-qx^2}H_n(x)H_{n^{'}}(x)e^{ilx}\right]
\eeqa

\beqa
\label{V31}
V_3^1=-\f{N^{3/2}}{g^2} G_3^1(l,n,n^{'}) \beta\delta_{w+m+m^{'}}\mbox{~~~~~~~(Figure \ref{v11pv13})}
\eeqa

\beqa
\label{G31}
G_3^1(l,n,n^{'})=\int dx e^{ilx}\left[q x A_n(x) A_{n^{'}}(x) + \partial_x\phi_{n^{'}}(x)A_n(x)-il A_n(x)\phi_{n^{'}}(x)\right]
\eeqa

\begin{figure}[h]
\begin{center}
\begin{psfrags}
\psfrag{c}[][]{$\Phi^3_1(m,l)$}
\psfrag{a1}[][]{$\tilde{C}_{m,n}$}
\psfrag{a3}[][]{$\tilde{C}_{m^{'},n^{'}}$}
\psfrag{v5}[][]{$\tilde{V}^1_3$}
\psfrag{c1}[][]{$C_{n,m}$}
\psfrag{c2}[][]{$\tilde{C}_{m^{'},n^{'}}$}
\psfrag{v5p}[][]{$\tilde{V}^{1'}_3$}
\includegraphics[width= 12cm,angle=0]{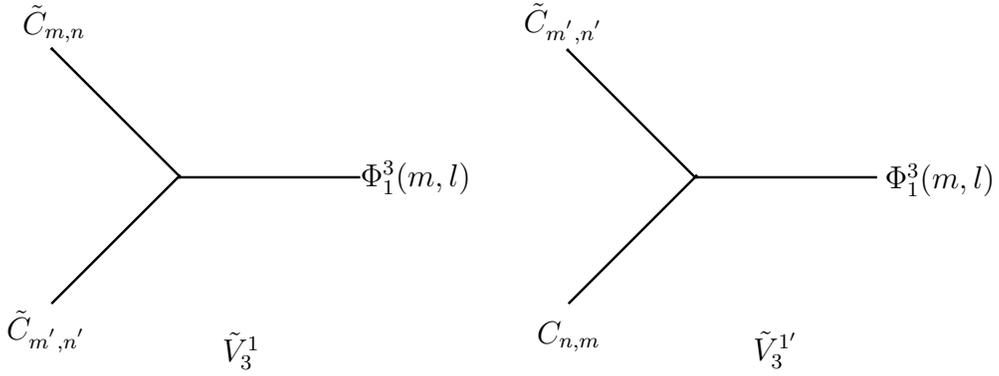} 
\caption{$\tilde{V}^1_3$ and $\tilde{V}^{1'}_3$ vertices}
\label{tv13tv13p}
\end{psfrags}
\end{center}
\end{figure}

\beqa
\label{V31t}
\tilde{V}_3^1=-\f{N^{3/2}}{g^2} \tilde{G}_3^1(l,n,n^{'}) \beta\delta_{w+m+m^{'}}\mbox{~~~~~~~(Figure \ref{tv13tv13p})}
\eeqa

\beqa
\label{G11t}
\tilde{G}_3^1(l,n,n^{'})=\int dx e^{ilx}\left[q x \tilde{A}_n(x) \tilde{A}_{n^{'}}(x) + \partial_x\tilde{\phi}_{n^{'}}(x)\tilde{A}_n(x)-
il \tilde{A}_n(x)\tilde{\phi}_{n^{'}}(x)\right]
\eeqa

\beqa
\label{V31tp}
\tilde{V}_3^{1'}=-\f{N^{3/2}}{g^2} \tilde{G}_3^{1'}(l,n,n^{'}) \beta\delta_{w+m+m^{'}}\mbox{~~~~~~~(Figure \ref{tv13tv13p})}
\eeqa

\beqa
\label{G31p}
\tilde{G}_3^{1'}(l,n,n^{'})&=&\int dx e^{ilx}\left[2q x A_n(x) \tilde{A}_{n^{'}}(x) + \partial_x\tilde{\phi}_{n^{'}}(x)A_n(x)-il A_n(x)\tilde{\phi}_{n^{'}}(x)
\right.\nonumber\\
&+&\left.  \partial_x \phi_{n}(x)\tilde{A}_{n^{'}}(x)e^{ilx}-il \tilde{A}_{n^{'}}(x)\phi_{n}(x)\right]
\eeqa

\begin{figure}[h]
\begin{center}
\begin{psfrags}
\psfrag{c}[][]{$\Phi^3_1(m,l)$}
\psfrag{a1}[][]{$\theta_i(m,n)$}
\psfrag{a3}[][]{$\theta_j(m^{'},n^{'})$}
\psfrag{v5}[][]{$V^1_f$}
\psfrag{c1}[][]{$\theta_i(m,n)$}
\psfrag{c2}[][]{$\theta^{*}_{j}(m^{'},n^{'})$}
\psfrag{v5p}[][]{$V^2_f$}
\includegraphics[width= 12cm,angle=0]{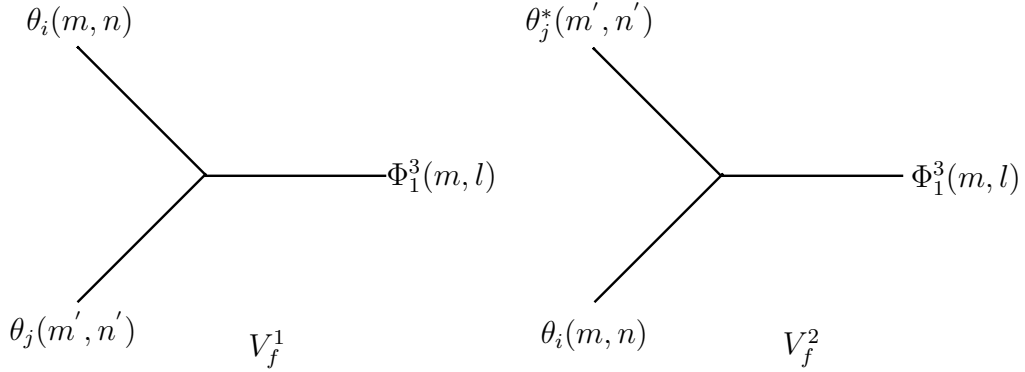} 
\caption{The fermionic vertices $V^1_f$ and $V^2_f$ for $\Phi^3_1-\Phi^3_1$ amplitudes}
\label{v1fv2f}
\end{psfrags}
\end{center}
\end{figure}

\beqa
\label{Vf1}
V_f^1=i\f{N}{g^2} G_f^1(l,n,n^{'}) \delta_{w+m+m^{'}}\mbox{~~~~~~~(Figure \ref{v1fv2f})}
\eeqa

\beqa
\label{Gf1}
G_f^1(l,n,n^{'})= \sqrt{q}\int dx e^{ilx}R_n(x)L_{n^{'}}(x)
\eeqa

\beqa
\label{Vf2}
V_f^2=i\f{N}{g^2} G_f^2(l,n,n^{'}) \delta_{w+m+m^{'}}\mbox{~~~~~~~(Figure \ref{v1fv2f})}
\eeqa

\beqa
\label{Gf2}
G_f^{2}(l,n,n^{'})= \sqrt{q}\int dx e^{ilx}\left[R_n(x)L_{n^{'}}^{*}-L_n(x)R_{n^{'}}^{*}(x)\right]
\eeqa

\subsection{$\Phi_I^3$, $I\ne 1$ vertices}\label{vphii3}
We list here the vertices for the one-loop finite temperature mass-squared corrections for the massless 
field $\Phi^3_I(m,l), I \ne 1$.  

\begin{figure}[h]
\begin{center}
\begin{psfrags}
\psfrag{c1}[][]{$\Phi^3_I(m,l)$}
\psfrag{c2}[][]{$\Phi^3_I(m^{'},l^{'})$}
%\psfrag{c}[][]{$\tilde{C}_{w,k}$}
\psfrag{a1}[][]{$\Phi^1_I{m^{'},n}, \Phi^2_I{m^{'},n}$}
\psfrag{a3}[][]{$A^3_x(m,l)$}
\psfrag{a31}[][]{$\Phi^{(1,2)}_I{m,n}$}
\psfrag{a32}[][]{$\Phi^{(1,2)}_I{m^{'},n^{'}}$}
\psfrag{p13}[][]{$C_{m,n}$}
\psfrag{p23}[][]{$C_{m^{'},n^{'}}$}
\psfrag{v3}[][]{$V^I_1$}
\psfrag{v3p}[][]{$V^I_2$}
\includegraphics[width=10cm,angle=0]{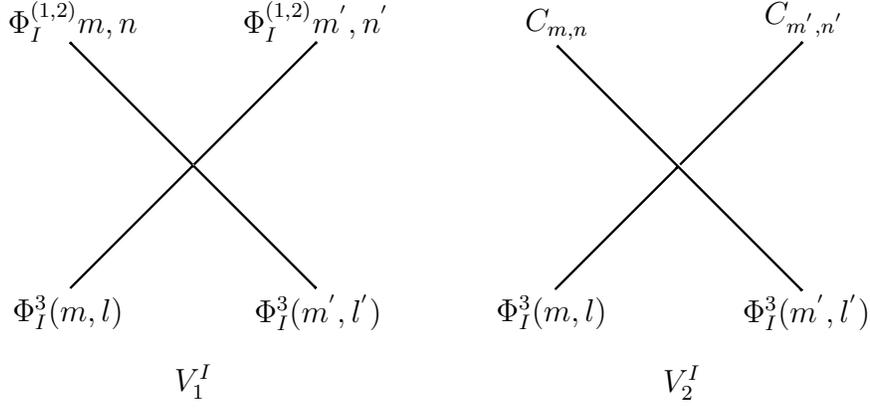} 
\caption{$V^I_1$ and $V^I_2$ vertices}
\label{vi1vi2}
\end{psfrags}
\end{center}
\end{figure}

\beqa
\label{V1I}
V_1^I=-\f{N}{2g^2} G_1^I(l,l^{'},n,n^{'}) \delta_{w+w^{'}+m+m^{'}}\mbox{~~~~~~~(Figure \ref{vi1vi2})}
\eeqa

\beqa
\label{G1I}
G_1^I(l,l^{'},n,n^{'})=\sqrt{q}\int dx e^{-qx^2}H_n(x)H_{n^{'}}(x)e^{ilx}e^{il^{'}x}
\eeqa

\beqa
\label{V2I}
V_2^I=-\f{N}{2g^2} G_2^I(l,l^{'},n,n^{'}) \delta_{w+w^{'}+m+m^{'}}\mbox{~~~~~~~(Figure \ref{vi1vi2})}
\eeqa

\beqa
\label{G2I}
G_2^I(l,l^{'},n,n^{'})=\sqrt{q}\int dx \left[A_n(x)A_{n^{'}}(x)+\phi_n(x)\phi_{n^{'}}(x)\right]e^{ilx}e^{il^{'}x}
\eeqa

\begin{figure}[h]
\begin{center}
\begin{psfrags}
\psfrag{c1}[][]{$\Phi^3_I(m,l)$}
\psfrag{c2}[][]{$\Phi^3_I(m^{'},l^{'})$}
\psfrag{a11}[][]{$\tilde{C}_{m,n}$}
\psfrag{a12}[][]{$\tilde{C}_{m^{'},n^{'}}$}
\psfrag{v2}[][]{$\tilde{V}_2^I$}
\includegraphics[width= 4cm,angle=0]{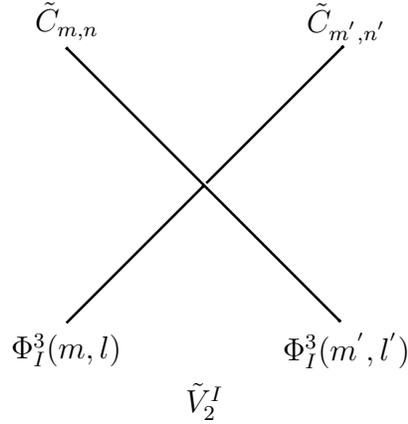}
\caption{$\tilde{V}_2^I$ vertex}
\label{tvi2}
\end{psfrags}
\end{center}
\end{figure}

\beqa
\label{V2It}
\tilde{V}_2^I=-\f{N}{2g^2} \tilde{G}_2^I(l,l^{'},n,n^{'}) \delta_{w+w^{'}+m+m^{'}}\mbox{~~~~~~~(Figure \ref{tvi2})}
\eeqa

\beqa
\label{G2It}
\tilde{G}_2^I(l,l^{'},n,n^{'})=\sqrt{q}\int dx \left[\tilde{A}_n(x)\tilde{A}_{n^{'}}(x)+\tilde{\phi}_n(x)\tilde{\phi}_{n^{'}}(x)\right]e^{ilx}e^{il^{'}x}
\eeqa

\begin{figure}[h]
\begin{center}
\begin{psfrags}
\psfrag{c}[][]{$~~\Phi^3_I(m,l)$}
\psfrag{a1}[][]{$~~\Phi^{(1,2)}_I(m,n)$}
\psfrag{a3}[][]{$C_{m^{'},n^{'}}$}
\psfrag{v5}[][]{$V^I_3$}
\psfrag{c1}[][]{$\Phi^{(1,2)}_I(m,n)$}
\psfrag{c2}[][]{$\tilde{C}_{m^{'},n^{'}}$}
\psfrag{v5p}[][]{$\tilde{V}^I_3$}
\includegraphics[width= 10cm,angle=0]{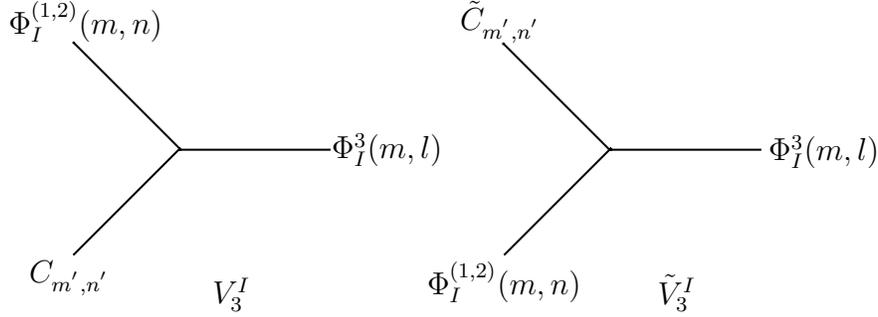} 
\caption{$V^I_3$ and $\tilde{V}^{I}_3$ vertices}
\label{vi3tvi3}
\end{psfrags}
\end{center}
\end{figure}

\beqa
\label{V3I}
V_3^I=-\f{N^{3/2}}{g^2} G_3^I(l,n,n^{'}) \beta\delta_{w+m+m^{'}}\mbox{~~~~~~~(Figure \ref{vi3tvi3})}
\eeqa

\beqa
\label{G3I}
G_3^I(l,n,n^{'})&=&\int dx e^{ilx}\left[\partial_x (e^{-qx^2/2} H_{n^{'}}(x))A_n(x)-il e^{-qx^2/2}A_n(x)H_{n^{'}}(x)\right.\non
&&\left.-qx\phi_n(x)e^{-qx^2/2}H_{n^{'}}(x)\right]
\eeqa

\beqa
\label{V3It}
\tilde{V}_3^I=-\f{N^{3/2}}{g^2} \tilde{G}_3^I(l,n,n^{'}) \beta\delta_{w+m+m^{'}}\mbox{~~~~~~~(Figure \ref{vi3tvi3})}
\eeqa

\beqa
\label{G3It}
\tilde{G}_3^I(l,n,n^{'})&=&\int dx e^{ilx}\left[\partial_x (e^{-qx^2/2} H_{n^{'}}(x))\tilde{A}_n(x)-il e^{-qx^2/2}\tilde{A}_n(x)H_{n^{'}}(x)\right.\non
&&\left.-qx\tilde{\phi}_n(x)e^{-qx^2/2}H_{n^{'}}(x)\right]
\eeqa

\begin{figure}[h]
\begin{center}
\begin{psfrags}
\psfrag{c1}[][]{$\Phi^3_I(m,l)$}
\psfrag{c}[][]{$\Phi^3_I(m^{'},l^{'})$}
\psfrag{a1}[][]{$\theta_i(m,n)$}
\psfrag{a3}[][]{$\theta^{*}_j(m^{'},n^{'})$}
\psfrag{v4}[][]{$V^I_f$}
\includegraphics[width= 6cm,angle=0]{vertex1.eps}
\caption{{$V^I_f$ vertex}}
\label{vif}
\end{psfrags}
\end{center}
\end{figure}

\beqa
\label{VfI}
V_f^I=i\f{N}{g^2} G_f^I(l,n,n^{'}) \delta_{w+m+m^{'}}\mbox{~~~~~~~(Figure \ref{vif})}
\eeqa

\beqa
\label{GfI}
G_f^{I}(l,n,n^{'})= \sqrt{q}\int dx e^{ilx}\left[R_n(x)L_{n^{'}}+L_n(x)R_{n^{'}}(x)\right]
\eeqa

\subsection{$A^3_x$ vertices}\label{va13}
In this section we write down the various vertices needed for computing the two point $A^3_x$ amplitude.

\begin{figure}[h]
\begin{center}
\begin{psfrags}
\psfrag{c1}[][]{$A^3_x(m,l)$}
\psfrag{c2}[][]{$A^3_x(m^{'},l^{'})$}
%\psfrag{c}[][]{$\tilde{C}_{w,k}$}
\psfrag{a1}[][]{$\Phi^{(1,2)}1_I(m^{'},n)$}
\psfrag{a3}[][]{$A^3_x(m)$}
\psfrag{a31}[][]{$\Phi^{(1,2)}_I(m,n)$}
\psfrag{a32}[][]{$\Phi^{(1,2)}_I(m^{'},n^{'})$}
\psfrag{p13}[][]{$C_{m,n}$}
\psfrag{p23}[][]{$C_{m^{'},n^{'}}$}
\psfrag{v3}[][]{$V^A_1$}
\psfrag{v3p}[][]{$V^A_2$}
\includegraphics[width=10cm,angle=0]{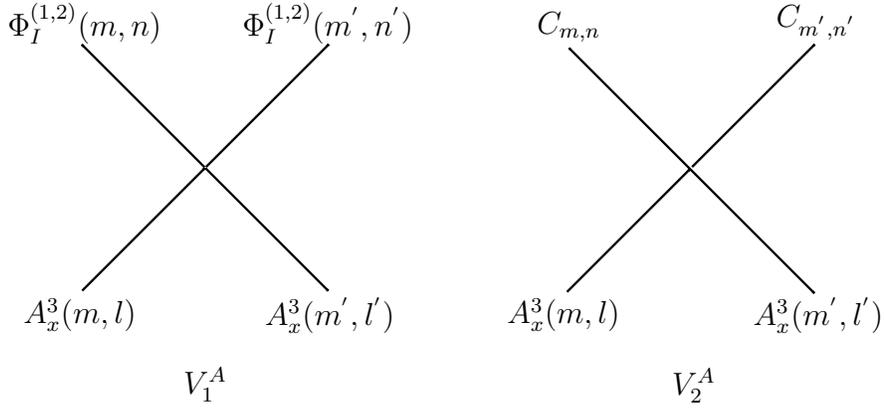} 
\caption{$V^A_1$ and $V^A_2$ vertices}
\label{va1va2}
\end{psfrags}
\end{center}
\end{figure}

\beqa
\label{V1A}
V_1^A=-\f{N}{2g^2} G_1^A(n,n^{'},l,l^{'}) \delta_{w+w^{'}+m+m^{'}}\mbox{~~~~~~~(Figure \ref{va1va2})}
\eeqa

\beqa
\label{G1A}
G_1^A(n,n^{'}l,l^{'})= \sqrt{q} \int dx e^{-q x^2} H_n(\sqrt{q}x) H_{n^{'}}(\sqrt{q}x) e^{i(l+ l^{'})x}
\eeqa

\beqa
\label{V2A}
V_2^A=-\f{N}{2g^2} G_2^A(n,n^{'},l,l^{'}) \delta_{w+w^{'}+m+m^{'}}\mbox{~~~~~~~(Figure \ref{va1va2})}
\eeqa

\beqa
\label{G2A}
G_2^A(n,n^{'},l,l^{'})=\sqrt{q}\int dx  \phi_n(x)\phi_{n^{'}}(x) e^{i(l+ l^{'})x} 
\eeqa

\begin{figure}[h]
\begin{center}
\begin{psfrags}
\psfrag{c1}[][]{$A^3_x(m,l)$}
\psfrag{c2}[][]{$A^3_x(m^{'},l^{'})$}
\psfrag{a11}[][]{$\tilde{C}_{m,n}$}
\psfrag{a12}[][]{$\tilde{C}_{m^{'},n^{'}}$}
\psfrag{v2}[][]{$\tilde{V}_2^A$}
\includegraphics[width= 4cm,angle=0]{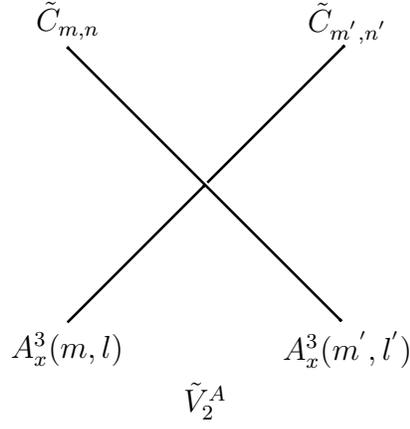}
\caption{$\tilde{V}_2^A$ vertex}
\label{tva2}
\end{psfrags}
\end{center}
\end{figure}

\beqa
\label{V2A}
\tilde{V}_2^A=-\f{N}{2g^2} \tilde{G}_2^A(n,n^{'},l,l^{'}) \delta_{w+w^{'}+m+m^{'}}\mbox{~~~~~~~(Figure \ref{tva2})}
\eeqa

\beqa
\label{G2At}
\tilde{G}_2^A(n,n^{'},l,l^{'})=\sqrt{q}\int dx \tilde{\phi}_n(x)\tilde{\phi}_{n^{'}}(x) e^{i(l+ l^{'})x} 
\eeqa

\begin{figure}[h]
\begin{center}
\begin{psfrags}
\psfrag{c}[][]{$~~A^3_x(m,l)$}
\psfrag{a1}[][]{$C_{m,n}$}
\psfrag{a3}[][]{$C_{m^{'},n^{'}}$}
\psfrag{v5}[][]{$V^A_3$}
\psfrag{c1}[][]{$\tilde{C}_{m,n}$}
\psfrag{c2}[][]{$\tilde{C}_{m^{'},n^{'}}$}
\psfrag{v5p}[][]{$\tilde{V}^A_3$}
\includegraphics[width= 10cm,angle=0]{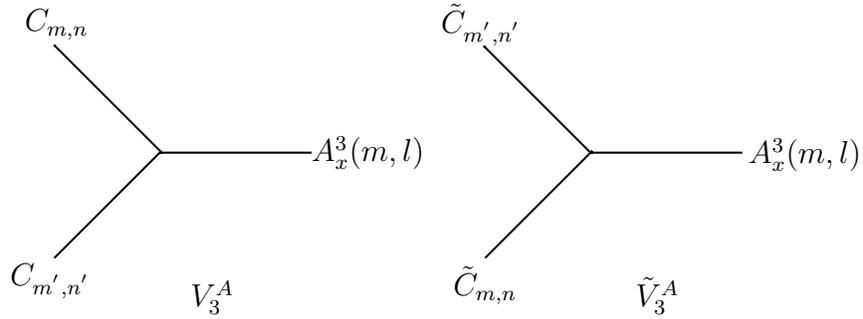} 
\caption{$V^A_3$ and $\tilde{V}^{A}_3$ vertices}
\label{va3tva3}
\end{psfrags}
\end{center}
\end{figure}

\beqa
\label{V3A}
V_3^A=-\f{N^{3/2}}{g^2} G_3^A(n,n^{'},l) \beta\delta_{w+m+m^{'}}\mbox{~~~~~~~(Figure \ref{va3tva3})}
\eeqa

\beqa
\label{G3A}
G_3^A(n,n^{'},l)&=& \int dx \left[\partial_x (\phi_{n^{'}}(x))\phi_n(x)-\partial_x (\phi_n(x))\phi_{n^{'}}(x)\right.\\\nonumber
&+& \left.qx A_{n^{'}}(x)\phi_n(x)-qx A_{n}(x)\phi_{n^{'}}(x)\right] e^{ilx}
\eeqa

\beqa
\label{V3At}
\tilde{V}_3^A=-\f{N^{3/2}}{g^2} \tilde{G}_3^A(n,n^{'},l) \beta\delta_{w+m+m^{'}}\mbox{~~~~~~~(Figure \ref{va3tva3})}
\eeqa

\beqa
\label{G3At}
\tilde{G}_3^A(n,n^{'},l)&=& \int dx \left[\partial_x (\tilde{\phi}_{n^{'}}
(x))\tilde{\phi}_n(x)-\partial_x (\tilde{\phi}_n(x))\tilde{\phi}_{n^{'}}(x)\right.\\\nonumber
&+& \left.qx \tilde{A}_{n^{'}}(x)\tilde{\phi}_n(x)-qx \tilde{A}_{n}(x)\tilde{\phi}_{n^{'}}(x)
\right] e^{ilx}
\eeqa

\begin{figure}[h]
\begin{center}
\begin{psfrags}
\psfrag{c}[][]{~~$A^3_x(m,l)$}
\psfrag{a1}[][]{$C_{m,n}$}
\psfrag{a3}[][]{$\tilde{C}_{m^{'},n^{'}}$}
\psfrag{v5}[][]{$\tilde{V}^{A'}_3$}
\psfrag{c1}[][]{$\Phi^{(1,2)}_I(m,n)$}
\psfrag{c2}[][]{$\Phi^{(1,2)}_I(m^{'},n^{'})$}
\psfrag{v5p}[][]{$V^A_4$}
\includegraphics[width= 10cm,angle=0]{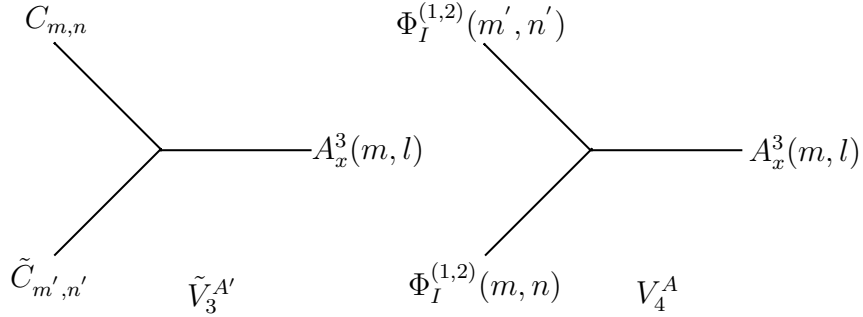} 
\caption{$\tilde{V}^{A'}_3$ and $V^A_4$ vertices}
\label{tvap3va4}
\end{psfrags}
\end{center}
\end{figure}

\beqa
\label{V3At}
\tilde{V}_3^{A'}=-\f{N^{3/2}}{g^2} \tilde{G}_3^{A'}(n,n^{'},l) \beta\delta_{w+m+m^{'}}\mbox{~~~~~~~(Figure \ref{tvap3va4})}
\eeqa

\beqa
\label{G3At}
\tilde{G}_3^{A'}(n,n^{'},l)&=& \int dx \left[\partial_x (\tilde{\phi}_{n^{'}}(x))\phi_n(x) - \partial_x (\phi_n(x))\tilde{\phi}_{n^{'}}(x)\right.\\\nonumber 
&+& \left. qx \tilde{A}_{n^{'}}(x)\phi_n(x)- \tilde{\phi}_{n^{'}}(x)A_n(x)\right] e^{ilx}
\eeqa

\beqa
\label{V4A}
V_4^A=-\f{N^{3/2}}{g^2} G_4^A(n,n^{'},l) \beta\delta_{w+m+m^{'}}\mbox{~~~~~~~(Figure \ref{tvap3va4})}
\eeqa

\beqa
\label{G4A}
G_4^A(n,n^{'},l)&=&\int dx e^{-qx^2/2}\left[\partial_x (e^{-qx^2/2}H_{n^{'}}(x))H_n(x)\right.\\\nonumber
&-& \left.\partial_x (e^{-qx^2/2}H_n(x))H_{n^{'}}(x)\right] e^{ilx}
\eeqa

\begin{figure}[h]
\begin{center}
\begin{psfrags}
\psfrag{c}[][]{~~$A^3_x(m,l)$}
\psfrag{a1}[][]{$\theta_i(m,n)$}
\psfrag{a3}[][]{$\theta_j{m^{'},n^{'}}$}
\psfrag{v5}[][]{$\tilde{V}^{A1}_f$}
\psfrag{c1}[][]{$\theta_i(m,n)$}
\psfrag{c2}[][]{$\theta^{*}_j(m^{'},n^{'})$}
\psfrag{v5p}[][]{$V^{A2}_f$}
\includegraphics[width= 10cm,angle=0]{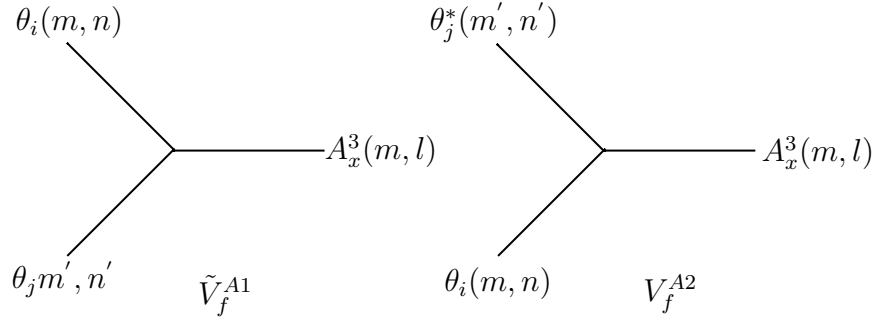} 
\caption{$V^{A1}_f$ and $V^{A2}_f$ vertices}
\label{va1fva2f}
\end{psfrags}
\end{center}
\end{figure}

\beqa
\label{VfA1}
V_f^{A1}=i\f{N}{g^2} G_4^{A1}(n,n^{'},l) \beta\delta_{w+m+m^{'}}\mbox{~~~~~~~(Figure \ref{va1fva2f})}
\eeqa

\beqa
\label{GA1f}
G_f^{A1}(n,n^{'},l)=\sqrt{q}\int dx \left[L_n(x) L_{n^{'}}(x) - R_n(x) R_{n^{'}}(x)\right] e^{ilx}
\eeqa

\beqa
\label{VfA2}
V_f^{A2}=i\f{N}{g^2} G_4^{A2}(n,n^{'},l) \beta\delta_{w+m+m^{'}}\mbox{~~~~~~~(Figure \ref{va1fva2f})}
\eeqa

\beqa
\label{GA2f}
G_f^{A2}(n,n^{'},l)=\sqrt{q}\int dx \left[L_n(x) L_{n^{'}}^{*}(x) - R_n(x) R_{n^{'}}^{*}(x)\right]e^{ilx}
\eeqa

\section{Matsubara Sums}\label{matsubara}

In this appendix we show some sample computations showing sums over Matsubara frequencies. Let us evaluate the sum over $m$ in the propagator

\begin{equation}\label{summ}
\f{1}{\beta}\sum^{\infty}_{n=2,m = -\infty} \f{1}{\o^2_m + \lambda_n}   
\end{equation}

where $\o_m=2m\pi/\beta$

Following \cite{kapusta} we convert the sum over $m$ into a contour integral as follows. 
By writing $p_0=i\o_m$, define the function

\begin{equation}
f(p_0) = -\f{1}{p_0^2 - \lambda_n} 
\end{equation}

The function $f(p_0)$ does not have poles on the imaginary axis. We multiply it by a function with simple poles on the imaginary axis at values $p_0= \f{2 im\pi}{\beta}$ and analytic and bounded otherwise. A function with this property is $\coth(p_0 \beta/2)$.  
The sum over $m$ in (\ref{summ}) can now be reproduced from the contour integral

\beqa
\label{intsum}
 \f{1}{2 \pi i \beta}\oint dp_0 \left(\f{\beta}{2}\right)\coth\left(\f{p_0 \beta}{2}\right)f(p_0)
\eeqa

\begin{figure}[h]
\begin{center}
\begin{psfrags}
\psfrag{p}[][]{$p_0$}
\psfrag{g}[][]{$\Gamma$}
\psfrag{c1}[][]{$\Gamma_1$}
\psfrag{c2}[][]{$\Gamma_2$}
\psfrag{c}[][]{$c_n$}
\psfrag{a}[][]{(a)}
\psfrag{b}[][]{(b)}
\psfrag{c7}[][]{$ $}
\psfrag{c8}[][]{$ $}
%\psfrag{c6}[][]{C1}
%\psfrag{c7}[][]{C2}
\includegraphics[width= 14cm,angle=0]{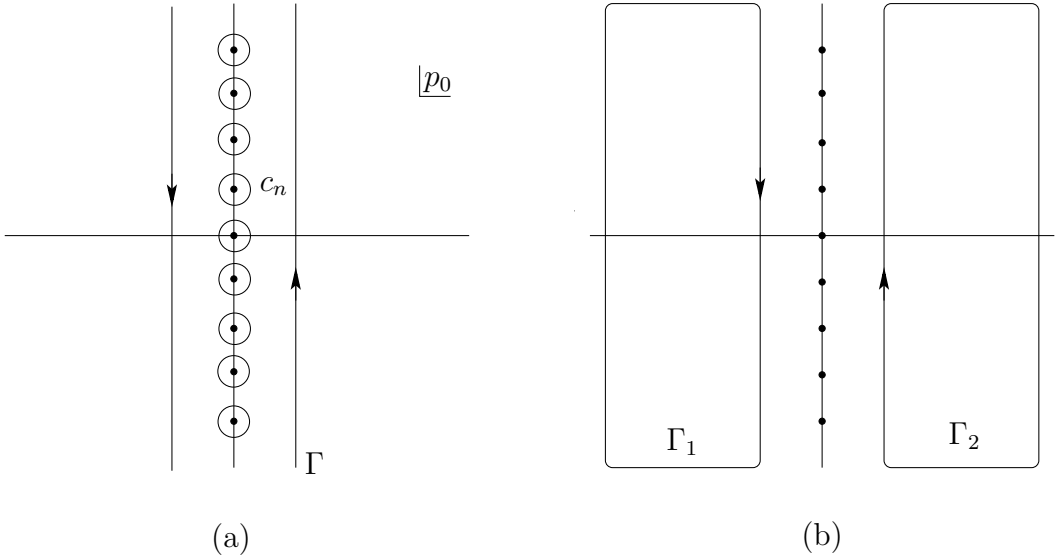} 
\caption{Contours for doing the Matsubara sums.}
\label{contours}
\end{psfrags}
\end{center}
\end{figure}

where the contour is the sum over the contours $C_n$ shown in Figure \ref{contours}(a). The $C_n$'s can now be deformed into the contour $\Gamma$. The contour integral can then be written in terms of line integrals as

\beqa
\label{contint1}
\f{1}{2 \pi i}\int^{-i \infty - \epsilon}_{i \infty-\epsilon}dp_0 f(p_0)\left(-\hf-\f{1}{e^{-p_0 \beta} -1}\right)+ 
\f{1}{2 \pi i}\int^{i \infty + \epsilon}_{-i \infty+\epsilon}dp_0 f(p_0)\left(\hf + \f{1}{e^{p_0 \beta} -1}\right)
\eeqa

since the function $f(p_0)$ vanishes for $p_0=\pm i\infty$.
%Putting $p_0 \rightarrow -p_0$ in the first integral yeilds

%\beqa
%\label{contint2}
%\sum^{\infty}_{m = -\infty} \f{1}{\o^2_m + \lambda_n} &=& \f{1}{4 \pi i}\int^{i \infty}_{-i \infty} dp_0  \left(f(p_0) + f(-p_0)\right)\nonumber%\\
%&+& \f{1}{4 \pi i}\int^{i \infty+ \epsilon}_{-i \infty+ \epsilon} dp_0  \left(f(p_0) + f(-p_0)\right)\f{1}{e^{p_0 \beta} - 1}
%\eeqa

In (\ref{contint1}) the frequency sum separates into the zero-temperature part and the temperature-dependent part for both the integrals. The line integrals above can now be evaluated using the contour integrals over the contours $\Gamma_1$ and $\Gamma_2$ in Figure \ref{contours}(b). This is because the line integral over rest of the rectangle vanishes when the length of the sides are taken to infinity. These contour integrals now have contributions only from the poles of $f(p_0)$ at $p_0=\pm \sqrt{\lambda_n}$. Thus, 

\beqa
\label{res}
\f{1}{\beta}\sum^{\infty}_{n=2,m = -\infty} \f{1}{\o^2_m + \lambda_n}  
= \sum_{n=2} \f{1}{\sqrt{\lambda_n}}\left(\hf + \f{1}{e^{\sqrt{\lambda_n}\beta}-1}\right)
\eeqa

Similarly using the above formula, the several bosonic propagators upon being summed over the Matsubara frequency $\o_m$ are,

\beqa
\label{propgamma}
\f{1}{\beta}\sum^{\infty}_{n=2,m = -\infty} \f{1}{\o^2_m + \gamma_n}
=\sum^{\infty}_{n=2}\f{1}{\sqrt{\gamma_n}} \left(\hf + \f{1}{e^{\sqrt{\gamma_n}\beta} - 1}\right)
\eeqa

\beqa
\label{propl}
\f{1}{\beta}\sum^{\infty}_{m = -\infty} \int^\infty_{-\infty} \f{dl}{2 \pi \sqrt{q}}\f{1}{\o^2_m + l^2}
=\int^\infty_{-\infty} \f{dl}{2 \pi \sqrt{q}}\f{1}{l} \left(\hf + \f{1}{e^{l\beta} - 1}\right)
\eeqa

\beqa
\label{propgammalambda}
&&\f{1}{\beta}\sum^{\infty}_{n=2,n^{'}=2,m = -\infty} \f{1}{(\o^2_m + \gamma_n)(\o^2_m + \lambda_{n^{'}})}=\nonumber\\
&&\sum^{\infty}_{n=2, n^{'}=2}
\f{1}{{\gamma_n - \lambda_{n^{'}}}} \left(\f{1}{\lambda_{n^{'}}}\left(\hf + \f{1}{e^{\sqrt{\lambda_{n^{'}}}\beta} - 1}\right)
-\f{1}{\sqrt{\gamma_n}}\left(\hf + \f{1}{e^{\sqrt{\gamma_n}\beta} - 1}\right)\right)
\eeqa

\beqa
\label{proplambdal}
&&\f{1}{\beta}\sum^{\infty}_{n=2,m = -\infty}\int^\infty_{-\infty} \f{dl}{2 \pi \sqrt{q}} \f{1}{(\o^2_m + l^2)(\o^2_m + \lambda_n)}=\nonumber\\
&&\sum^{\infty}_{n=2} \int^\infty_{-\infty} \f{dl}{2 \pi \sqrt{q}}
\f{1}{{l^2 - \lambda}} \left(\f{1}{\lambda_n}\left(\hf + \f{1}{e^{\sqrt{\lambda_n}\beta} - 1}\right)
-\f{1}{l}\left(\hf + \f{1}{e^{l \beta} - 1}\right)\right)
\eeqa

\beqa
\label{proplgamma}
&&\f{1}{\beta}\sum^{\infty}_{n=2,m = -\infty}\int^\infty_{-\infty} \f{dl}{2 \pi \sqrt{q}} \f{1}{(\o^2_m + l^2)(\o^2_m + \gamma_n)}=\nonumber\\
&&\sum^{\infty}_{n=2} \int^\infty_{-\infty} \f{dl}{2 \pi \sqrt{q}}
\f{1}{{l^2 - \gamma_n}} \left(\f{1}{\gamma_n}\left(\hf + \f{1}{e^{\sqrt{\gamma_n}\beta} - 1}\right) 
-\f{1}{l}\left(\hf + \f{1}{e^{l \beta} - 1}\right)\right)
\eeqa

The last three propagators in \label{propmatsu} are mixed propagators. So we have decomposed them into partial fractions and then and have done the sum separately. 

The fermions due to their anti-periodic boundary conditions along the Euclidean time direction have have their propagators with $\o_m = \f{(2m+1)\pi}{\beta}$. The sum over the odd integers can be performed by converting the sum into a contour integral as above. The only change here is that we must introduce $\tanh(p_0\beta/2)$. Thus,

\begin{eqnarray}\label{summf}
\f{1}{\beta}\sum^{\infty}_{m = -\infty} \f{1}{\o^2_m + \lambda_n^{'}}&=& \f{1}{2 \pi i \beta}\oint dp_0 \left(\f{\beta}{2}\right)\tanh\left(\f{p_0 \beta}{2}\right)f(p_0)\non
&=&\f{1}{\sqrt{\lambda^{'}_n}}\left(\hf - \f{1}{e^{\sqrt{\lambda^{'}_n}\beta}+1}\right)
\end{eqnarray}

Similarly using this result we can do the sum over the Matsubara frequencies for the following,

\beqa
\label{propferm}
&&\sum^{\infty}_{n=0,m=-\infty}\int^{\infty}_{-\infty} \f{dl}{2 \pi \sqrt{q}}\f{1}{(i\o_m + \sqrt{\lambda{'}_n})(i\o_m \pm l)}= \nonumber\\ 
&&-\sum^{\infty}_{n=0,m=-\infty}\int^{\infty}_{-\infty} \f{dl}{4 \pi \sqrt{q}} \left(\f{1}{(\o^2_m + \lambda^{'}_n)} + \f{1}{(\o^2_m + l^2)}
-\f{l^2 + \lambda_n}{(\o^2_m + \lambda^{'}_n)(\o^2_m + l^2)}\right)\nonumber\\
\eeqa

We can now do the sum over each of the terms separately, which gives

\begin{equation}
\label{fermpropagator}
\sum_n \int \frac{dl}{2\pi \sqrt{q}} 
\left(\f{-\beta \tanh \left(\frac{\beta l}{2}\right)+\beta \tanh \left(\f{1}{2} \beta \sqrt{\lambda^{'}_n}\right)}{2 \left(l-\sqrt{\lambda^{'}_n}\right)}\right) 
\end{equation}

\end{document}